%% file: main.tex
\documentclass[11pt]{article}
\usepackage[T1]{fontenc}
\usepackage{lmodern}
\usepackage[margin=1in]{geometry}
\usepackage{amsmath,amssymb,amsthm,mathtools,mathrsfs,booktabs,array}
\usepackage[dvipsnames]{xcolor}
\usepackage{microtype}
\usepackage{xurl}
\usepackage{hyperref}
\hypersetup{colorlinks=true,linkcolor=MidnightBlue,citecolor=MidnightBlue,
  urlcolor=MidnightBlue,pdftitle={Determinant boundaries and gaps in determinantal complexity},
  pdfauthor={Karthik Sheshadri}}
\newtheorem{theorem}{Theorem}[section]
\newtheorem{lemma}[theorem]{Lemma}
\newtheorem{proposition}[theorem]{Proposition}
\newtheorem{corollary}[theorem]{Corollary}
\theoremstyle{definition}

\theoremstyle{remark}

\newcommand{\C}{\mathbb C}
\newcommand{\Mat}{\operatorname{Mat}}
\newcommand{\Sym}{\operatorname{Sym}}
\newcommand{\Skew}{\operatorname{Skew}}
\newcommand{\GL}{\operatorname{GL}}
\newcommand{\adj}{\operatorname{adj}}
\newcommand{\tr}{\operatorname{tr}}
\newcommand{\rank}{\operatorname{rank}}

\newcommand{\Sing}{\operatorname{Sing}}
\newcommand{\sym}{\operatorname{sym}}
\newcommand{\Trans}{\operatorname{Trans}}
\newcommand{\dc}{\operatorname{dc}}
\newcommand{\bdc}{\underline{\operatorname{dc}}}
\newcommand{\rdc}{\operatorname{rdc}}
\newcommand{\Stab}{\operatorname{Stab}}
\newcommand{\Orb}{\mathcal O}
\newcommand{\Pf}{\operatorname{Pf}}
\newcommand{\per}{\operatorname{per}}
\newcommand{\coker}{\operatorname{coker}}
\newcommand{\depth}{\operatorname{depth}}
\newcommand{\pd}{\operatorname{pd}}
\newcommand{\Cl}{\operatorname{Cl}}
\newcommand{\Spec}{\operatorname{Spec}}
\newcommand{\Ext}{\operatorname{Ext}}
\newcommand{\Hom}{\operatorname{Hom}}
\newcommand{\im}{\operatorname{im}}

\title{Determinant boundaries and gaps\\in determinantal complexity}
\author{Karthik Sheshadri}
\date{2 November 2026}
\begin{document}
\maketitle

\begin{abstract}
In 2013, Landsberg, Manivel, and Ressayre found a component of the
boundary of the determinant orbit with the following property: for odd
$n\ge3$, its polynomial is a limit of $n\times n$ determinants, but cannot
itself be written as an $n\times n$ determinant. Its border determinantal
complexity (DC) is therefore smaller than its exact DC.

This paper proves that the difference can grow with $n$. For every
$n\ge10$, we give an explicit homogeneous polynomial $f_n$ of degree
$n$ whose orbit closure is a boundary component and for which
\[
 \underline{\operatorname{dc}}(f_n)=n,
 \qquad \operatorname{dc}(f_n)\ge2n-1.
\]
An exact representation thus needs at least $2n-1$ rows, while $n$ rows
suffice if limits are allowed. The same bound holds for every odd
$n\ge3$ using the unpadded odd-skew forms.

The usual orbit-closure formulation of geometric complexity theory
allows these limits. We also prove a bound for every degree-$n$
boundary point that essentially uses all $n^2$ variables: its exact DC
is at least $n+2$. We construct three families of pairwise distinct
boundary components. Together with the traceless determinant, they
give at least $\lfloor(n-1)/2\rfloor+\lfloor n/3\rfloor$ components
for every $n\ge3$, with further components counted explicitly below.

The $2n-1$ bound comes from one degree lemma. Suppose an affine matrix
represents an irreducible homogeneous polynomial $f$ of degree $d$, and
its cokernel on $f=0$ is free of rank one. Its adjugate then factors
as the product of a right-kernel vector and a left-kernel vector.
Each vector needs a coordinate of degree at least $d-1$, whereas a
cofactor of a size-$m$ matrix has degree at most $m-1$. Hence
$m\ge2d-1$. Algebraic branching programs give representations that
attain this lemma's bound. Whether the exact-versus-border gap can
be superlinear remains open.
\end{abstract}

\section{Introduction}\label{sec:introduction}

The permanent of an $n\times n$ matrix $Y$ is the polynomial
\[
 \per_n(Y)=\sum_{\sigma\in S_n}\prod_{i=1}^nY_{i,\sigma(i)},
\]
where $S_n$ is the set of permutations of $\{1,\ldots,n\}$.
The determinant is given by the same sum with the sign of $\sigma$
attached to each term. The determinant can be computed by a polynomial
number of arithmetic operations. The arithmetic complexity of the
permanent is a central open problem in algebraic complexity theory.

Valiant~\cite{Val79} formulated this problem in terms of determinantal
representations. A determinantal representation of a polynomial $F$ is
a square matrix whose determinant is $F$ and whose entries are
affine-linear forms in the variables of $F$, that is, polynomials of
degree at most one. The \emph{determinantal complexity} (DC) of $F$,
denoted by $\dc(F)$, is the smallest size of such a matrix. Valiant
showed that a polynomial computed by an arithmetic formula of size
$s$ has a determinantal representation of size linear in $s$. He also
proved that the permanent is complete for VNP, an algebraic analogue
of NP, in characteristic different from two. Valiant's determinantal
complexity conjecture asks whether $\dc(\per_n)$ grows faster than
every polynomial in $n$.

The \emph{border determinantal complexity} $\bdc(F)$ permits limits.
It is the least $m$ such that $F$ is a coefficientwise limit of
determinants of $m\times m$ matrices with affine-linear entries.
The size $m$ remains fixed during the limit. Every exact representation
is also such a limit, and neither a size-$m$ determinant nor a limit
of size-$m$ determinants can have degree above $m$. Consequently,
for every nonzero polynomial $F$,
\begin{equation}\label{eq:basic-dc-inequalities}
 \deg F\le\bdc(F)\le\dc(F).
\end{equation}
In particular, a lower bound on border DC also gives the same lower
bound on exact DC. Throughout the paper the ground field is $\C$,
and a \emph{form} means a homogeneous polynomial: all of its monomials
have the same total degree.

Mignon and Ressayre~\cite{MR04} proved the quadratic lower bound
$\dc(\per_n)\ge n^2/2$ in characteristic zero. Landsberg, Manivel,
and Ressayre~\cite{LMR} proved that the same bound holds for border DC.
More recently, OpenAI's September 2026 preprint~\cite{OAI26}
reports a cubic lower bound, $\bdc(\per_n)=\Omega(n^3)$ over $\C$.
Related lower bounds from smooth restrictions and polar counts
appear in the author's earlier work~\cite{She26conormal,She26slices}.
In addition to their lower bound for the permanent, Landsberg,
Manivel, and Ressayre exhibit forms for which the second inequality
in~\eqref{eq:basic-dc-inequalities} is strict. The present paper
concerns this distinction between exact and border complexity.

Geometric complexity theory (GCT), introduced by Mulmuley and
Sohoni~\cite{MS01}, studies orbit closures associated with
polynomials such as the determinant. Let
\[
 \Orb_n=\GL_{n^2}\cdot\det_n
\]
be the orbit of the determinant under invertible linear changes of
its $n^2$ variables. Its closure $\overline{\Orb_n}$ also contains
their limits, and the \emph{determinant boundary} is
\[
 \partial\overline{\Orb_n}=\overline{\Orb_n}\setminus\Orb_n.
\]
Closures are taken in the Zariski topology; over $\C$ the orbit
closure also agrees with the closure in the usual complex topology.
A polynomial is \emph{concise} in its ambient variables if no
invertible linear change of coordinates reduces their number.
Equivalently, its first partial derivatives are linearly independent.

Every nonzero degree-$n$ form in $\overline{\Orb_n}$ has border DC
$n$. If it is concise and lies on the boundary, its exact DC is
larger than $n$. Indeed, a size-$n$ affine representation would have
homogeneous degree-$n$ part $\det L$, where $L$ is its linear part.
Conciseness in $n^2$ variables would force this linear substitution
to be invertible, placing the form in $\Orb_n$ itself. Thus the
orbit-closure formulation distinguishes border complexity from the
size required by an exact representation.

Landsberg, Manivel, and Ressayre~\cite[Proposition~3.5.1]{LMR}
constructed such a concise boundary component for every odd
$n\ge3$. Their construction starts with odd skew-symmetric
matrices. Chen~\cite[Theorem~5.5.3]{Chen} proved that adding one row
and column gives another component. H\"uttenhain~\cite{Hut} treats
the traceless-determinant component and a further concise component
at size four. The leading determinant coefficient of a singular
matrix space gives a more general source of boundary points
\cite[Section~6.7.1]{LandsbergBook}. The bounded-rank spaces of
Huang and Landsberg~\cite{HL} and the exterior-multiplication spaces
studied by Landsberg and Manivel~\cite{LMConstant} provide further
starting spaces for this construction.

These examples lead to two questions. How large can the difference
between exact and border DC be on an explicit boundary component?
How many distinct components can be obtained by varying the starting
matrix space and adding rows and columns? We prove a linear lower
bound for the first question and a uniform component count for the
second.

\subsection{Our contributions}

Our main result is the following.

\begin{theorem}\label{thm:intro-gap}
For every $n\ge10$, there is an explicit homogeneous form $f_n$ of
degree $n$, concise in $n^2$ variables, whose orbit closure is a
determinant-boundary component and which satisfies
\[
 \bdc(f_n)=n,\qquad \dc(f_n)\ge2n-1.
\]
One may take $f_n=\mathcal K_{2,n-10}$, defined in
Section~\ref{subsec:ten-construction}. For every odd $n\ge3$,
the unpadded odd-skew form $Q_{n,0}$ also has these properties.
\end{theorem}

The defining determinant limits give $\bdc(f_n)\le n$, and the
degree gives the reverse inequality. The exact lower bound comes
from a property of the whole hypersurface $f_n=0$: every polynomial
matrix with determinant $f_n$ has a free rank-one cokernel over its
coordinate ring. For the unpadded odd-skew forms, this follows from
unique factorization in that ring. For the Koszul forms, it follows
from a calculation of the divisor class group and the depth of the
two possible nonfree cokernels.

The passage from freeness to a size bound is a general degree
argument. Suppose an affine matrix $M$ represents an irreducible
homogeneous polynomial of degree $d$ and has a free rank-one
cokernel on its zero set. Its adjugate factors there as $uv^T$,
where $u$ and $v$ are nonzero right- and left-kernel vectors. Each
vector has a coordinate of degree at least $d-1$. Since a cofactor
of a size-$m$ affine matrix has degree at most $m-1$, it follows
that $m\ge2d-1$. This is Theorem~\ref{thm:degree}.
The hypothesis is proved for every exact representation of the forms
in Theorem~\ref{thm:intro-gap}; the resulting bound therefore applies
to unrestricted affine determinantal representations.

In particular, the exact-minus-border gap is at least $n-1$ and
\[
 \frac{\dc(f_n)}{\bdc(f_n)}\ge2-\frac1n.
\]
For example, the degree-ten construction has border DC ten and exact
DC at least nineteen. These are lower bounds on the exact complexity;
the theorem does not determine its value.

We next construct three families of boundary components. They arise
from odd skew-symmetric matrices, the block spaces of Huang and
Landsberg, and middle Koszul maps. Each starting space can be padded
by any number of rows and columns.

\begin{theorem}\label{thm:intro-components}
Every form in the three families defined in
Section~\ref{sec:construction-guide} is concise in $n^2$ variables,
and its orbit closure is a determinant-boundary component. For fixed
$n$, the components obtained from different listed parameters are
pairwise distinct. Including the traceless-determinant component,
the boundary has at least
\begin{equation}\label{eq:intro-count}
 \left\lfloor\frac{n-1}{2}\right\rfloor
 +\left\lfloor\frac n3\right\rfloor+\kappa(n),
 \qquad
 \kappa(n)=\#\left\{p\ge2:\binom{2p+1}{p}\le n\right\},
\end{equation}
irreducible components for every $n\ge3$.
\end{theorem}

The odd-skew family includes every odd starting size $r$ from three
to $n$, with $n-r$ added rows and columns. It therefore unifies the
unpadded construction of~\cite{LMR}, Chen's one-row extension, and
arbitrary padding of the smallest skew block. The block family has
starting sizes $6,9,12,\ldots$, and the Koszul family has starting
sizes $10,35,126,\ldots$. The count in~\eqref{eq:intro-count} is a
lower bound; it does not classify the entire boundary.
The proofs identify singular or higher-multiplicity loci determined
by each polynomial and then compute its stabilizer, the group of
linear changes of variables preserving that polynomial. The same
loci distinguish the components. We also obtain general criteria
for componenthood and for preservation of componenthood under
padding.

Our final general lower bound excludes exact complexity $n+1$ for
every concise boundary point.

\begin{theorem}\label{thm:intro-forbidden}
Let $n\ge3$, and let $f$ be a concise degree-$n$ form on the
determinant boundary in $n^2$ variables. Then
\[
 \bdc(f)=n,\qquad \dc(f)\ge n+2.
\]
\end{theorem}

Theorem~\ref{thm:gap} proves the stronger statement that a
homogeneous degree-$n$ form with at least $n^2-n+2$ essential
variables cannot have exact DC equal to $n+1$. For the padded
odd-skew forms with starting size three, an explicit upper bound
gives
\[
 n+2\le\dc(Q_{3,n-3})\le n+4.
\]
Thus the bound $2n-1$ in Theorem~\ref{thm:intro-gap} applies to the
unpadded odd-skew forms and all padded Koszul forms, rather than to
every padded odd-skew form.

\subsection{Organization of the paper}

Section~\ref{sec:overview} gives a proof overview using the smallest
odd-skew example. Section~\ref{sec:construction-guide} defines the
three families, and Section~\ref{sec:kernel-degree} proves the
general degree bound. Sections~\ref{sec:factorial}
and~\ref{sec:koszul-universal-regularity} establish the cokernel
property needed for Theorem~\ref{thm:intro-gap}.
Section~\ref{sec:dc} treats general concise boundary points.
The component proofs occupy Sections~\ref{sec:odd-family},
\ref{sec:higher}, and~\ref{sec:koszul}, followed by the component
count and general padding criteria.
The remaining sections develop bounds from affine level sets,
a reduction to regular representations of the determinant, and a
criterion for rational matrix realizations. They also explain the
limitations of these methods for obtaining a superlinear gap.
Section~\ref{sec:questions} records the open questions.

\input{proof_overview}
\input{construction_guide}
\input{kernel_degree}
\input{factoriality}
\input{koszul_cokernels}
\input{forbidden_value}
\input{geometry}

\input{affine_planes}
\input{regular_determinant}
\input{residue_criterion}
\input{local_limits}

\section{Open questions}\label{sec:questions}

The exact complexities of the forms in Theorem~\ref{thm:intro-gap}
remain undetermined. Even the padded odd-skew family with starting
size three leaves the interval
\[
 n+2\le\dc(Q_{3,n-3})\le n+4.
\]
Determining these values would give a concrete comparison between
the geometry of a boundary component and the size of its smallest
affine representation.

The principal asymptotic question is whether there are explicit
forms with
\[
 \bdc(f_n)=n,\qquad \dc(f_n)=\omega(n).
\]
The kernel-degree bound is sharp for individual representations
with free cokernel, so its hypotheses alone do not imply
superlinear growth. A stronger argument must use additional
properties of the target forms, such as the number of coordinates
required by compatible kernel vectors or relations among them.

The rational-matrix criterion in Theorem~\ref{thm:residue} gives an
obstruction for a prescribed realization of the Koszul projector.
An arbitrary scalar determinantal representation need not induce
that realization. Moreover, linear-size constructions can agree
with the Koszul form in its base value, all first coefficient
derivatives, and, in the stated range, its bidegree. These data
therefore do not suffice for a superlinear scalar lower bound.
An extension would need to use further simultaneous coefficient
identities or a quantitative constraint on the corresponding
kernel modules.

A sufficient intermediate result would be a superlinear lower
bound on $\rdc(\det_q)$, the minimum size of a representation of
$\det_q$ that is regular at the zero matrix.
Section~\ref{sec:regular-det} transfers such a bound to padded
Koszul forms and gives an equivalent spectral formulation.
This problem remains open.

The component count also leaves a separate classification problem.
Theorems~\ref{thm:all-corank} and~\ref{thm:general-padding} give
criteria for further singular matrix spaces to produce boundary
components. Finding additional spaces satisfying their recognition
and symmetry hypotheses may improve the count. The count does not provide a classification of the entire
determinant boundary.

\section*{Acknowledgment of assistance}
ChatGPT (OpenAI) assisted with mathematical exploration, proof drafting,
symbolic and finite-field checks, and preparation of the manuscript.

\appendix
\input{certificates}

\input{references}
\end{document}

%% file: proof_overview.tex
\section{Proof overview}\label{sec:overview}

The smallest odd-skew form illustrates both the determinant limit and
the degree argument for exact complexity. Let $x=(x_1,x_2,x_3)^T$,
let $S$ be a symmetric $3\times3$ matrix with six independent entries,
and put
\[
 A(x)=\begin{pmatrix}
 0&-x_3&x_2\\
 x_3&0&-x_1\\
 -x_2&x_1&0
 \end{pmatrix}.
\]
Then $A(x)x=0$ and $\adj A(x)=xx^T$. Expanding the determinant gives
\begin{equation}\label{eq:intro-small-example}
 \det(A(x)+tS)=t\,x^TSx+t^3\det S.
\end{equation}
For nonzero $t$, the entries of $A(x)+tS$ are an invertible linear
function of the nine variables, since a matrix decomposes uniquely
into its skew-symmetric and symmetric parts. Multiplying one row by
$t^{-1}$ therefore gives a polynomial in the determinant orbit
$\Orb_3$, and
\[
 f=x^TSx=\lim_{t\to0}t^{-1}\det(A(x)+tS).
\]
Thus $\bdc(f)=3$: the limit proves the upper bound, and the degree
proves the lower bound.

We next outline why every exact representation of $f$ has size at
least five. Let $M$ be an $m\times m$ affine matrix with $\det M=f$,
and put $R=\C[x,S]/(f)$. The matrix $M$ is singular over $R$.
The main algebraic input is that
\[
 \coker_R M=R^m/MR^m\simeq R.
\]
This means that the cokernel has one global generator, with no
nonzero relation over $R$. It implies that every row annihilating
$M$ is a multiple of one row $v^T$. Since the rows of the adjugate
annihilate $M$ modulo $f$, we obtain
\[
 \adj M=uv^T\quad\text{over }R,
 \qquad Mu=0,\quad v^TM=0,
\]
with $u,v$ nonzero.

Suppose all coordinates of $u$ had degree at most one. Lift them to
polynomials of degree at most one before taking the quotient by $f$.
The entries of $Mu$ then have degree at most two and are divisible
by the cubic $f$. Hence they vanish identically. This is impossible,
since a matrix with nonzero determinant has no nonzero kernel over
the field of rational functions. Therefore $u$ has a coordinate of
degree at least two. The same holds for $v$.

Degrees add in the graded domain $R$. Consequently some entry of
$uv^T$, and hence of $\adj M$, has degree at least four. A cofactor
of an affine matrix of size $m$ has degree at most $m-1$, so
$m\ge5$. For an irreducible homogeneous polynomial of degree $d$,
the same argument gives a coordinate of degree at least $d-1$ in
each vector and yields $m\ge2d-1$. Section~\ref{sec:kernel-degree}
proves this statement in full.

It remains to justify the cokernel hypothesis for every exact
representation. The polynomial $x^TSx$ is linear in the entries of
$S$, and its coefficient polynomials are the products $x_ix_j$.
They vanish simultaneously only at $x=0$, a set of codimension three.
Section~\ref{sec:factorial} proves that, more generally, a polynomial
linear in one group of variables defines a factorial coordinate
ring when its coefficients have a common zero set of codimension
at least three. Every determinant cokernel on such a hypersurface
is free. This proves the hypothesis for $x^TSx$ and for all unpadded
odd-skew forms.

The Koszul forms require a different freeness argument. Away from
the zero value of their starting variables, they become ordinary
determinants in the coefficient variables. A classification for the
generic determinant leaves three possible rank-one cokernels. One
is free. The other two extend to modules whose depths are too small
on the omitted set, so neither can be the cokernel of a determinant
representation. Section~\ref{sec:koszul-universal-regularity}
proves this assertion and establishes the lower bound at every
$n\ge10$.

Finally, the component statements require an independent geometric
argument. The defining limits give points of the determinant orbit
closure. To show that their orbit closures are entire boundary
components, we compute their stabilizer dimensions. We distinguish
the components using singular and multiplicity loci that are
preserved under every invertible linear change of variables.

%% file: construction_guide.tex
\section{Three families of boundary forms}\label{sec:construction-guide}

We define the forms in Theorems~\ref{thm:intro-gap}
and~\ref{thm:intro-components}. Each family is obtained from a
singular matrix space by taking a leading determinant coefficient.
The construction of this coefficient is standard
\cite[Section~6.7.1]{LandsbergBook}. The component and distinctness
statements are proved in Sections~\ref{sec:odd-family},
\ref{sec:higher}, and~\ref{sec:koszul}.

\subsection{Leading coefficients and padding}\label{subsec:padding-guide}

Let $L\subset\Mat_r$ be a linear space whose generic rank is
$r-c$, where $c\ge1$. Choose a complementary space $S_L$, so
$\Mat_r=L\oplus S_L$, and let $A$ and $S$ be the generic matrices
of these two spaces. We call $L$ the starting space, or seed space.
For $k\ge0$, put $n=r+k$ and introduce independent matrices
\[
 U\in\Mat_{r,k},\qquad V\in\Mat_{k,r},\qquad Z\in\Mat_k.
\]
Define
\begin{equation}\label{eq:guide-padding-matrix}
 M_t=\begin{pmatrix}A+tS&U\\tV&Z\end{pmatrix},
 \qquad F_{L,k}=[t^c]\det M_t.
\end{equation}
Here $[t^c]$ denotes the coefficient of $t^c$, and the empty blocks
are omitted when $k=0$. There are
$r^2+2rk+k^2=n^2$ independent variables.
The addition of these blocks is called padding.

All coefficients of $t^j$ for $j<c$ vanish. On $\det Z\ne0$, Schur
elimination gives
\[
 \det M_t=\det Z\,\det\bigl(A+t(S-UZ^{-1}V)\bigr).
\]
The vanishing follows from $\rank A\le r-c$, and extends to every
value of the variables because it is a polynomial identity.
The leading coefficient is nonzero. Indeed, at a rank-$r-c$ point
of $L$, every direction in $L$ induces the zero map from the kernel of $A$ to its
cokernel. Since $S_L$ complements $L$ in the full
matrix space, its induced maps between those $c$-dimensional spaces
are arbitrary. A perturbation invertible between them has a nonzero
coefficient of $t^c$.

For $t\ne0$, the entries of $M_t$ are an invertible linear
parametrization of $\Mat_n$. The factor $t^{-c}$ can be absorbed
into one row, so
\[
 F_{L,k}=\lim_{t\to0}t^{-c}\det M_t\in\overline{\Orb_n}.
\]
Every nonzero term of $F_{L,k}$ has total degree $n$. Hence
$\bdc(F_{L,k})=n$. The three choices of $L$ below give the families
used in this paper.

\subsection{Odd skew-symmetric matrices}\label{subsec:odd-construction-guide}

Let $r=2s+1\ge3$, take $L=\Skew_r$, and take the symmetric matrices
as its complement. An odd skew-symmetric matrix is singular, since
$\det A=\det(-A)=-\det A$, and its generic rank is $r-1$.
Thus $c=1$.

The signed maximal Pfaffians form a polynomial kernel vector:
\[
 p_i(A)=(-1)^{i+1}\Pf(A_{\widehat i}),
 \qquad p(A)=(p_1(A),\ldots,p_r(A))^T.
\]
Here $A_{\widehat i}$ denotes the matrix obtained by deleting row
and column $i$. The Pfaffian of an even-size skew matrix is the
signed sum over pairings of its indices; its square is the
determinant. Each $p_i$ has degree $s$, and the Pfaffian identities
give
\[
 Ap=0,\qquad \adj A=pp^T.
\]
Consequently the unpadded form is
\begin{equation}\label{eq:guide-odd}
 Q_{r,0}(A,S)=p(A)^TSp(A)=[t]\det(A+tS).
\end{equation}
For $k\ge1$, the padded form is
\begin{equation}\label{eq:guide-padding-form}
 Q_{r,k}
 =\det Z\,p^TSp-p^TU\adj Z\,Vp.
\end{equation}
This identity follows from Schur elimination of $Z$ and
$\adj A=pp^T$. Although elimination initially requires $Z$ to be
invertible, the resulting polynomial identity holds everywhere.
For $r=3$ and $k=0$, this is the cubic $x^TSx$ from
Section~\ref{sec:overview}.

Theorem~\ref{thm:main} proves that every $Q_{r,k}$ has an orbit
closure that is a boundary component, and that distinct values of
$r$ give distinct components at a fixed final size $n=r+k$.
Section~\ref{sec:factorial} proves the bound
$\dc(Q_{r,0})\ge2r-1$ for the unpadded forms. No such bound is
asserted for every padding: Section~\ref{sec:dc} gives an upper
bound $n+4$ when $r=3$.

\subsection{Block matrix spaces}\label{subsec:block-construction-guide}

For $h\ge2$, let $a,b$ be scalar variables and let $F$ be a generic
$h\times h$ matrix. Set
\[
 A_h(a,b,F)=
 \begin{pmatrix}
 0&aI_h&bI_h\\
 -aI_h&0&F\\
 -bI_h&-F&0
 \end{pmatrix}.
\]
These spaces were constructed by Huang and Landsberg~\cite{HL}.
Their dimension is $h^2+2$, their matrix size is $r=3h$, and their
generic rank is $2h$. The matrices
\[
 K_R=\begin{pmatrix}F\\-bI_h\\aI_h\end{pmatrix},
 \qquad K_L=(F,-bI_h,aI_h)
\]
satisfy $A_hK_R=0$ and $K_LA_h=0$. For a complementary matrix
$S$ and padding blocks $U,V,Z$, define
\[
 H_{h,k}=[t^h]\det\begin{pmatrix}A_h+tS&U\\tV&Z\end{pmatrix}.
\]
Equivalently, the coefficient calculation in Section~\ref{sec:higher}
gives
\[
 H_{h,k}=
 \det\begin{pmatrix}K_LSK_R&K_LU\\VK_R&Z\end{pmatrix}.
\]
This is a polynomial of total degree $n=3h+k$. The last display is
not an affine representation: the degrees of its four blocks are
$3,2,2,1$.
Theorem~\ref{thm:higher} proves componenthood and pairwise
distinctness for every $h\ge2$ and $k\ge0$.
The complexity bounds for this family are given in
Section~\ref{sec:growing-dc-gap}.

\subsection{Middle Koszul maps}\label{subsec:general-koszul-guide}

Let $p\ge2$, put $V=\C^{2p+1}$, and let $E=\Lambda^pV$.
Choose a volume form on $V$. Exterior multiplication by $v\in V$
defines a square matrix $A(v):E\to E^*$ through the pairing
\[
 A(v)(\alpha,\beta)
 =\operatorname{vol}(v\wedge\alpha\wedge\beta),
 \qquad \alpha,\beta\in E.
\]
In the standard basis, rows and columns are indexed by increasing
$p$-element subsets of $\{1,\ldots,2p+1\}$. An entry is zero if
the two subsets intersect. Otherwise it is the missing coordinate
$v_a$, with the sign of the permutation obtained by listing $a$,
the row subset, and the column subset. This gives an explicit
matrix-entry definition of the exterior-multiplication map.

Its size, rank at every nonzero $v$, and corank are
\begin{equation}\label{eq:guide-koszul-parameters}
 r=\binom{2p+1}{p},\qquad
 R_0=\binom{2p}{p},\qquad
 c=\binom{2p}{p-1},\qquad r=R_0+c.
\end{equation}
For the rank calculation, take $v=e_1$. The basis vectors indexed
by subsets containing $1$ lie in the kernel. The remaining basis
vectors are paired invertibly with their complementary subsets in
$\{2,\ldots,2p+1\}$. A change of coordinates in $V$ induces
invertible changes in the exterior powers, so the rank is the same
for every nonzero $v$.

Choose a complement $\mathcal S_p$ to $A(V)$ in $\Mat_r$, let $S$
range over this complement, and define
\[
 \mathcal K_{p,k}
 =[t^c]\det\begin{pmatrix}A(v)+tS&U\\tV_0&Z\end{pmatrix},
 \qquad n=r+k.
\]
The notation $V_0$ distinguishes the lower-left block from the
vector space $V$. The polynomial has degree $R_0$ in $v$ and degree
$c+k$ in the coefficient variables $S,U,V_0,Z$, hence total degree
$n$. Theorem~\ref{thm:koszul-family} proves that its orbit closure
is a boundary component. Section~\ref{sec:koszul-universal-regularity}
proves
\[
 \bdc(\mathcal K_{p,k})=n,
 \qquad \dc(\mathcal K_{p,k})\ge2n-1
\]
for every $p\ge2$ and $k\ge0$.

\subsection{The size-ten construction}\label{subsec:ten-construction}

For $p=2$, the preceding matrix has size ten and depends on
$v_1,\ldots,v_5$. Its rows and columns are indexed by
\[
 12,13,14,15,23,24,25,34,35,45.
\]
For example, the entry rule gives
\[
 A(v)_{12,34}=v_5,\qquad
 A(v)_{12,35}=-v_4,\qquad
 A(v)_{12,13}=0.
\]
The matrix is symmetric and has rank six at every nonzero $v$.
Its four-dimensional kernel at $v=e_1$ is spanned by the pair
coordinates containing $1$. An invertible change of the five
coordinates induces an invertible change of the ten pair coordinates,
with coefficients given by the corresponding $2\times2$ minors.
This verifies rank invariance directly in this case.

An explicit complement is
\begin{equation}\label{eq:guide-koszul-complement}
 \mathcal S=
 \left\{S\in\Mat_{10}:\tr\bigl(A(e_i)^TS\bigr)=0
 \text{ for }i=1,\ldots,5\right\}.
\end{equation}
The five matrices $A(e_i)$ have disjoint supports and six entries
in $\{1,-1\}$ each, so
$\tr(A(e_i)^TA(e_j))=6\delta_{ij}$.
Thus $\mathcal S$ is indeed a complement and has dimension $95$.
One can choose coordinates on it by solving each of the five
equations for one entry with nonzero coefficient and taking the
other $95$ entries freely. The disjoint supports make these choices
independent. The matrix $S$ is not required to be symmetric.

For any $n\ge10$, put $k=n-10$ and set
\begin{equation}\label{eq:guide-fn}
 f_n=\mathcal K_{2,k}
 =[t^4]\det\begin{pmatrix}A(v)+tS&U\\tV_0&Z\end{pmatrix}.
\end{equation}
The number of variables is
\[
 5+95+20k+k^2=(10+k)^2=n^2.
\]
This is an explicit polynomial definition by coefficient extraction
from a finite determinant. It has degree six in $v$ and degree
$k+4$ in the other variables. The general construction gives
\[
 f_n=\lim_{t\to0}t^{-4}
 \det\begin{pmatrix}A(v)+tS&U\\tV_0&Z\end{pmatrix},
 \qquad \bdc(f_n)=n.
\]
The global cokernel theorem and the component calculation complete
the proof of Theorem~\ref{thm:intro-gap}.

\subsection{Component statements}\label{subsec:component-proof-guide}

The parameters of the three families are summarized below. In each
case the final size is the starting size plus $k$.
\begin{center}
\begin{tabular}{llll}
\toprule
Starting space & Size $r$ & Generic corank $c$ & Parameters\\
\midrule
Odd skew-symmetric & odd $r$ & $1$ & $r\ge3$, $k\ge0$\\
Block matrices & $3h$ & $h$ & $h\ge2$, $k\ge0$\\
Middle Koszul & $\binom{2p+1}{p}$ & $\binom{2p}{p-1}$
 & $p\ge2$, $k\ge0$\\
\bottomrule
\end{tabular}
\end{center}

For every member of these families, the component proofs establish
\[
 \dim\Stab(F)=2n^2-1,
 \qquad \dim(\GL_{n^2}\cdot F)=n^4-2n^2+1.
\]
The stabilizer consists of the invertible linear changes of variables
preserving $F$. Since the determinant stabilizer has dimension
$2n^2-2$, its orbit has dimension $n^4-2n^2+2$.
Each constructed orbit closure is therefore a proper irreducible
closed subset of codimension one in $\overline{\Orb_n}$, and hence
a boundary component.

The component distinction is obtained from loci preserved by linear
equivalence. For the odd-skew family, the translation dimension of
the largest singular component recovers the starting size $r$.
For the block family, the span of the corresponding singular
component has a different codimension. The Koszul family has a
singular locus of different codimension, and a higher-multiplicity
locus recovers $p$. The detailed calculations also prove conciseness
in all $n^2$ variables.

%% file: kernel_degree.tex
\section{A lower bound from kernel degrees}\label{sec:kernel-degree}

We prove the general degree bound used in Theorem~\ref{thm:intro-gap}.
Let $f\in T=\C[x_1,\ldots,x_N]$ be irreducible and homogeneous,
and let $R=T/(f)$ be the coordinate ring of its zero hypersurface.
Elements of $R$ are polynomial classes modulo multiples of $f$.
Since $f$ is irreducible, $R$ is a domain.

For a square matrix $M$ with $\det M=f$, its cokernel over $R$ is
\[
 \coker_R M=R^m/MR^m.
\]
It is free of rank one if every class can be written uniquely as
$a b$ for a fixed class $b$ and some $a\in R$. This is a global
condition. In particular it is stronger than requiring the
specialized matrix to have corank one at every point of $f=0$.
The following theorem concerns one representation. Sections~\ref{sec:factorial}
and~\ref{sec:koszul-universal-regularity} show that its hypothesis
holds for every representation of the forms in the main result.

\begin{theorem}[Kernel-degree bound]\label{thm:degree}
Let $f\in\C[x_1,\ldots,x_N]$ be homogeneous and irreducible of degree
$d\ge2$. Let $M$ be an $m\times m$ matrix of affine-linear forms with
$\det M=f$. If its cokernel over $R=\C[x_1,\ldots,x_N]/(f)$ is free
of rank one, then
\[
 m\ge2d-1.
\]
\end{theorem}

\begin{proof}
We first factor the adjugate over $R$. The adjugate is the transpose
of the matrix of cofactors and satisfies
\[
 M\adj M=(\adj M)M=fI_m.
\]
Thus its columns are right-kernel vectors, and its rows are
left-kernel vectors, when we work modulo $f$.

The assumed isomorphism $\coker_R M\simeq R$ gives the
quotient map
\[
 \pi:R^m\longrightarrow R,
 \qquad \ker\pi=MR^m.
\]
There is a column vector $v\in R^m$ with $\pi(z)=v^Tz$.
A row vector $w^T$ satisfies $w^TM=0$ exactly when its map $R^m\to R$
vanishes on $MR^m$. It then factors uniquely through $\pi$.
Every $R$-linear map $R\to R$ is multiplication by one element of
$R$. Consequently every such row has the form $a v^T$.
In particular the rows of the adjugate have this form, so
\begin{equation}\label{eq:adj-factor}
 \adj M=u v^T\quad\text{over }R
\end{equation}
for a column vector $u\in R^m$.

The adjugate is nonzero modulo $f$. Indeed, differentiating
$\det M=f$ expresses each partial derivative of $f$ as a linear
combination of cofactors with constant coefficients.
If every cofactor were divisible by $f$, every derivative of $f$
would also be divisible by $f$. This is impossible: at least one
partial derivative is nonzero, and its degree is $d-1$.
Thus both $u$ and $v$ in~\eqref{eq:adj-factor} are nonzero.
Since $R$ is a domain, the identities
\[
 (Mu)v^T=0,\qquad v^TM=0
\]
give $Mu=0$ and $v^TM=0$.

We next bound the degrees of $u$ and $v$. Because $f$ is homogeneous,
$R$ is graded by total degree. For a
nonzero element of $R$, use its largest nonzero homogeneous degree.
This is also the smallest possible degree of a polynomial representing
that element. Set
\[
 \alpha=\max_{u_i\ne0}\deg_R u_i,
 \qquad
 \beta=\max_{v_j\ne0}\deg_R v_j.
\]
Suppose, for a contradiction, that $\alpha\le d-2$. Choose polynomial
representatives $\widetilde u_i$ of degree at most $d-2$.
Every coordinate of $M\widetilde u$ then has degree at most $d-1$.
It vanishes modulo $f$, so it is a multiple of $f$. A polynomial of
degree below $d$ cannot be a nonzero multiple of $f$. Therefore
\[
 M\widetilde u=0\quad\text{already over }T.
\]
But $M$ has nonzero determinant over $T$, so it has no nonzero
polynomial kernel vector. The same argument
for the left kernel gives
\[
 \alpha\ge d-1,\qquad \beta\ge d-1.
\]
This argument applies to any nonzero kernel vectors over $R$.

Finally, choose coordinates $u_i$ and $v_j$ whose degrees are $\alpha$ and
$\beta$. Degrees add for nonzero products in a graded domain. Thus
\eqref{eq:adj-factor} gives
\[
 \deg_R(\adj M)_{ij}=\alpha+\beta\ge2d-2.
\]
A cofactor of an affine-linear $m\times m$ matrix has degree at most
$m-1$. Passing to $R$ cannot increase that degree. Hence
\[
 m-1\ge2d-2,
\]
which is the claimed inequality.
\end{proof}

Normality of $R$ is not a hypothesis of Theorem~\ref{thm:degree}.
It will be used later to establish the required freeness for the
particular forms under consideration.

\begin{corollary}\label{cor:entry-degree}
Keep the hypotheses on $f$ and the cokernel, but allow entries of $M$
to have degree at most $e\ge1$. Then
\[
 (m+1)e\ge2d.
\]
\end{corollary}
\begin{proof}
If $d>e$, the same lifting argument gives
$\alpha,\beta\ge d-e$. A cofactor has degree at most $e(m-1)$, so
$e(m-1)\ge2(d-e)$. If $d\le e$, the claimed inequality follows from
$m\ge1$. The adjugate is again nonzero modulo $f$: differentiating
$\det M=f$ gives the same divisibility contradiction, even though
the derivatives of the entries are no longer constant.
\end{proof}

\subsection{Regularity and global freeness}

An affine representation is \emph{regular at the origin} if $M(0)$
has rank $m-1$. This condition alone does not imply the conclusion
of Theorem~\ref{thm:degree}. The matrix
\[
 \begin{pmatrix}
 0&x&y\\
 -y&1+a&b\\
 x&c&1+d
 \end{pmatrix}
\]
has rank two at the origin and determinant
\[
 b x^2+(d-a)xy-cy^2.
\]
This is an irreducible homogeneous cubic represented at size three.
Its size is below $2\deg f-1$, and its cokernel is not globally
free. Thus regularity at one point cannot replace the global
hypothesis in the theorem.

One useful sufficient condition is that $R$ be a normal unique
factorization domain. Standard facts about hypersurface modules then
force the cokernel of every polynomial representation of $f$ to be
free. Section~\ref{sec:factorial} explains those facts and proves the
condition for the unpadded odd-skew forms. The Koszul forms require a
different argument, given in Section~\ref{sec:koszul-universal-regularity}.

\subsection{Sharpness}

The bound is sharp for individual representations with free cokernel.
The following explicit matrix is the standard construction from a
two-path algebraic branching program (ABP)~\cite[Section~3]{IL}.

For $d\ge2$, consider
\[
 g_d=x_1\cdots x_d+y_1\cdots y_d.
\]
This polynomial is irreducible. Indeed, viewed as a polynomial in
$x_d$, it has degree one and its two coefficients
$x_1\cdots x_{d-1}$ and $y_1\cdots y_d$ are relatively prime.
Gauss's lemma reduces irreducibility to the degree-one polynomial
over the fraction field of the other variables.

Let $H_x$ be the $(d-1)\times(d-1)$ matrix with ones on the diagonal,
entries $-x_2,\ldots,-x_{d-1}$ immediately above the diagonal, and
zero elsewhere. Define $H_y$ in the same way, and put
\[
 H=\begin{pmatrix}H_x&0\\0&H_y\end{pmatrix},\qquad
 a^T=(x_1e_1^T,\ y_1e_1^T),\qquad
 b=\begin{pmatrix}x_de_{d-1}\\y_de_{d-1}\end{pmatrix}.
\]
Here $e_i$ is the $i$th coordinate vector of $\C^{d-1}$.
The matrix $H$ has determinant one, and its inverse sums products
along the two chains. In particular
\[
 a^TH^{-1}b=x_1\cdots x_d+y_1\cdots y_d=g_d.
\]
Therefore
\begin{equation}\label{eq:sharp-abp-matrix}
 M_d=\begin{pmatrix}0&a^T\\-b&H\end{pmatrix}
 \quad\text{satisfies}\quad
 \det M_d=g_d.
\end{equation}
It has size $2d-1$. Since $H$ is invertible over the polynomial ring,
polynomial row and column operations reduce $M_d$ to the block matrix
$\operatorname{diag}(g_d,H)$. Its cokernel over $g_d=0$ is consequently
free of rank one. Equality holds in Theorem~\ref{thm:degree}.

The corresponding ABP consists of two directed paths of
length $d$, with common starting and ending vertices and no shared
internal vertices. A path contributes the product of its edge labels;
the program adds the two products. The entries of the two kernel
vectors are products along initial and final portions of these paths.
Their degrees are therefore at most $d-1$.

The polynomial $g_d$ also has a
homogeneous determinant of size $d$: use diagonal entries $x_i$,
entries $y_1,\ldots,y_{d-1}$ immediately above the diagonal, and
$(-1)^{d-1}y_d$ in the bottom-left corner. Only the diagonal and the
full cycle contribute to its determinant, giving $g_d$. Thus
$\dc(g_d)=d$, and this smaller representation has nonfree cokernel.
The example proves sharpness for representations satisfying the lemma;
it does not determine the exact complexity of our boundary families.

A superlinear lower bound therefore requires information beyond the
hypotheses of Theorem~\ref{thm:degree}. Possible additional data include
the number of coordinates required by compatible kernel vectors and
the relations among them for a specific family of forms.

%% file: factoriality.tex
\section{Odd-skew forms and factorial hypersurfaces}\label{sec:factorial}

We prove that every polynomial representation of an unpadded odd-skew
form has free rank-one cokernel. The kernel-degree theorem will then
give a lower bound on its determinantal complexity. The proof uses a
general criterion for a polynomial that is linear in one group of
variables. Write
\[
 F(x,y)=y_1g_1(x)+\cdots+y_sg_s(x).
\]
For each $x$ at which some $g_i(x)$ is nonzero, the equation $F(x,y)=0$
can be solved for one coordinate of $y$. Its fiber is therefore an
$(s-1)$-dimensional vector space. The exceptional fibers lie over the
common zero set of the $g_i$. We will show that codimension at least
three for this set implies factoriality of the hypersurface ring,
and hence freeness of every determinant cokernel.

\subsection{Preliminaries on determinant cokernels}

We recall the standard facts about modules over a hypersurface ring
that will be used in this section and the next.

The coordinate ring of the hypersurface $F=0$ is
$R=\C[x,y]/(F)$. In this ring, two polynomials are identified when
their difference is divisible by $F$. For a matrix $M$ of size $m$, its
cokernel over this ring is
\[
 \coker_R M=R^m/MR^m:
\]
it consists of vectors of functions, with vectors in the image of $M$
identified with zero. It is \emph{free of rank one} if a single element
generates it with no relation, so that it is isomorphic to $R$.
Throughout, freeness means global freeness over $R$. A one-dimensional
cokernel at each individual point does not by itself imply this
condition.

A domain is \emph{normal} if it is integrally closed in its fraction
field. It is \emph{factorial} if it is a unique factorization domain
(UFD). Over a normal factorial ring, every rank-one reflexive module
is free. Determinant cokernels on an irreducible normal hypersurface
are rank-one reflexive modules, so this fact applies to them.

A finite module $C$ is \emph{reflexive} if the
natural map
\[
 C\longrightarrow C^{**},\qquad
 C^*=\Hom_R(C,R),
\]
is an isomorphism. Over a normal domain, rank-one reflexive modules
are recorded by the \emph{divisor class group} $\Cl(R)$; the zero
class is represented by $R$ itself. Thus $\Cl(R)=0$ says precisely
that every such module is free. Normality allows us to compute this
group after removing a set of codimension at least two.
Here codimension is the ambient dimension minus the dimension of the
set removed.

At a local ring, the \emph{depth} of a module is the largest length
of a sequence from the maximal ideal that acts successively without
zero divisors on the module and its quotients. Depth is at most the
dimension of the local ring. A module is \emph{maximal
Cohen--Macaulay}, abbreviated MCM, when equality holds at every prime
in its support. For a normal hypersurface ring, a rank-one MCM module
is reflexive. A polynomial matrix whose determinant is an irreducible
polynomial supplies a rank-one MCM module through its cokernel.
We verify that last assertion below. These standard facts, including
the depth facts used in the next section, are treated
in~\cite{BHBook}.

\subsection{A factoriality criterion}

\begin{theorem}\label{thm:universal-form}
Let $g_1,\ldots,g_s\in\C[x_1,\ldots,x_d]$. Suppose their common
zero set has codimension at least three. Set
\[
 F(x,y)=\sum_{i=1}^s y_i g_i(x),\qquad
 R=\C[x,y]/(F).
\]
Then $R$ is a normal factorial domain. Every square polynomial matrix
$M(x,y)$ with $\det M=F$ satisfies
\[
 \coker_R M\simeq R.
\]
Consequently, at every point of $F=0$, the matrix $M$ has corank one.
The entries of $M$ need not be linear, and the $g_i$ need not be
homogeneous.
\end{theorem}

\begin{proof}
We first prove normality and factoriality, and then apply these
properties to the cokernel of an arbitrary representation.

\paragraph{Normality.}
Let
\[
 Z=V(g_1,\ldots,g_s),\qquad B=\mathbb A^d\setminus Z,
 \qquad X=V(F).
\]
There is no common nonconstant factor of the $g_i$: such a factor
would give a codimension-one subset of $Z$. Over the fraction field
$\C(x)$, the polynomial $F$ is a nonzero linear polynomial in the
$y_i$, hence irreducible. Gauss's lemma now shows that $F$ is
irreducible in $\C[x,y]$ as well.

For $x\in B$, the equation $F(x,y)=0$ cuts out an $(s-1)$-dimensional
vector space. These vector spaces vary as the kernel of the
surjective map
\[
 \mathcal O_B^s\longrightarrow\mathcal O_B,
 \qquad (y_i)\longmapsto\sum_i g_i y_i.
\]
Its kernel is a vector bundle: on the open set $D(g_i)$, where $g_i$
is invertible, solve for $y_i$ to obtain a product with affine
$(s-1)$-space. Thus the part $X_B$ of $X$ over $B$ is smooth.

The part we have omitted is $Z\times\mathbb A^s$. If $Z$ has
codimension $c\ge3$ in $\mathbb A^d$, this part has codimension
$c-1\ge2$ in $X$. Hence $X$ is regular in codimension one. A
hypersurface ring is Cohen--Macaulay, so it also satisfies the depth
condition $S_2$: at each prime its depth is at least the smaller of
two and the local dimension. Serre's normality criterion combines
these two properties to show that $R$ is normal.

\paragraph{The divisor class group.}
Because the omitted part has codimension at least two, it does not
change the divisor class group:
\[
 \Cl(X)=\Cl(X_B).
\]
Likewise, removing $Z$ from affine space leaves $\Cl(B)=0$.
On a smooth space, the divisor class group is the Picard group,
which records line bundles. The total space of a vector bundle over
this smooth base has the same Picard group as the base. In the
present case this fact is also visible on the cover $D(g_i)$.
There the coordinate ring of $X_B$ is a polynomial extension of the
UFD $\C[x]_{g_i}$, and every invertible function on an overlap comes
from the base. The transition functions for a line bundle on $X_B$
therefore give exactly a line bundle on $B$.
It follows that
\[
 \Cl(X_B)=\operatorname{Pic}(X_B)
 =\operatorname{Pic}(B)=\Cl(B)=0.
\]
A normal domain with zero divisor class group is factorial. This
proves the first assertion.

\paragraph{Determinant cokernels.}
Let $T=\C[x,y]$, let $M$ have size $m$, and set $C=\coker_T M$.
Since $\det M=F\ne0$, the matrix is injective over $T$. We have the
exact sequence
\[
 0\longrightarrow T^m\xrightarrow{M}T^m
 \longrightarrow C\longrightarrow0.
\]
The identity $M\adj(M)=F I_m$ shows that $F$ annihilates $C$, so $C$
is an $R$-module and equals the cokernel computed over $R$.
At the generic point of $F=0$, the determinant vanishes to order
one. The matrix therefore has corank one there, and $C$ has rank
one over $R$.

The displayed two-term free resolution, localized at any prime in
its support, gives $C$ depth equal to the dimension of the local
hypersurface ring. Thus $C$ is MCM. Over the normal ring $R$, it is
reflexive; factoriality then gives $C\simeq R$. Finally, specialize
at any point of $X$. The fiber of $C$ has dimension one, and this
fiber is the cokernel of the specialized matrix. Its dimension is
exactly the corank of that matrix.
\end{proof}

The same theorem gives a lower bound in terms of the number of
essential variables.

\begin{corollary}\label{cor:universal-size}
If $F$ in Theorem~\ref{thm:universal-form} depends essentially on all
$N$ of its variables, then every affine determinantal representation
of size $m$ satisfies
\[
 m(m+1)\ge2N.
\]
\end{corollary}
\begin{proof}
Essential dependence makes the linear part of the representing affine
map injective: a vector in its kernel would give a direction along
which the polynomial never changes. The image is therefore an
$N$-dimensional affine space of matrices. On $F=0$ every matrix has
rank $m-1$, by the theorem; elsewhere it has rank $m$.
Meshulam's bound for an affine space with this minimum
rank~\cite[Theorem~2]{Meshulam} gives
\[
 N\le m^2-\binom m2=\frac{m(m+1)}2.
\]
\end{proof}

\subsection{Application to unpadded odd-skew forms}

Let $r\ge3$ be odd. Take a generic skew-symmetric matrix $A$ of size
$r$, its signed maximal-Pfaffian vector $p(A)$, and an independent
symmetric matrix $S$. The polynomial is
\[
 Q_{r,0}(A,S)=p(A)^TSp(A).
\]
It is linear in the entries of $S$. The coefficients of those entries
are the products $p_i p_j$, with a factor of two on off-diagonal
entries. Thus their common zero set is exactly $p(A)=0$. Equivalently,
$A$ has rank at most $r-3$. This set has codimension three in the
space of skew-symmetric matrices. In the language of ideals, the
coefficient ideal is $(p_1,\ldots,p_r)^2$.
Theorem~\ref{thm:universal-form} now applies.

\begin{corollary}\label{cor:odd-regularity}
Every square polynomial matrix with determinant $Q_{r,0}$ has free
rank-one cokernel over $\C[A,S]/(Q_{r,0})$, and therefore has corank
one at every zero of $Q_{r,0}$. In particular, every affine
representation is regular at the origin, so
$\dc(Q_{r,0})=\rdc(Q_{r,0})$. Since $Q_{r,0}$ is homogeneous of
degree $r$, the kernel-degree lemma gives
\[
 \dc(Q_{r,0})\ge2r-1.
\]
\end{corollary}

The freeness assertion holds for every polynomial representation.
Thus its application to affine representations imposes no restriction
on their size or matrix form.

\paragraph{The degree-one hypothesis.}
The hypothesis concerns the degree of the whole polynomial in $S$.
For example, the polynomial $\det(A+S)$ has $Q_{r,0}$ as its first
$S$-term, and hence the same coefficient ideal for that term. It is
itself an exact determinant of size $r$. Its higher $S$-terms change
the hypersurface: the equation is no longer the linear equation in
$S$ used above. The proof relies on the degree-one dependence of the
\emph{entire} polynomial on $S$.

The conclusion applies to the unpadded forms $Q_{r,0}$. No corresponding
freeness assertion for every padded odd-skew form is used here.

%% file: koszul_cokernels.tex
\section{Cokernels of the padded Koszul forms}
\label{sec:koszul-universal-regularity}

We prove that every polynomial representation of a padded Koszul form
has free rank-one cokernel. Unlike the factorial rings of the preceding
section, the hypersurface rings here have nontrivial rank-one reflexive
modules. We will show that none of the nonfree modules is MCM, which
excludes them as determinant cokernels.

Away from $v=0$, the polynomial is locally a generic determinant in a
smaller set of variables. The classification of rank-one MCM modules
over a generic determinant gives three possibilities: the free module
and two nonfree modules. We first prove that a global reflexive class
is determined by its restriction to this open set. We then construct
the two nonfree candidates across $v=0$ and compute their depths.
Both fail the MCM condition there. The only remaining possibility is
the free module. We use the definitions and standard module-theoretic
facts recalled in Section~\ref{sec:factorial}.

\subsection{Local determinant form}

Fix $p\ge2$ and set
\[
 d=2p+1,\qquad r=\binom{2p+1}{p},\qquad
 R_0=\binom{2p}{p},\qquad c=\binom{2p}{p-1}.
\]
Here $d$ is the number of $v$-variables, $r$ is the size of the seed
matrix, $R_0$ is its rank at nonzero $v$, and $c=r-R_0$ is its
corank there. For the smallest case $p=2$, these numbers are
$d=5$, $r=10$, $R_0=6$, and $c=4$. Padding this case gives every final size $n\ge10$.

Let $V=\C^d$ and $E=\bigwedge^pV$. Fix a volume form on $V$.
The seed matrix records the alternating-product pairing
\[
 A(v):E\longrightarrow E^*,\qquad
 A(v)(\alpha,\beta)=\operatorname{vol}(v\wedge\alpha\wedge\beta).
\]
For $v\ne0$, its left and right kernels are both
\[
 U_v=v\wedge\bigwedge^{p-1}V,
\]
a space of dimension $c$. The first nonzero determinant coefficient
is the restriction of the perturbation to these left and right
kernel spaces.

To add $k$ rows and columns, put $n=r+k$ and define
\[
 f=\mathcal K_{p,k}
 =[t^c]\det\begin{pmatrix}A(v)+tS&U\\tV_0&Z\end{pmatrix}.
\]
The notation $[t^c]$ means the coefficient of $t^c$. The matrix $S$
ranges over a linear complement to $A(V)$ in $\Mat_r$; its role is
to provide the matrix directions that the seed does not already
use. The independent matrices $U,V_0,Z$ have sizes $r\times k$,
$k\times r$, and $k\times k$. Write $Y$ for all these coefficient
variables and put $q=c+k$. There are $n^2$ variables altogether.

On any sufficiently small open set with $v\ne0$, choose bases in
the kernel spaces. The determinant coefficient then has the form
\begin{equation}\label{eq:regularity-local-fiber}
 f(v,Y)=u(v)\det_q\bigl(\kappa_v(Y)\bigr),\qquad u(v)\ne0.
\end{equation}
Here $u$ is invertible on the chosen open set, and $\kappa_v$ is the
surjective map that restricts the coefficient bilinear form to
$U_v\oplus\C^k$ on both sides. Thus, after a linear change of the
$Y$-coordinates, $f$ is an ordinary determinant of size $q$, times
an invertible factor. The remaining $Y$-coordinates are independent
variables that do not enter that determinant.
We can therefore apply the classification of rank-one MCM modules
over a generic determinant.

\subsection{Rank-one MCM modules over the generic determinant}

Let $F$ be a field, let $X$ be a generic $q\times q$ matrix with
$q\ge2$, and let $z$ be any list of independent extra variables.
Set
\[
 A_F=F[z,X]/(\det X).
\]
The ring $A_F$ is normal. Its class group is $\mathbb Z$, with
opposite generators represented by $\coker X$ and $\coker X^T$.
Although the class group has infinitely many elements, only three
of them are represented by rank-one MCM modules. Up to isomorphism,
those modules are
\[
 A_F,\qquad \coker X,\qquad \coker X^T.
\]
This is the generic-determinant classification of Bruns and
Bruns--Vetter, in the form recorded by
Buchweitz--Leuschke~\cite[Section~4.4]{BL}. The cokernels can be
computed over $F[z,X]$ or over $A_F$, since $\det X$ annihilates
them. We use this classification only over a field with polynomial
extra variables; no version over an arbitrary coefficient ring is
being assumed.

The classification has the following consequence for polynomial
matrix representations.

\begin{lemma}\label{lem:generic-fiber-corank-dichotomy}
Let $F$ be a field, $q\ge2$, and let $z$ denote any list of additional
variables. Suppose a square polynomial matrix $M(X,z)$ satisfies
\[
 \det M=a\det_qX,\qquad a\in F^*.
\]
Then, along $X=0$, either $M(0,z)$ has corank one for every $z$, or
it has corank $q$ for every $z$.
\end{lemma}
\begin{proof}
The matrix gives an injective map between free $F[z,X]$-modules,
whose cokernel is annihilated by $\det X$. Its two-term free
resolution makes this cokernel MCM over $A_F$. It has rank one there,
because the determinant vanishes to order one at the generic point
of $\det X=0$. The classification therefore applies.

Now set $X=0$. The free module has a one-dimensional fiber. Each of
$\coker X$ and $\coker X^T$ has a $q$-dimensional fiber, since the
specialized matrix is zero. These dimensions do not depend on $z$.
The dimension of the cokernel fiber is exactly the corank of the
specialized matrix $M(0,z)$.
\end{proof}

Applying the lemma over $F=\C(v)$ gives corank one or $q$ along
$Y=0$ at generic $v$. Here the seed variables are treated as rational
scalars, and the coefficient variables remain polynomial variables.
The behavior at $v=0$ will exclude the two nonfree possibilities.

\subsection{Global freeness}

\begin{theorem}\label{thm:koszul-universal-regularity}
For $p\ge2$ and $k\ge0$, set
\[
 \mathscr R=\C[v,Y]/(\mathcal K_{p,k}).
\]
Every rank-one MCM module over $\mathscr R$ is free. Consequently,
every square polynomial matrix $M$ with $\det M=\mathcal K_{p,k}$
satisfies
\[
 \coker_{\mathscr R}M\simeq\mathscr R.
\]
At every zero of the polynomial, the matrix has corank one. In
particular, every affine representation is regular at the origin.
\end{theorem}

\begin{proof}
Write $B=\C[v_1,\ldots,v_d]$ and $T=B[Y]$.

\paragraph{The divisor class group.}
We first prove that $\mathscr R$ is a normal domain. Equation~\eqref{eq:regularity-local-fiber}
shows that $f$ is irreducible over $\C(v)$. There is no nonconstant
factor of $f$ that lies in $B$. Indeed, such a factor would vanish
at some nonzero $v$, forcing $f(v,Y)$ to be zero for every $Y$,
contrary to the local determinant description. Gauss's lemma shows
that $f$ is irreducible over $B$.

On $v\ne0$, the singular locus has codimension four in the ambient
space, just as it does for a generic determinant. The remaining set
$v=0$ has ambient codimension $d\ge5$. Thus the hypersurface is
regular in codimension one. As a hypersurface it satisfies $S_2$,
so Serre's criterion makes $\mathscr R$ normal.

Let $\mathscr X=\Spec\mathscr R$, and remove the exceptional set:
\[
 \mathscr X^\circ=\mathscr X\setminus\{v=0\}.
\]
The polynomial vanishes identically when $v=0$, so the set removed
has codimension $d-1\ge4$ in $\mathscr X$. It does not change the
class group:
\[
 \Cl(\mathscr X)=\Cl(\mathscr X^\circ).
\]
Projection onto the $v$-coordinates makes $\mathscr X^\circ$ a
family over $\mathbb A^d\setminus\{0\}$. Locally this family is a
generic determinant hypersurface, with independent affine
coordinates. In particular, the family is flat and its fibers are
integral. Passing from the base to its fraction field gives a map
\begin{equation}\label{eq:koszul-class-injection}
 \Cl(\mathscr R)\longrightarrow
 \Cl\bigl(\C(v)[Y]/(f)\bigr)=\mathbb Z.
\end{equation}
We claim that this map is injective.

The only divisor classes lost on passing to the fraction field are
generated by \emph{vertical} prime divisors: divisors lying over a
proper subset of the $v$-space. Flatness and the dimension formula
show that each such divisor lies over a divisor of the base, rather
than over a subset of higher codimension. A base divisor is defined
by an irreducible polynomial $h(v)$. In a local product chart, its
inverse image is a generic determinant hypersurface over the domain
$B/(h)$, together with extra affine coordinates. It is therefore
integral and generically reduced. Its multiplicity in the divisor
of $h$ is one. Thus the whole vertical divisor is principal: it is
already the divisor of the single function $h$.
Every class in the kernel is consequently zero, proving injectivity.

Consequently, the class of a rank-one reflexive module is determined
by its class after passage to $\C(v)$.

\paragraph{The nonfree reflexive classes.}
Let $K=\ker A(v)$ as a $B$-module. Its restriction away from the
origin is the kernel bundle $U_v$. The exact Koszul complex gives
a free resolution of $K$ over the whole base:
\begin{equation}\label{eq:koszul-kernel-resolution}
 0\longrightarrow B(-p)\longrightarrow
 B(-(p-1))\otimes V\longrightarrow\cdots\longrightarrow
 B(-1)\otimes\bigwedge^{p-1}V\longrightarrow K\longrightarrow0.
\end{equation}
Here $B(-j)$ denotes the grading shift with generator in degree $j$.
We will use the length and ranks of this resolution and the fact
that its differentials are linear in $v$.

The pairing on the ambient vector spaces gives a map
$\coker A\to K^*$, where $K^*=\Hom_B(K,B)$. It is an isomorphism
away from the origin. We need it at the origin as well. A truncation
of the same Koszul complex shows that $\coker A$ has projective
dimension $p+1$ and depth $p$ there. Here projective dimension is
the shortest possible length of a projective resolution; the
resolutions in use are free. The module is locally free elsewhere.
Its positive depth excludes torsion supported at the origin. It is
therefore torsion-free, and it satisfies $S_2$ because $p\ge2$.
It follows that $\coker A$ is reflexive. The dual $K^*$ is
reflexive too. Isomorphisms between reflexive modules extend across
sets of codimension at least two, so the map extends to an
isomorphism on all of $\Spec B$:
\[
 K^*\simeq\coker A,
\]
up to the grading shift fixed by the convention for $A$.

We record the numerical information supplied by the two Koszul
resolutions. The depths here are at the origin of the $v$-space:
\begin{align}
 \pd_B K&=p-1,& \depth_B K&=p+2,
 \label{eq:koszul-depths-a}\\
 \pd_B K^*&=p+1,& \depth_B K^*&=p,
 \label{eq:koszul-depths-b}\\
 \Ext_B^i(K^*,B)&=0\qquad(1\le i\le p).
 \label{eq:koszul-ext-vanishing}
\end{align}
The depth formulas use the equality between depth plus projective
dimension and the dimension $2p+1$ of the regular local ring.
The last line follows by dualizing the truncated Koszul complex:
its positive-degree cohomology occurs only at the last term.

Add the padding directions by setting
\[
 P=K\oplus B^k,
 \qquad P^*=K^*\oplus B^k,
\]
and extend both modules to $T=B[Y]$. The coefficient variables
induce a map
\[
 \phi_Y:P\longrightarrow P^*.
\]
When $v\ne0$, this is precisely the restriction of the coefficient
bilinear form to $U_v\oplus\C^k$. It is invertible over the full
fraction field, hence injective. Define
\[
 C=\coker\phi_Y.
\]
Away from $v=0$, this is the cokernel module of the smaller generic
determinant. We next verify that it extends to a reflexive module
on the entire hypersurface.

Let $\mathfrak q$ be any prime of $T$ containing $(v)$, and put
$h=\dim T_{\mathfrak q}$. The Koszul differentials have entries in
$(v)$, so their resolutions remain minimal after this localization.
The displayed projective dimensions therefore give
\[
 \depth_{T_{\mathfrak q}}P_{\mathfrak q}=h-p+1,
 \qquad
 \depth_{T_{\mathfrak q}}P^*_{\mathfrak q}=h-p-1.
\]
Apply the depth lemma to
$0\to P\to P^*\to C\to0$. Since the first module has strictly
larger depth than the second, the cokernel has the second depth:
\begin{equation}\label{eq:koszul-natural-depth}
 \depth_{T_{\mathfrak q}}C_{\mathfrak q}
   =h-p-1\ge p\ge2.
\end{equation}
The inequality uses $h\ge2p+1$, since $\mathfrak q$ contains all
$2p+1$ seed variables.

This lower bound has two immediate uses. First, $C$ has no nonzero
submodule supported on $v=0$. Such a submodule would have an
associated prime containing $(v)$, and at an associated prime the
module has depth zero. Equation~\eqref{eq:koszul-natural-depth}
excludes that possibility. Off $v=0$, the local determinant
calculation gives $fC=0$. The submodule $fC$ is therefore supported
on $v=0$, so it must be zero everywhere. We can now regard $C$ as
an $\mathscr R$-module, not just a $T$-module. Second, any torsion
of $C$ would be supported on $v=0$, since the generic determinant
cokernel is torsion-free off that set. The same argument excludes
this torsion.

Off $v=0$, the module $C$ is rank-one MCM. At primes over $v=0$,
Equation~\eqref{eq:koszul-natural-depth} gives $S_2$. The depth is
unchanged when calculated over $T_{\mathfrak q}$ or its quotient
by $f$, because $f$ annihilates the module. Thus $C$ is torsion-free
and satisfies $S_2$ over the normal ring $\mathscr R$: it is a
rank-one reflexive module.

Its class over $\C(v)$ is the generator represented by the cokernel
of the generic determinant. Hence the injection
\eqref{eq:koszul-class-injection} is also surjective. We have now
identified $\Cl(\mathscr R)$ with $\mathbb Z$, with $C$ representing
one generator and its dual representing the opposite generator.

\paragraph{The depth of the nonfree classes.}
Localize at the prime $(v)$. At this prime the coefficient variables
are generic, and $\dim T_{(v)}=2p+1$. Formula
\eqref{eq:koszul-natural-depth} becomes
\[
 \depth C_{(v)}=p
 <2p=\dim\mathscr R_{(v)}.
\]
The module $C$ therefore fails the MCM condition. For $p=2$, this is
the concrete inequality $2<4$.

We need the same conclusion for the opposite class. Its reflexive
representative is the dual
\[
 C^*=\Hom_{\mathscr R}(C,\mathscr R).
\]
The hypersurface change-of-rings isomorphism identifies it with
$\Ext_T^1(C,T)$, with a grading shift if we retain gradings.
To compute this module, dualize the exact sequence
$0\to P\xrightarrow{\phi_Y}P^*\to C\to0$ over $T$.
Here $\Hom_T(C,T)=0$, since $C$ is annihilated by the nonzero
polynomial $f$. Reflexivity gives $\Hom_T(P^*,T)=P$, and
\eqref{eq:koszul-ext-vanishing} gives $\Ext_T^1(P^*,T)=0$.
The resulting sequence is
\[
 0\longrightarrow P\xrightarrow{\phi_Y^T}P^*
 \longrightarrow\Ext_T^1(C,T)\longrightarrow0.
\]
Thus $C^*\simeq\coker\phi_Y^T$. The same depth calculation applies
to this transposed map and gives
\[
 \depth_{\mathscr R_{(v)}}C^*_{(v)}
 =p<2p=\dim\mathscr R_{(v)}.
\]
The opposite class is not MCM either.

Now let $D$ be any rank-one MCM $\mathscr R$-module. It is
reflexive, and it remains MCM after passage to $\C(v)$.
The generic-determinant classification permits only the classes
$0$, $1$, and $-1$. Our class-group isomorphism says that globally
these are represented by $\mathscr R$, $C$, and $C^*$, respectively.
Rank-one reflexive modules in the same class are isomorphic. Since
$C$ and $C^*$ are not MCM, the only possibility for $D$ is
$D\simeq\mathscr R$.

Finally, a matrix with determinant $f$ has a two-term free resolution
over $T$, so its cokernel is rank-one MCM over $\mathscr R$, exactly
as in Section~\ref{sec:factorial}. It is therefore free. At any zero
of $f$, its one-dimensional fiber is the cokernel of the specialized
matrix, proving the corank assertion.
\end{proof}

\subsection{The determinantal complexity bound}

The freeness theorem verifies the hypothesis of the kernel-degree
bound for every exact representation of a padded Koszul form.

\begin{corollary}\label{cor:koszul-degree-gap}
For $p\ge2$, $k\ge0$, and $n=\binom{2p+1}{p}+k$,
\[
 \bdc(\mathcal K_{p,k})=n,
 \qquad \dc(\mathcal K_{p,k})\ge2n-1.
\]
In particular, $f_n=\mathcal K_{2,n-10}$ has exact-minus-border gap
at least $n-1$ for every $n\ge10$.
\end{corollary}
\begin{proof}
The polynomial has degree $n$, so its border complexity is at least
$n$. Its defining determinant limit gives the matching upper bound.
It is homogeneous and irreducible, and
Theorem~\ref{thm:koszul-universal-regularity} shows that every exact
representation has free cokernel. The kernel-degree lemma,
Theorem~\ref{thm:degree}, then gives size at least $2n-1$.
\end{proof}

\subsection{Betti numbers of the natural cokernel}

The Betti numbers of the natural nonfree module $C$ grow faster than
the seed size. Since $C$ has been excluded as a determinant cokernel,
this growth does not directly give a larger lower bound for exact
representations, whose cokernels belong to the free class.
The calculation is as follows.
At generic coefficient values, $C$ has a minimal linear mapping-cone
resolution. Its Betti numbers, the numbers of free generators in
successive terms of a minimal resolution, are
\[
 \beta_0(C)=r,\qquad
 \beta_i(C)=\binom{2p+2}{p+1-i}\quad(1\le i\le p+1).
\]
Their sum is
\[
 \sum_i\beta_i(C)=2^{2p+1}=\Theta(r\sqrt{\log r}).
\]
To obtain these numbers, combine the free resolutions of $K$ and
$K^*$ using the map $\phi_Y$. This is the mapping-cone construction:
the terms of the two resolutions appear in adjacent degrees.
Over the fraction field of the coefficient variables, the invertible
padding block cancels. Every remaining comparison map has positive
$v$-degree, so the resulting resolution is minimal at $(v)$.
Adjacent binomial ranks add by Pascal's identity, giving the formula
for $\beta_i$. Symmetry of the binomial coefficients of order
$2p+2$ gives the sum.

The projective dimension $p+1$ is the depth defect that excludes
$C$ and its dual from being determinant cokernels. It does not
transfer their growing Betti sum to the size of a representation
using the free class. In fact, the seed module $\ker A(v)$ already
has total Betti number
\[
 4^p-r=\Theta(r\sqrt{\log r}),
\]
even though the matrix $A(v)$ has size only $r$.
A superlinear lower bound would need an additional quantitative
restriction on the representations that remain possible.

%% file: forbidden_value.tex
\section{Exact complexity of general boundary points}
\label{sec:dc}

We now prove that every concise degree-$n$ point of the determinant
boundary has exact complexity at least $n+2$. The argument excludes
size $n+1$ under the weaker hypothesis of at least $n^2-n+2$ essential
variables.

A size-$(n+1)$ representation of a degree-$n$ form has a homogeneous
linear part consisting entirely of singular matrices. A theorem of
de Seguins Pazzis classifies such matrix spaces when their dimension
is sufficiently large. The resulting block form allows the original
polynomial to be written as a homogeneous determinant of size $n$.

Write $\operatorname{ess}(f)$ for the number of essential variables
of a homogeneous form $f$. This is the smallest number of variables
left after an invertible linear change of coordinates, or,
equivalently, the dimension of the span of the first partial
derivatives. Throughout this section, matrix entries may be
affine-linear, as required by the usual definition of $\dc$.

\begin{theorem}\label{thm:gap}
Let $f$ be a nonzero homogeneous complex form of degree $n\ge3$.
If $\operatorname{ess}(f)\ge n^2-n+2$, then
\[
 \dc(f)\le n+1\quad\Longrightarrow\quad
 f=\det_n(L)
\]
for a matrix $L$ of homogeneous linear forms. Thus exact complexity
$n+1$ is impossible in this range of essential dimensions.
\end{theorem}

\begin{proof}
Work in $N=\operatorname{ess}(f)$ essential variables. If a
representation has size at most $n$, its size must be exactly $n$,
because its determinant has degree $n$. Taking the degree-$n$
homogeneous part then gives a homogeneous size-$n$ representation.
It remains to consider size $n+1$.

\paragraph{The homogeneous linear part.}
Write
\[
 f(x)=\det(A_0+L(x)),
\]
where $A_0$ is constant and $L$ is homogeneous linear, both of size
$n+1$. The linear map $L$ is injective. Indeed, if $L(w)=0$ for a
nonzero direction $w$, then the determinant and hence $f$ would be
unchanged on every line $x+\lambda w$. That would contradict our
choice of essential variables.

Homogenizing the determinant identity gives
\[
 \det(L(x)+tA_0)=t f(x).
\]
Setting $t=0$ shows that every matrix in $L(\C^N)$ is singular.
This is an $N$-dimensional linear space in $\Mat_{n+1}$. Its
maximum rank is $n$: if all ranks were at most $n-1$, the coefficient
of $t$ would vanish, contradicting $f\ne0$.

\paragraph{The compression-space classification.}
Theorem~1.7 of de Seguins Pazzis~\cite{Pazzis} classifies sufficiently
large spaces of this kind. For upper rank $n$ in square size $n+1$,
its dimension threshold is
\[
 (n+1)n-2n+2=n^2-n+2.
\]
Our hypothesis meets that threshold, and its exceptional finite-field
case does not occur over $\C$. After constant row and column
changes, the space lies in a \emph{compression space}, meaning a
space with a prescribed rectangular zero block:
\[
 L=\begin{pmatrix}A&C\\B&0\end{pmatrix},\qquad
 A\in\Mat_{s,q},\quad
 C\in\Mat_{s,s+1},\quad
 B\in\Mat_{q+1,q},\quad s+q=n.
\]
The row and column changes may multiply the determinant by a
nonzero constant. We absorb that constant into one row of the final
representation.

Consider the coefficient of $t$ in $\det(L+tA_0)$. A determinant
term contributing to that coefficient uses exactly one entry from
$A_0$. It must use that entry in the zero block. Of the remaining
entries, $s$ come from $C$ and $q$ from $B$. No entry comes from
$A$. Thus the entire polynomial $f$ depends only on the entries of
$B$ and $C$.

If $2\le s\le n-2$, the number of those entries is at most
\[
 \begin{aligned}
 s(s+1)+q(q+1)
 &=n^2+n-2s(n-s)\\
 &\le n^2-3n+8<n^2-n+2\le N.
 \end{aligned}
\]
This contradicts essential dependence on $N$ variables. The interior
range occurs only for $n\ge4$, which is exactly the range in which
the displayed inequality is strict. We are left with
$s=0,1,n-1,n$.

\paragraph{Reduction to size $n$.}
If $s=0$, the linear matrix has the form $L=[B,0]$. Its coefficient
of $t$ is $\det[B,d]$, where $d$ is a fixed column from $A_0$.
This column is nonzero since $f\ne0$. A constant row change makes
it a coordinate vector. Expansion along it leaves a homogeneous
determinant of size $n$. The case $s=n$ is the transpose of this
argument.

If $s=1$, write $C=(c_1,c_2)$, and denote the two constant columns
in the lower-right block of $A_0$ by $d_1,d_2$. Expanding along the
first row gives
\[
 \begin{aligned}
 f&=\pm\bigl(c_1\det[B,d_2]-c_2\det[B,d_1]\bigr)\\
  &=\pm\det[B,c_1d_2-c_2d_1].
 \end{aligned}
\]
This is again a homogeneous determinant of size $n$. Transposing
handles $s=n-1$. These four cases include all possibilities, also
when $n=3$.
\end{proof}

\begin{corollary}\label{cor:all-concise-gap}
For $n\ge3$, every degree-$n$ point of the determinant boundary
that is concise in its $n^2$ variables satisfies
\[
 \bdc(f)=n,\qquad\dc(f)\ge n+2.
\]
Its orbit closure need not itself be a boundary component.
\end{corollary}
\begin{proof}
The degree gives $\bdc(f)\ge n$, and membership in the determinant
orbit closure gives the reverse inequality. Suppose that $f$ had
an exact representation of size at most $n+1$. The theorem would
give a homogeneous size-$n$ representation. Conciseness makes its
linear map from $\C^{n^2}$ to $\Mat_n$ injective, hence an
isomorphism. The polynomial would then be an invertible linear
change of variables of $\det_n$, putting it in the orbit itself.
That contradicts its being a boundary point.
\end{proof}

\subsection{Bounds for padded odd-skew and block forms}

The theorem applies to the padded odd-skew forms $Q_{r,k}$ and the
block forms $H_{h,k}$. For these families, explicit affine matrices
also give the following upper bounds.

For $r=3$, the Pfaffian vector is linear. Write it as $x\in\C^3$.
Then
\[
 Q_{3,k}=\det\begin{pmatrix}
 0&x^T&0&0\\
 0&I_3&S&U\\
 x&0&I_3&0\\
 0&0&V&Z
 \end{pmatrix},
 \qquad\text{block sizes }1,3,3,k.
\]
The two identity blocks allow ordinary row and column elimination;
the remaining block determinant is
$\det Z\,x^TSx-x^TU\adj(Z)Vx$, as required. The total matrix size
is $k+7=n+4$. Therefore
\[
 n+2\le\dc(Q_{3,n-3})\le n+4.
\]

The same construction works for the block family because its kernel
matrices $K_L$ and $K_R$ are linear; these matrices are defined in
Section~\ref{sec:higher}. With $r=3h$, it gives
\[
 H_{h,k}=\det\begin{pmatrix}
 0&K_L&0&0\\
 0&I_r&S&U\\
 K_R&0&I_r&0\\
 0&0&V&Z
 \end{pmatrix},
 \qquad\text{block sizes }h,r,r,k.
\]
After eliminating the identity blocks, the two off-diagonal signs
can be removed simultaneously. The result is exactly
\eqref{eq:higher-form}. The matrix size is $h+2r+k=n+4h$, so
\[
 n+2\le\dc(H_{h,k})\le n+4h.
\]
These intervals do not determine the exact complexities. For odd
$r>3$, the entries of the Pfaffian vector have degree greater than
one. Substituting them into the first block matrix would no longer
give an affine representation, so it would not prove an $n+4$
upper bound.

\paragraph{The dimension threshold.}
The threshold $n^2-n+2$ comes from the matrix-space classification.
We do not claim that it is optimal for determinantal complexity.
For a hypothetical representation of size $n+2$, the same
classification would require at least $n^2+n+2$ essential variables.
The boundary polynomials have $n^2$ variables, so this argument does
not exclude exact size $n+2$.

\subsection{Regularity of the smallest odd-skew form}

For the smallest odd-skew form, regularity at the origin also follows
from an elementary argument about linear spaces in its zero set.

\begin{proposition}\label{prop:cubic-regularity}
Every affine determinantal representation of the nine-variable cubic
$Q_{3,0}=x^TSx$ has constant matrix of corank one.
\end{proposition}

\begin{proof}
First observe that $Q_{3,0}=0$ contains no linear subspace of
codimension at most two. To see this, let a linear zero subspace
project to a $t$-dimensional space $T$ in the three $x$-coordinates.
Its kernel in the six-dimensional space $\Sym_3$ must consist of
quadratic forms vanishing on $T$. Restriction to $T$ imposes
$t(t+1)/2$ independent conditions, so that kernel has dimension at
most $6-t(t+1)/2$. The whole subspace consequently has dimension
at most
\[
 t+6-\frac{t(t+1)}2\le6.
\]
A subspace of codimension at most two in nine variables would have
dimension at least seven, which is impossible.

Now make constant row and column changes so that the constant matrix
of a representation has an identity block and a zero block of size
$a$. Since the polynomial vanishes to order three at the origin,
$1\le a\le3$.

If $a=3$, the degree-three term is the determinant of the corresponding
$3\times3$ linear block. Conciseness in nine variables would make
this an invertible linear change of variables of $\det_3$. But
$Q_{3,0}$ lies on the boundary and is not in that orbit, a
contradiction.

If $a=2$, the $2\times2$ linear block is identically singular:
otherwise its determinant would give a nonzero quadratic term.
A linear space of singular $2\times2$ matrices has a common kernel
or a common cokernel. Indeed, two rank-one matrices with independent
column factors and independent row factors have an invertible sum;
excluding that possibility throughout the space forces one of the
two common factors. After transposing if necessary, the block has
the form
\[
 \begin{pmatrix}\ell_1&\ell_2\\0&0\end{pmatrix}.
\]
Eliminate the invertible constant block by a Schur complement. The
cubic term must then lie in the ideal $(\ell_1,\ell_2)$, so it
vanishes on the linear subspace $\ell_1=\ell_2=0$. This subspace
has codimension at most two, contrary to the first paragraph.

Both $a=2$ and $a=3$ have been excluded. Hence $a=1$.
\end{proof}

The factoriality argument in Section~\ref{sec:factorial} strengthens
this conclusion: every representation has corank one at \emph{every}
zero, and its cokernel is globally free of rank one. The exact
complexity of $Q_{3,0}$ remains undetermined; the bounds above leave
the values $5$, $6$, and $7$.

%% file: geometry.tex
\section{Odd-skew boundary components}\label{sec:odd-family}

We prove that every odd-skew form remains a boundary component after
arbitrary padding, and that distinct skew sizes give distinct components
at a fixed final size. The degeneration follows from a determinant
coefficient identity. The component and distinctness statements follow
from the singular locus and the dimension of the linear stabilizer.

Let $r=2m+1\ge3$, let $k\ge0$, and put $n=r+k$. The padded form
uses the independent entries of the following five blocks:
\[
 A\in\Skew_r,\quad S\in\Sym_r,\quad
 U\in\Mat_{r,k},\quad V\in\Mat_{k,r},\quad Z\in\Mat_k.
\]
Here $A$ is skew-symmetric, $S$ is symmetric, and $U,V,Z$ are unrestricted.
The skew and symmetric entries together give $r^2$ variables. Including
the padding blocks, there are $r^2+2rk+k^2=n^2$ variables.

The signed maximal Pfaffians of $A$ give a polynomial kernel vector.
For an even-size skew matrix, the Pfaffian is a polynomial whose square
is its determinant. Define
\[
 p_i(A)=(-1)^{i+1}\Pf(A_{\widehat i}),\qquad 1\le i\le r.
\]
Write $p(A)$ for the column of these $r$ polynomials. Two standard
Pfaffian identities give both a kernel vector and a factorization of
the adjugate:
\begin{equation}\label{eq:pf}
 Ap(A)=0,\qquad \adj A=p(A)p(A)^T.
\end{equation}
For $k\ge1$, define the padded odd-skew form by
\begin{equation}\label{eq:Q}
 Q_{r,k}=\det Z\,p(A)^TSp(A)-p(A)^TU\adj Z\,Vp(A).
\end{equation}
For $k=0$, set $Q_{r,0}=P_r=p(A)^TSp(A)$ and omit the empty blocks.
Each $p_i$ has degree $m$, so $Q_{r,k}$ is homogeneous of degree $n$.
For $r=3$, the entries of $p$ are the three skew variables, up to
signs. Thus $Q_{3,0}=p^TSp$ is the cubic consisting of a quadratic
form in $p$ with independent coefficients.

Write $\Orb_n=\GL_{n^2}\cdot\det_n$ for the affine orbit under invertible
linear changes of variables. Its boundary is
$\partial\overline{\Orb_n}=\overline{\Orb_n}\setminus\Orb_n$.
A form is \emph{concise} if no invertible linear change of variables
can make one variable disappear. Equivalently, its first partial
derivatives are linearly independent. Its \emph{stabilizer} is the group
of invertible linear changes of all $n^2$ variables that leave the
polynomial unchanged. The orbit--stabilizer dimension formula gives
\[
 \dim(\GL_{n^2}\cdot Q)=n^4-\dim\Stab(Q).
\]
Since $\dim\Stab(\det_n)=2n^2-2$, a boundary orbit with stabilizer
dimension $2n^2-1$ has codimension one in $\overline{\Orb_n}$. Its
closure is therefore a boundary component.

\begin{theorem}\label{thm:main}
For every odd $r\ge3$ and every $k\ge0$, with $n=r+k$, the form $Q_{r,k}$
is concise and
\[
 \dim\Stab_{\GL_{n^2}}(Q_{r,k})=2n^2-1,
 \qquad
 \dim(\GL_{n^2}\cdot Q_{r,k})=n^4-2n^2+1.
\]
Its orbit closure is an irreducible component of $\partial\overline{\Orb_n}$.
For fixed $n$, these components are pairwise distinct as odd $r$ ranges
from $3$ to $n$.
\end{theorem}

\begin{corollary}\label{cor:number}
For every $n\ge3$, the determinant boundary has at least
$\lfloor(n-1)/2\rfloor$ pairwise distinct concise components. Together
with the traceless-determinant component, it has at least
$\lfloor(n-1)/2\rfloor+1$ irreducible components.
\end{corollary}

This is a lower bound on the number of components, not a classification of
the whole boundary. The odd-skew theorem includes the following cases:
\begin{center}
\begin{tabular}{lll}
\toprule
Parameters & Construction & Prior case\\
\midrule
$k=0$ & Full odd-skew block & Landsberg--Manivel--Ressayre\\
$k=1$ & One additional row and column & Chen\\
$r=3$ & Fixed skew block, arbitrary padding & Arbitrary padding of the smallest seed\\
\bottomrule
\end{tabular}
\end{center}

\subsection{Pfaffian identities and degeneration}

The map $A\mapsto p(A)$ attains every nonzero vector of $\C^r$.
This permits the kernel vector to be varied freely in the calculations
below.
Indeed, for $x\ne0$, choose a skew form of rank $r-1$ with kernel $\C x$.
Its Pfaffian vector is a nonzero multiple of $x$. Rescaling $A$ rescales
$p(A)$ by an $m$-th power, so any nonzero multiple can be obtained.

For later use, put
\[
 L_x=\{A\in\Skew_r:Ax=0\},\qquad x\ne0.
\]
This is a linear space. A dense open subset has rank $r-1$, since the
induced skew form on the quotient by $\C x$ has even dimension $r-1$.
On that open subset, $p(A)$ is a nonzero multiple of $x$. Polynomial
identities proved there therefore extend to every $A\in L_x$.

For the degeneration, introduce a parameter $t$ and consider
\[
 M_t=\begin{pmatrix}A+tS&U\\tV&Z\end{pmatrix}.
\]
For $t\ne0$, its entries are an invertible linear parametrization of
$\Mat_n$: the upper-left block separates into arbitrary skew and symmetric
parts, and the other blocks are independent. On $\det Z\ne0$,
\begin{align*}
 \det M_t
 &=\det Z\,\det\bigl(A+t(S-UZ^{-1}V)\bigr)\\
 &=t\det Z\,\tr\bigl(\adj A\,(S-UZ^{-1}V)\bigr)+O(t^2)\\
 &=tQ_{r,k}+O(t^2).
\end{align*}
The calculation used $Z^{-1}$ only to simplify the algebra. Both
sides of the resulting coefficient identity are polynomials, and the
identity holds on the dense open set $\det Z\ne0$, so it holds everywhere.
Multiplying one row by $t^{-1}$ gives a determinant in $\Orb_n$ whose
limit is $Q_{r,k}$. For $k=0$, use $t^{-1}\det(A+tS)$ instead.
Thus every $Q_{r,k}$ lies in $\overline{\Orb_n}$. The singular-locus
calculation below excludes membership in $\Orb_n$, and the stabilizer
calculation then proves componenthood.

\subsection{Singular locus and translation invariants}\label{sec:odd-singular}

The singular locus of $Q=0$ is the common zero set of its first
partial derivatives. Every linear stabilizer of $Q$ preserves this
locus. For $Q_{r,k}$, its unique largest component determines the
coefficient directions and recovers the skew size $r$.

Put $a=\binom r2$, the number of skew variables, and $N=n^2$, the total
number of variables. Define
\[
 Y_r=\{A\in\Skew_r:\rank A\le r-3\},\qquad
 M_{r,k}=Y_r\times\C^{N-a}.
\]
The set $Y_r$ is the zero set of the maximal Pfaffians. The skew
normal form shows that it is irreducible of codimension three in
$\Skew_r$. In particular, $\dim M_{r,k}=N-3$.

\begin{proposition}\label{prop:sing}
For $k=0$, the singular locus of $Q_{r,0}$ is exactly $M_{r,0}$.
For $k\ge1$, it has exactly two irreducible components:
\[
 \Sing V(Q_{r,k})=M_{r,k}\cup C_{r,k},
 \qquad \dim M_{r,k}=N-3,\quad \dim C_{r,k}=N-4.
\]
Thus $M_{r,k}$ is always the unique largest singular component.
\end{proposition}

\begin{proof}
If $p(A)=0$, all first derivatives vanish, since every term of $Q_{r,k}$
contains two entries of $p(A)$. When $k=0$, the $S_{ii}$ derivatives are
$p_i(A)^2$, so no other points are singular.

On $p(A)\ne0$, the form is a determinant in a smaller set of
coefficient coordinates. For $k\ge1$, put
\[
 W=\begin{pmatrix}p^TSp&p^TU\\Vp&Z\end{pmatrix},\qquad Q_{r,k}=\det W.
\]
Fix $A$ with $p(A)\ne0$. The map $(S,U,V,Z)\mapsto W$ is a surjective
linear map onto $\Mat_{k+1}$. Consequently the derivatives in these blocks
vanish precisely when $\rank W\le k-1$. At such a point the entire
differential of $\det W$ is zero, so the $A$-derivatives vanish too.

These singular points form an irreducible set of dimension $N-4$.
Indeed, on a chart $p_i(A)\ne0$ use one symmetric entry, one entry in each column
of $U$, and one entry in each row of $V$ to replace coordinates by the
entries of $W$. This is a regular invertible coordinate change on the chart.
The fibers are an affine space times the irreducible rank-$\le k-1$
matrix variety. The base is the irreducible open set of rank-$r-1$ skew
matrices. The dimension is
\[
 a+\bigl(N-a-(k+1)^2\bigr)+\bigl((k+1)^2-4\bigr)=N-4.
\]
Its closure is $C_{r,k}$. All other singular points are in $M_{r,k}$.
The two varieties do not contain one another: $C_{r,k}$ has points outside
$M_{r,k}$ and has smaller dimension. This proves the claim.
\end{proof}

The determinant singular locus has dimension $N-4$, whereas
$\Sing V(Q_{r,k})$ contains $M_{r,k}$ of dimension $N-3$. Therefore
$Q_{r,k}$ is not linearly equivalent to $\det_n$. Together with the
degeneration, this proves boundary membership.

The coefficient directions in $M_{r,k}$ are translation directions.
The skew directions contribute no translations, as the next lemma
shows. Define the \emph{translation space} of a variety $X$ by
\[
 \Trans(X)=\{v:X+tv=X\text{ for every }t\in\C\}.
\]
It is a vector subspace, and its dimension is preserved by linear equivalence.

\begin{lemma}\label{lem:trans}
The translation space of $Y_r$ is zero. Consequently
\[
 \Trans(M_{r,k})=\{0\}\times\C^{N-a},\qquad
 \dim\Trans(M_{r,k})=n^2-\binom r2.
\]
\end{lemma}

\begin{proof}
For $r=3$, $Y_r=\{0\}$. Otherwise suppose $B\ne0$ is a translation
direction. Since $0\in Y_r$, we have $B\in Y_r$. Write $\rank B=2s$.
After skew congruence, its support is a $2s$-dimensional block. On the
complement choose a skew matrix $A$ of maximal rank $r-2s-1$, and extend
it by zero on the support of $B$. Then $\rank A\le r-3$, but
\[
 \rank(A+B)=(r-2s-1)+2s=r-1.
\]
This contradicts translation invariance. The product assertion follows.
\end{proof}

These facts already prove pairwise inequivalence of the polynomials for
fixed $n$. A linear equivalence must preserve the unique largest singular
component and the dimension of its translation space. That dimension
$n^2-\binom r2$ determines $r$.

They also simplify the stabilizer argument. Any linear stabilizer preserves
$M_{r,k}$, hence its translation space, which is exactly the space of all
$(S,U,V,Z)$ variables. Therefore every infinitesimal stabilizer has
\begin{equation}\label{eq:no-mixing}
 \delta A=T(A).
\end{equation}
Here an \emph{infinitesimal stabilizer} is a linear vector field
$\delta$ satisfying $\delta Q=0$. Equation~\eqref{eq:no-mixing}
therefore follows from the polynomial's singular locus, without
assuming that the stabilizer preserves the original block decomposition.

\subsection{Linear Pfaffian syzygies}

The stabilizer calculation uses two spaces of linear polynomial
relations, or \emph{syzygies}. The first consists of vector-valued
changes annihilated by $p^T$. The second consists of symmetric
matrix-valued changes annihilated by $B\mapsto p^TBp$.

\begin{lemma}\label{lem:syzygies}
Let $u:\Skew_r\to\C^r$ and $F:\Skew_r\to\Sym_r$ be linear.
Then
\begin{align*}
 p(A)^Tu(A)=0\text{ for all }A
 &\quad\Longleftrightarrow\quad u(A)=Ab\text{ for a constant }b,\\
 p(A)^TF(A)p(A)=0\text{ for all }A
 &\quad\Longleftrightarrow\quad F(A)=AB-B^TA\text{ for a constant }B.
\end{align*}
The two spaces have dimensions $r$ and $r^2-1$, respectively.
\end{lemma}

\begin{proof}
Fix $x\ne0$. On the dense rank-$r-1$ subset of $L_x$, the hypotheses
imply respectively $x^Tu(A)=0$ and $x^TF(A)x=0$. They extend to all of
$L_x$. For fixed $x$, either expression is a linear functional of the skew
matrix $A$. The standard nondegenerate pairing identifies such a
functional with a two-form, which we call $K(x)$.
Its polynomial degree in $x$ is one or two, respectively.
The annihilator of $L_x$ is $x\wedge\C^r$, so $x\wedge K(x)=0$.

The identity $x\wedge K(x)=0$ gives a polynomial factorization
with a factor $x$. The following contraction identity produces the
other factor with the required degree.
On polynomial-valued exterior forms define
\[
 D=\sum_{i=1}^r\partial_{x_i}\,\iota_{e_i^*}.
\]
For an exterior $q$-form $\omega$ whose coefficients have degree $d$,
the contraction rule and Euler's identity give
\[
 D(x\wedge\omega)+x\wedge D\omega=(r+d-q)\omega.
\]
Apply this with $q=2$ to obtain
\[
 K(x)=x\wedge\frac{DK(x)}{r+d-2}.
\]
For $d=1$ the second factor is a constant vector. For $d=2$ it is $Bx$
for a constant matrix. Translating the pairing back, and absorbing a
nonzero scalar into the parameters, gives the displayed formulas.
Their converse follows from $Ap(A)=0$.

The vector parameter has zero kernel. The kernel of
$B\mapsto[A\mapsto AB-B^TA]$ consists exactly of scalar matrices:
diagonal entries first force all off-diagonal entries of $B$ to vanish,
and off-diagonal entries then force its diagonal entries to agree.
This proves the dimensions.
\end{proof}

\subsection{The unpadded stabilizer}

The unpadded stabilizer is one of the quotients in the padded
stabilizer calculation. We prove the required upper bound here;
boundary membership will give the reverse inequality.

\begin{proposition}\label{prop:base}
The Lie algebra of the full linear stabilizer of $P_r(A,S)=p(A)^TSp(A)$
has dimension at most $2r^2-1$.
\end{proposition}

\begin{proof}
By~\eqref{eq:no-mixing}, a stabilizer vector field has
$\delta A=T(A)$ and $\delta S=F(A)+R(S)$.
The change $F(A)$ replaces one $S$ by an $A$, so its contribution
has one more $A$-degree than the contributions from $T(A)$ and $R(S)$.
Terms of different degrees cannot cancel. Lemma~\ref{lem:syzygies}
therefore treats $F$ separately and gives at most $r^2-1$ parameters. For the other part put $h(A)=dp_A[T(A)]$.
Its equation is
\[
 2p^TSh+p^TR(S)p=0.
\]
Let $R^*$ be the transpose map for the trace pairing on symmetric matrices.
Then
\begin{equation}\label{eq:Rstar}
 R^*(pp^T)=-(ph^T+hp^T).
\end{equation}
The diagonal entry $(R^*(xx^T))_{ii}$ is a quadratic polynomial in $x$.
It vanishes when $x_i=0$, since every nonzero $x$ can be realized as
$p(A)$. Thus it is divisible by $x_i$, and we can write
\[
 (R^*(xx^T))_{ii}=-2x_i(Cx)_i
\]
for one constant matrix $C$. Substitute $x=p(A)$ in~\eqref{eq:Rstar}
and cancel the nonzero polynomial $p_i$ to obtain $h(A)=Cp(A)$.
The matrix identity determines $R$ from $C$, since rank-one symmetric
matrices span $\Sym_r$. Projection $(T,R)\mapsto C$ therefore has image
dimension at most $r^2$.

It remains to prove that the projection $(T,R)\mapsto C$ is
injective. Suppose $C=0$. Then $h=0$ and $R=0$.
Differentiating $Ap(A)=0$ gives $T(A)p(A)=0$.
For fixed $x\ne0$, density in $L_x$ implies $T(A)x=0$ whenever $Ax=0$.
In particular, if $A$ has rank two, the skew form $T(A)$ kills $\ker A$.
Skew forms on the two-dimensional quotient by that kernel form a
one-dimensional space, so $T(A)$ is a scalar multiple of $A$.
Every decomposable two-form is therefore an eigenvector of $T$.

Elementary wedges form a basis. Sums of two elementary wedges sharing
a vector are decomposable, and force the corresponding eigenvalues to
agree. The graph of pairs joined when they share a vertex is connected
for $r\ge3$. Hence $T=c\,\mathrm{Id}$. Euler's identity gives
$h(A)=dp_A[cA]=cm\,p(A)$, so $h=0$ and $m\ge1$ force $c=0$.
The projection is injective. Adding the $F$-space proves the bound
$r^2+(r^2-1)=2r^2-1$.
\end{proof}

\subsection{Infinitesimal determinant preservers}

The linear changes of a generic matrix that preserve its determinant
up to scalar are infinitesimal row and column operations. We recall
an elementary proof for use in the padding calculation.

\begin{lemma}\label{lem:det}
For $k\ge1$, the space of linear maps $J:\Mat_k\to\Mat_k$ satisfying
$D(\det)_Z[J(Z)]=c\det Z$ for some constant $c$ has dimension $2k^2-1$.
Every such map is $J(Z)=LZ+ZR$, with $c=\tr L+\tr R$.
The subspace with $c=0$ has dimension $2k^2-2$.
\end{lemma}

\begin{proof}
The case $k=1$ is immediate. Write
$J_{ij}(Z)=\sum_{a,b}c_{ij}^{ab}z_{ab}$ and expand
$\sum_{i,j}J_{ij}(Z)\operatorname{cof}_{ij}(Z)$.
If $a\ne i$ and $b\ne j$, choose a cofactor monomial containing $z_{ab}$.
Multiplication by $z_{ab}$ gives a square, with row $i$ missing, row $a$
doubled, column $j$ missing and column $b$ doubled. These degrees uniquely
identify its coefficient, which must vanish. Thus $J_{ij}$ uses only
entries in row $i$ or column $j$.

Monomials with column $j$ missing and column $b$ doubled have exactly
two possible cofactor contributions, with opposite signs. They force
the coefficient of $z_{ib}$ in $J_{ij}$ to be independent of $i$.
The row version gives the corresponding independence of $j$.
These supply the off-diagonal parts of $LZ+ZR$.
After subtracting them, write $J_{ij}=d_{ij}z_{ij}$.
Permutation monomials give $\sum_i d_{i,\sigma(i)}=c$ for every $\sigma$.
Permutations differing on two rows imply the rectangle relations, hence
$d_{ij}=a_i+b_j$. These supply the diagonal parts.
The kernel of $(L,R)\mapsto J$ is $(aI,-aI)$, proving dimension
$2k^2-1$. The trace functional is nonzero, so $c=0$ removes one dimension.
\end{proof}

We also use linear independence of the entries of $\adj Z$. For $k\ge2$
their distinct missing row and column degrees prove this. For $k=1$,
the sole entry is $1$. Thus an identity linear in $\adj Z$ can be tested
by replacing $\adj Z$ with an arbitrary constant matrix.

\subsection{The stabilizer of a padded odd-skew form}

For $k\ge1$, we filter the stabilizer by its action on $Z$, then
by its induced action on the unpadded form, and then by the
$A$-dependent parts of $U,V$. The residual action involves only the
coefficient variables. The dimensions of these successive images
and kernels add to the full stabilizer dimension.

Let $\delta$ be an arbitrary linear vector field annihilating $Q_{r,k}$.
Put $Y=(S,U,V,Z)$.
Equation~\eqref{eq:no-mixing} gives $\delta A=T(A)$.
Terms from $\delta Y(A)$ have degree $2m+1$ in $A$; terms from
$\delta A(A)$ and $\delta Y(Y)$ have degree $2m$.
These groups satisfy separate identities.

\subsubsection{The action on the padding block}

Throughout this calculation, $\delta Z(U)$ means the part of
$\delta Z$ that is linear in the variables $U$; the same notation
is used for every pair of blocks. Comparing the different degrees
in the independent blocks lets us isolate these parts.

The degree-$2m+1$ terms containing $S$ force $\delta Z(A)=0$, by
cofactor independence. Terms quadratic in $S$ force $\delta Z(S)=0$.
The terms containing one $S$ and one $U$ give
\begin{equation}\label{eq:SU}
 (p^TSp)\tr(\adj Z\,\delta Z(U))
 =p^TU\adj Z\,\delta V(S)p.
\end{equation}
Fix $x=p(A)\ne0$ and restrict to $U$ with $x^TU=0$.
Choosing $S$ with $x^TSx\ne0$ shows that the linear functional
$U\mapsto\tr(\adj Z\,\delta Z(U))$ vanishes on this kernel for every $x$.
Every elementary matrix $U$ belongs to one of these kernels, by choosing
$x$ orthogonal to its nonzero column. Hence the functional is zero.
Cofactor independence gives $\delta Z(U)=0$.
Then~\eqref{eq:SU}, using arbitrary $U$ and invertible $Z$, forces
$\delta V(S)=0$. The $S,V$ terms similarly give
$\delta Z(V)=0$ and $\delta U(S)=0$.

Thus $\delta Z$ depends only on $Z$. The terms involving $S$ and no $U,V$ give
\[
 \det Z\bigl(2p^TS\,dp_A[T(A)]+p^T\delta S(S)p\bigr)
 +D(\det)_Z[\delta Z(Z)]\,p^TSp=0.
\]
The first bracket depends only on $A,S$. Choosing them with $p^TSp\ne0$
shows that $D(\det)_Z[\delta Z(Z)]=c\det Z$ for a constant $c$.
By Lemma~\ref{lem:det}, the image of the $Z$-projection has dimension
at most $2k^2-1$.

\subsubsection{Projection to the unpadded stabilizer}

Take the kernel of this projection, so $\delta Z=0$.
The induced changes
\[
 \delta A=T(A),\qquad \delta S=\delta S(A)+\delta S(S)
\]
annihilate $P_r$. Indeed, the degree-$2m$ equation above and the separate
degree-$2m+1$ equation $p^T\delta S(A)p=0$ are precisely its stabilizer
equations. This second projection has image dimension at most $2r^2-1$
by Proposition~\ref{prop:base}.

In the next kernel,
\[
 \delta A=\delta Z=0,\quad
 \delta S(A)=\delta S(S)=0,\quad \delta U(S)=\delta V(S)=0.
\]
The parts $\delta U(A),\delta V(A)$ satisfy
$p^T\delta U(A)=0$ and $\delta V(A)p=0$.
The vector syzygy lemma gives $r$ dimensions per column or row,
for at most $2rk$ dimensions in total.

\subsubsection{The residual coefficient action}

In the kernel of the preceding projections, all $A$-dependent
parts vanish. Since $p(A)$ attains every nonzero vector, the remaining
identity in $U,V,Z$ is
\begin{equation}\label{eq:resid}
 \det Z\,\delta S
 =\sym\bigl(\delta U\adj ZV+U\adj Z\delta V\bigr),
 \qquad \sym B=(B+B^T)/2.
\end{equation}
Terms involving $Z$ alone give $\delta S(Z)=0$.
Terms linear in $V$ give
\[
 \det Z\,\delta S(V)=\sym\bigl(\delta U(Z)\adj ZV\bigr).
\]
On $\det Z\ne0$, let $F(Z)=\delta U(Z)Z^{-1}$.
Then $\delta S(V)=\sym(F(Z)V)$.
The map $F\mapsto[V\mapsto\sym(FV)]$ is injective: take $V$ with one
nonzero column and inspect the corresponding diagonal entry.
The left side does not depend on $Z$, so $F(Z)=B$ is constant. Hence
\[
 \delta U(Z)=BZ,\qquad \delta S(V)=\sym(BV),
\]
with at most $rk$ parameters. Similarly,
\[
 \delta V(Z)=ZC,\qquad \delta S(U)=\sym(UC),
\]
with at most another $rk$ parameters.

Terms quadratic in $V$ give $x^T\delta U(V)\adj ZVx=0$ for all $x$.
Cofactor independence implies
\[
 (x^T\delta U(V))_a(Vx)_b=0\qquad\text{for every }a,b.
\]
The polynomial ring is a domain, so $\delta U(V)=0$.
Similarly $\delta V(U)=0$.

Write $P(U)=\delta U(U)$ and $R(V)=\delta V(V)$.
The remaining equation, for every constant $T\in\Mat_k$, is
\[
 (x^TP(U))T(Vx)+(x^TU)T(R(V)x)=0.
\]
It forces preservation of the kernels of $U\mapsto x^TU$ and $V\mapsto Vx$.
Taking $x=e_i$ shows that each output row of $P$ depends only on the
corresponding input row. Taking $x=e_i+e_j$ makes these row maps agree.
Thus $P(U)=UK$ and similarly $R(V)=JV$ for constant matrices $K,J$.
The equation becomes $KT+TJ=0$ for every $T$. Therefore
$K=\lambda I$ and $J=-\lambda I$, leaving one dimension.

\subsubsection{Stabilizer dimension and componenthood}

Adding the dimensions of the four successive images and kernels gives
\begin{align*}
 \dim\operatorname{LieStab}(Q_{r,k})
 &\le(2k^2-1)+(2r^2-1)+2rk+(2rk+1)\\
 &=2(r+k)^2-1=2n^2-1.
\end{align*}
For $k=0$, Proposition~\ref{prop:base} gives the same bound.

By Lemma~\ref{lem:det}, the determinant orbit has dimension
$n^4-(2n^2-2)$. An orbit is open in its closure, so its boundary has
dimension at most one less. We already proved that $Q_{r,k}$ is in this
boundary. In characteristic zero the stabilizer dimension equals its
Lie-algebra dimension. Boundary membership gives the reverse bound
$\dim\Stab(Q_{r,k})\ge2n^2-1$.
Equality follows, and the orbit dimension is $n^4-2n^2+1$.
Its irreducible closure must be a boundary component: any component
containing it has at most that dimension and hence must equal it.

\subsection{Conciseness and distinctness}

We complete the proof by establishing conciseness and distinctness
of the orbit closures.

A polynomial loses a variable after a linear coordinate change exactly
when some nonzero constant directional derivative annihilates it.
Let such a derivative annihilate $Q_{r,k}$.
Its $A$-part has degree $2m-1$ in $A$, while all other parts have degree
$2m$. The coefficient involving the arbitrary $S$ forces
$dp_A[B]=0$ for its constant skew direction $B$.
Differentiating $Ap(A)=0$ gives $Bp(A)=0$; surjectivity of $p$ gives $B=0$.
For $k\ge1$, the $S$ coefficient then forces the constant $Z$ direction
to be zero by cofactor independence. The $V$ and $U$ coefficients force
the constant $U,V$ directions to vanish. Finally, the symmetric direction
$C$ in $S$ satisfies $p^TCp=0$ for every $p$, so $C=0$.
For $k=0$ only the first and last steps are needed. This proves conciseness.

Section~\ref{sec:odd-singular} proved that different odd $r$ give inequivalent representatives.
Their orbit closures are distinct as well. Each orbit has the full
dimension of a boundary component. If two closures were equal, both
orbits would be dense open subsets of one irreducible variety and would
intersect, which is impossible for distinct orbits.
This proves Theorem~\ref{thm:main}.

There are $\lfloor(n-1)/2\rfloor$ odd integers in $[3,n]$.
The traceless component is represented by a form in $n^2-1$ essential
variables~\cite[Section~8.3]{Hut}. The locus of nonconcise forms is closed,
being a rank condition on first derivatives. Thus it is different from
every component above. This proves Corollary~\ref{cor:number}.

For example, at $n=7$ the invariant is
\begin{center}
\begin{tabular}{cccc}
\toprule
$r$ & $k$ & Largest singular-component dimension & Translation-space dimension\\
\midrule
3&4&46&46\\
5&2&46&39\\
7&0&46&28\\
\bottomrule
\end{tabular}
\end{center}
It distinguishes the three concise components; the traceless component
supplies a fourth.

\section{Boundary components from block matrix spaces}
\label{sec:higher}

We next consider singular matrix spaces of higher generic corank.
Their leading determinant coefficient occurs at a higher power of
$t$. As in the odd-skew case, the singular locus determines the seed
directions, after which the stabilizer dimension can be computed.

The bounded-rank spaces in this section are due to Huang and Landsberg
\cite[Section~9]{HL}; their $6\times6$ member is case (III) of their
Theorem~3.1. The leading-coefficient construction is established in
\cite[Section~6.7.1]{LandsbergBook}. We prove that these leading
coefficients give boundary components after arbitrary padding, and
that the resulting components are pairwise distinct and different
from the odd-skew components.

Choose an integer $h\ge2$. The original matrix has size $r=3h$,
arranged in three blocks of size $h$. After adding $k\ge0$ rows and
columns, its size is $n=r+k$. Let $F\in\Mat_h$ be a matrix of
independent variables,
let $a,b$ be scalar variables, and set
\[
 A=A_h(a,b,F)=
 \begin{pmatrix}0&aI_h&bI_h\\-aI_h&0&F\\-bI_h&-F&0\end{pmatrix},
 \quad
 K_R=\begin{pmatrix}F\\-bI_h\\aI_h\end{pmatrix},
 \quad K_L=\begin{pmatrix}F&-bI_h&aI_h\end{pmatrix}.
\]
Direct multiplication gives $AK_R=0$ and $K_LA=0$. Thus $K_R$
and $K_L$ provide $h$ right and left kernel vectors, respectively.
They define the compression of a coefficient matrix to an $h\times h$
matrix. The space $\mathcal L_h$ has dimension $d=h^2+2$. Choose a linear complement $\mathcal S_h$ to $\mathcal L_h$ in
$\Mat_r$. Thus every $r\times r$ matrix is uniquely a sum of a
matrix in $\mathcal L_h$ and one in $\mathcal S_h$. Write $S$ for
a matrix whose independent coefficients run through this complement. The
remaining variables are $U\in\Mat_{r,k}$, $V\in\Mat_{k,r}$, and
$Z\in\Mat_k$. Together these give $n^2$ variables. Define
\begin{equation}\label{eq:higher-form}
 H_{h,k}=\det_{h+k}
 \begin{pmatrix}K_LSK_R&K_LU\\VK_R&Z\end{pmatrix}.
\end{equation}
The empty blocks are omitted when $k=0$. This formula is useful for studying the polynomial, but it is not
an affine determinantal representation of size $h+k$. Its entries
are not all linear: the degrees of the four blocks
are $3,2,2,1$. Every determinant term has degree $3h+k=n$ and degree $2h$
in the variables $x=(a,b,F)$.

\begin{theorem}\label{thm:higher}
For every $h\ge2$ and $k\ge0$, $H_{h,k}$ is concise in $n^2$ variables and
\[
 \dim\Stab(H_{h,k})=2n^2-1.
\]
Its orbit closure is a determinant-boundary component. For fixed $n$, these
components are pairwise distinct as $h$ ranges from $2$ to $\lfloor n/3\rfloor$,
and are different from every component in Theorem~\ref{thm:main}.
\end{theorem}

\subsection{Degeneration and singular locus}

The generic rank of $A$ is $2h$, so the leading determinant
coefficient occurs at $t^h$. On $a\ne0$, eliminate the leading
$2h\times2h$ block. That block has determinant $a^{2h}$; the normalized
left and right kernel matrices are $K_L/a$ and $K_R/a$. Consequently
\[
 [t^h]\det(A+tS)=\det(K_LSK_R).
\]
Schur elimination of $Z$ then proves, first on $a\det Z\ne0$ and hence
as a polynomial identity,
\begin{equation}\label{eq:higher-limit}
 [t^h]\det\begin{pmatrix}A+tS&U\\tV&Z\end{pmatrix}=H_{h,k}.
\end{equation}
All lower coefficients vanish. For $t\ne0$, the entries of the
displayed matrix give an invertible linear parametrization of $\Mat_n$.
Divide its determinant by $t^h$, which can be done by scaling one
row. Letting $t$ tend to zero proves that $H_{h,k}$ belongs to
$\overline{\Orb_n}$.

If $a\ne0$ or $b\ne0$, the rank of $A$ is $2h$. If $a=b=0$,
it is
$2\rank F$. Thus its rank-drop locus is the irreducible codimension-three
variety
\[
 Y_h=\{a=b=0,\ \det F=0\}\subset\C^d.
\]
The set $Y_h$ determines the largest singular component of
$H_{h,k}$. At a point of $Y_h$, both $K_L$ and $K_R$ lose rank. The matrix in
\eqref{eq:higher-form} then has a left and a right kernel. If its rank is
at most $h+k-2$, all first derivatives vanish. If its rank is $h+k-1$,
its adjugate pairs the indicated left and right kernels; differentiating
either kernel factor leaves the other factor zero. Thus all first
derivatives again vanish. This compression argument covers every $U,V,Z$,
including singular $Z$.

Explicitly, at $a=b=0$ choose nonzero $\ell,q$ with $\ell F=0$ and $Fq=0$. Then
$\ell K_L=0$ and $K_Rq=0$. These give a left kernel $(\ell,0)$ and
a right kernel $(q,0)$ of the compressed matrix. Its derivative paired
against these vectors is
\[
 \ell\,D(K_LSK_R)[\delta x,\delta S]q=0,
\]
because each of the three product-rule terms retains a zero outer factor.
This proves the asserted vanishing in the rank-$h+k-1$ case.

For fixed $x=(a,b,F)$ outside $Y_h$, the map
\[
 (S,U,V,Z)\longmapsto
 W_x=\begin{pmatrix}K_LSK_R&K_LU\\VK_R&Z\end{pmatrix}
\]
is onto $\Mat_{h+k}$. Indeed $K_L,K_R$ have full rank, and
$K_L\mathcal L_hK_R=0$ allows the full matrix space to be replaced by its
complement $\mathcal S_h$. The $S,U,V,Z$ derivatives vanish exactly when
$\rank W_x\le h+k-2$, which also kills the $x$ derivatives. This locus
is locally a product of the irreducible rank-$\le h+k-2$ determinantal
variety with an affine space, over the irreducible base $\C^d\setminus Y_h$.
It has dimension $n^2-4$. Therefore the unique largest singular component is
\begin{equation}\label{eq:higher-M}
 M_{h,k}=Y_h\times\C^{n^2-d},\qquad
 \dim M_{h,k}=n^2-3.
\end{equation}
The generic determinant has singular locus of codimension four. Hence
$H_{h,k}$ is outside its orbit and is a boundary point.

The determinant hypersurface in $\Mat_h$ has no nonzero translation
direction, since its first derivatives are independent. Hence the
translation space of $M_{h,k}$ consists exactly of the coefficient
directions. Its linear span is cut out by $a=b=0$. Therefore
\begin{equation}\label{eq:higher-invariants}
 \Trans(M_{h,k})=\{x=0\},\qquad
 \dim\Trans(M_{h,k})=n^2-(h^2+2),\qquad
 \operatorname{codim}\langle M_{h,k}\rangle=2.
\end{equation}
The last equality uses the fact that singular matrices span $\Mat_h$ for
$h\ge2$. Any linear stabilizer therefore preserves the $S,U,V,Z$
directions: its $x$-part has the form $\delta x=T(x)$.

The same calculation proves conciseness. A constant direction annihilating
$H_{h,k}$ must preserve its unique largest singular component, and thus has
zero $x$-part. The surjective compression above and conciseness of a generic
determinant force its compressed $S,U,V,Z$ direction to vanish for every $x$.
This kills $Z,U,V$, since the kernel matrices have no common linear
annihilators. Finally the common constant kernel of $S\mapsto K_LSK_R$
inside $\Mat_r$ is exactly $\mathcal L_h$. To see this, expand a constant
block matrix $B=(B_{ij})$:
\begin{align*}
 K_LBK_R={}&FB_{11}F-b(FB_{12}+B_{21}F)
 +a(FB_{13}+B_{31}F)\\
 &+b^2B_{22}-ab(B_{23}+B_{32})+a^2B_{33}.
\end{align*}
The coefficients give $B_{11}=B_{22}=B_{33}=0$, $B_{23}=-B_{32}$,
$B_{12}=cI=-B_{21}$, and $B_{13}=eI=-B_{31}$. These describe precisely
$\mathcal L_h$. The chosen complement has zero intersection with it, proving
conciseness.

\subsection{Linear syzygies and the action on the seed space}

The stabilizer splits into the action on the seed variables
$x=(a,b,F)$, the maps from seed variables to coefficient variables,
and the action among coefficient variables. We first compute the
first two contributions. The linear syzygies are matrix-valued
because the kernels have $h$ columns. They follow from the next
identity.
\begin{lemma}\label{lem:matrix-syzygy}
For $h\ge2$, linear maps $X,Y:\Mat_h\to\Mat_h$ satisfy
$FX(F)+Y(F)F=0$ identically if and only if
\[
 X(F)=BF+\lambda(F)I,\qquad Y(F)=-FB-\lambda(F)I
\]
for a unique constant matrix $B$ and a unique scalar linear form $\lambda$.
\end{lemma}
\begin{proof}
Put $C=X(I)$; evaluation at $I$ gives $Y(I)=-C$.
The linear coefficient after substituting $F=I+tM$ yields
$Y(M)=-X(M)-MC+CM$. The quadratic coefficient then gives
$[M,X(M)-CM]=0$ for all $M$. A linear map commuting with its argument
has the form $M\mapsto cM+\mu(M)I$. One elementary verification is as
follows. For a diagonal matrix $D$, its image is diagonal, first for
distinct diagonal entries and then for all $D$ by linearity. Polarization
with each matrix unit $E_{ij}$ shows that the off-diagonal part of its
image is $c_{ij}E_{ij}$ and that the differences between diagonal entries
of the image of $D$ are $c_{ij}(d_i-d_j)$. Comparing three indices forces
one common $c$; for $h=2$ the two opposite pairs give the same conclusion.
After subtracting $c$ times the identity operator, polarization between
distinct matrix units forces every remaining diagonal image to be scalar.
Commutation with $E_{ij}$ itself supplies the same conclusion in the
$2\times2$ case. This proves the stated form of a commuting map.
Substitution now gives the displayed formulas with $B=C+cI$ and
$\lambda=\mu$. Uniqueness follows since $BF$ cannot be a scalar matrix
for every $F$ unless $B=0$.
\end{proof}

\begin{lemma}\label{lem:higher-syzygies}
The space of linear maps $B:x\mapsto\mathcal S_h$ satisfying
$K_LB(x)K_R=0$ has dimension $14h^2-4$. The left and right vector
syzygy spaces $K_Lu(x)=0$ and $v(x)K_R=0$ each have dimension $r=3h$.
\end{lemma}
\begin{proof}
Combine $B_{23}+B_{32}$ into one block, since the opposite pair
belongs to $\mathcal L_h$. There are eight block-valued
linear functions remaining. Decompose each into its $a$, $b$, and $F$ parts.
We compare the independent groups of monomials separately. For
example, the $aF^2$ terms are linear in $a$ and quadratic in the
entries of $F$.
The $F^3$ coefficient forces $B_{11}(F)=0$. The $aF^2$ equation is
\[
 FC F+FX(F)+Y(F)F=0.
\]
By Lemma~\ref{lem:matrix-syzygy}, its solutions are
$X=(B-C)F+\lambda(F)I$, $Y=-FB-\lambda(F)I$: there are $3h^2$
parameters, including $C$. The $bF^2$ equation contributes another
$3h^2$. The $a^2F,b^2F,abF$ equations determine respectively the $F$
parts of $B_{33},B_{22},B_{23}+B_{32}$ from the $a,b$ parts of
$B_{12},B_{21},B_{13},B_{31}$; these supply $8h^2$ free parameters.
The $a^3,b^3$ coefficients set $B_{33}(a)=B_{22}(b)=0$, and the
$a^2b,ab^2$ equations leave $2h^2$ more parameters. Thus the count before
the remaining quotient is $16h^2$. The two scalar opposite block pairs
in $\mathcal L_h$ each permit an arbitrary linear form in $d=h^2+2$
variables. Quotienting subtracts $2d$, leaving $14h^2-4$.

For the left vector equation, write $u=(u_0,u_1,u_2)$, so that
$Fu_0-bu_1+au_2=0$. The part quadratic in $F$ forces $u_0(F)=0$
by invertibility of generic $F$. Coefficient comparison then gives
\[
 u_0=a p+b q,\qquad u_1=Fq+a w,\qquad
 u_2=-Fp+b w
\]
for arbitrary constant $p,q,w\in\C^h$. These are exactly the columns
$A c$ with constant $c$. They have dimension $3h$. Transposing the
calculation proves the right-vector statement.
\end{proof}

The induced action on $x$ must preserve $Y_h$ and its linear span.
Consequently $(\delta a,\delta b)$ is any linear function of $(a,b)$,
and
\begin{equation}\label{eq:higher-T}
 \delta F=CF+FD+aE+bG.
\end{equation}
The familiar determinant-similitude calculation gives dimension
$2h^2-1$ for $F\mapsto CF+FD$; it holds also for $h=2$. Including
the other terms, the space of possible $T$ has dimension
\begin{equation}\label{eq:higher-T-dim}
 4+(2h^2-1)+2h^2=4h^2+3.
\end{equation}
Every action in~\eqref{eq:higher-T} extends to a stabilizer of the
whole polynomial by a compensating action on the coefficient blocks.
Indeed, changes in
$(a,b)$ and the translations $F\mapsto F+aE+bG$ induce constant block
actions on $K_L,K_R$, which are canceled by opposite actions on $S,U,V$.
For the remaining term $CF+FD$, explicitly
\[
 \delta K_L=CK_L+K_L\operatorname{diag}(D,-C,-C),\qquad
 \delta K_R=\operatorname{diag}(C,-D,-D)K_R+K_RD.
\]
Cancel the internal block actions on $S,U,V$. The compressed matrix then
changes by left multiplication by $\operatorname{diag}(C,0_k)$ and right
multiplication by $\operatorname{diag}(D,0_k)$. Subtracting
$\tau S$ and $\tau U$ from $\delta S,\delta U$, with
$\tau=(\tr C+\tr D)/h$, cancels the resulting determinant character.
Whenever necessary project $\delta S$ to the chosen complement; this
does not change its compression. Hence the upper bound in
\eqref{eq:higher-T-dim} is attained.

\subsection{The coefficient stabilizer}

For fixed $x$, the polynomial is an ordinary determinant after
compression. A linear coefficient symmetry must induce a determinant
symmetry for every $x$. The following lemma determines the actions
satisfying these simultaneous conditions.

We retain the block notation of the construction.
Fix $h\geq2$ and $k\geq0$, and put $r=3h$ and $n=r+k$.
For $F\in\Mat_h$ and $a,b\in\C$, write
\[
 A(x)=\begin{pmatrix}
 0&aI_h&bI_h\\
 -aI_h&0&F\\
 -bI_h&-F&0
 \end{pmatrix},\qquad
 K_R(x)=\begin{pmatrix}F\\-bI_h\\aI_h\end{pmatrix},
 \qquad K_L(x)=\begin{pmatrix}F&-bI_h&aI_h\end{pmatrix}.
\]
Then $A(x)K_R(x)=0$ and $K_L(x)A(x)=0$.
Let $L\subset\Mat_r$ be the space of the matrices $A(x)$, and let
$L_{\mathrm{top}}\subset\Mat_n$ be its embedding in the upper-left block.
Adding an element of $L_{\mathrm{top}}$ does not change the
compressed coefficient matrix. We therefore identify the coefficient
space with the quotient
\[
 \mathcal Y=\Mat_n/L_{\mathrm{top}}.
\]
After choosing a complement to $L$, its coordinates are
$Y=(S,U,V,Z)$, with block sizes $r\times r$, $r\times k$,
$k\times r$, and $k\times k$, respectively. Set
\[
 B_R(x)=\operatorname{diag}(K_R(x),I_k),\qquad
 B_L(x)=\operatorname{diag}(K_L(x),I_k),
\]
and
\begin{equation}\label{eq:higher-compression}
 M_x(Y)=B_L(x)YB_R(x)
 =\begin{pmatrix}
 K_L(x)S K_R(x)&K_L(x)U\\
 VK_R(x)&Z
 \end{pmatrix},\qquad
 H_{h,k}(x,Y)=\det_{h+k}M_x(Y).
\end{equation}
These expressions are well defined on the quotient $\mathcal Y$.
The matrices $B_L$ and $B_R$ combine the seed kernel vectors with
the padding coordinates. We call $Y\mapsto B_LYB_R$ the
\emph{compression} map.

\begin{lemma}\label{lemma:higher-pure}
Let $\mathfrak p_{h,k}$ be the space of constant linear operators
$R:\mathcal Y\to\mathcal Y$ such that
\[
 D_YH_{h,k}(x,Y)[R(Y)]=0
 \qquad\text{for every }x,Y.
\]
Then
\[
 \dim\mathfrak p_{h,k}=2k^2+2rk.
\]
In particular, $\mathfrak p_{h,0}=0$.
More explicitly, after representing the quotient by a complement and
writing $\pi:\Mat_r\to\Mat_r/L$ for the quotient map, all its elements
are uniquely parametrized by
\[
 B_0\in\Mat_{r,k},\quad C_0\in\Mat_{k,r},\quad
 E_0,G_0\in\Mat_k,\quad \alpha\in\C,
 \qquad h\alpha+\tr E_0+\tr G_0=0,
\]
through the vector fields
\begin{align*}
 \delta S&=\pi(\alpha S+B_0V+UC_0),&
 \delta U&=\alpha U+B_0Z+UE_0,\\
 \delta V&=G_0V+ZC_0,&
 \delta Z&=G_0Z+ZE_0.
\end{align*}
For $k=0$, the empty blocks are omitted and the trace condition forces
$\alpha=0$.
\end{lemma}

\begin{proof}
We prove that a coefficient symmetry preserves the kernel of each
compression and hence induces an infinitesimal determinant preserver.
Comparing these induced actions as $x$ varies determines their
parameters.

We use the trace pairing to identify dual matrix spaces. First, the
common constant kernel of $Y\mapsto M_x(Y)$ is precisely
$L_{\mathrm{top}}$. Indeed, the seed compression expands as
\begin{align*}
 K_L S K_R={}&FS_{11}F-b(FS_{12}+S_{21}F)
       +a(FS_{13}+S_{31}F)\\
 &+b^2S_{22}-ab(S_{23}+S_{32})+a^2S_{33}.
\end{align*}
Its quadratic part in $F$ gives $S_{11}=0$. The coefficient of $b$
gives $FS_{12}+S_{21}F=0$ for every $F$, hence
$S_{12}=cI_h$ and $S_{21}=-cI_h$. The coefficient of $a$ gives the
analogous scalar pair $S_{13}=dI_h$, $S_{31}=-dI_h$.
The other coefficients give
$S_{22}=S_{33}=0$ and $S_{23}=-S_{32}$, exactly the matrices in $L$.
Furthermore, $K_LU=0$ for every $x$ forces the three block rows of
$U$ to vanish, and similarly $VK_R=0$ forces $V=0$; the $Z$ block is
unchanged. It follows, in particular, that the dual compression
spaces introduced below span $\mathcal Y^*$.

For the comparison it is enough to use values with $a\ne0$.
Rescale them to $a=1$, and write $t=b$. Polynomial identities obtained
on this dense set extend to all seed values. For every $F,t$, the map
$M_x:\mathcal Y\to\Mat_{h+k}$ is surjective: the blocks
$S_{33},U_3,V_3,Z$ already realize every compressed matrix.
We claim that $R\in\mathfrak p_{h,k}$ preserves $\ker M_x$.
For $Y_0\in\ker M_x$, substitute $Y+zY_0$ in its stabilizer identity
and take the coefficient of $z$. This gives
\[
 D(\det_{h+k})_T[M_x(R(Y_0))]=0
 \qquad\text{for every }T\in\Mat_{h+k}.
\]
The generic determinant is concise in its entries, so
$M_x(R(Y_0))=0$, proving the claim.

Dualizing the compression maps, the transpose $R^*$ preserves each
subspace
\[
 W_x=\{B_R(x)T B_L(x):T\in\Mat_{h+k}\}\subset\mathcal Y^*.
\]
Write $T=\left(\begin{smallmatrix}H&B\\C&D\end{smallmatrix}\right)$
in block sizes $h,k$. On the chosen chart its image is
\begin{equation}\label{eq:higher-dual-fiber}
 W_{F,t}(T)=
 \begin{pmatrix}
 FHF&-tFH&FH&FB\\
 -tHF&t^2H&-tH&-tB\\
 HF&-tH&H&B\\
 CF&-tC&C&D
 \end{pmatrix}.
\end{equation}
The submatrix on its third and fourth block rows and columns is
$T$ itself. Thus the uniquely induced operator $\Psi_{F,t}$, defined by
\[
 R^*(W_{F,t}(T))=W_{F,t}(\Psi_{F,t}(T)),
\]
has coefficients polynomial in $F,t$, of total degree at most two.
The induced primal operator preserves the determinant, and the dual
of a left--right determinant Lie action is again such an action.
The classical determinant stabilizer description therefore gives
\begin{equation}\label{eq:higher-fiber-action}
 \Psi_{F,t}(T)=E_{F,t}T+TG_{F,t},
 \qquad \tr E_{F,t}+\tr G_{F,t}=0.
\end{equation}
The left and right matrices are unique up to adding a scalar
identity to one and subtracting it from the other. We fix this
ambiguity before comparing degrees.
The kernel of $(E,G)\mapsto[T\mapsto ET+TG]$ consists of
$(cI,-cI)$. The condition $\tr E_{11}=0$, where the block $E_{11}$
has size $h$, removes this kernel and gives a fixed linear splitting
of this finite-dimensional map. Applying that splitting to
$\Psi_{F,t}$ makes all entries of $E_{F,t},G_{F,t}$ polynomial in
$F,t$. This also fixes the scalar ambiguity throughout the proof.

The repeated blocks in~\eqref{eq:higher-dual-fiber} impose degree
constraints on this action. In particular, a block occurring both
without a factor and with a factor $t^2$ must satisfy both degree
bounds.

Let $H'$ be the upper-left block of $\Psi_{F,t}(T)$. Its appearance
as the core $(3,3)$ block of \eqref{eq:higher-dual-fiber} shows that
it has total degree at most two in $F,t$. Its appearance as
$t^2H'$ in the core $(2,2)$ block shows that it is independent of
$F,t$. Since
\[
 H'=E_{11}H+E_{12}C+HG_{11}+BG_{21},
\]
the matrices $E_{12},G_{21}$ and the operator
$H\mapsto E_{11}H+HG_{11}$ are constant. The fixed gauge implies
that $E_{11},G_{11}$ themselves are constant.
For input $C$ alone, the left side of the defining equation for
$\Psi_{F,t}$ has degree at most one in $F,t$, whereas the output
core $(1,1)$ block is $FE_{12}CF$. Its quadratic degree in $F$
forces $E_{12}=0$. The same argument with input $B$ gives
$G_{21}=0$.

The upper-left action is constant. To prove that it is scalar,
put $\Theta(H)=E_{11}H+HG_{11}$.
For input $H$ and $t=0$, the only input block of degree two in
$F$ is $FHF$. Comparing degree two in the core $(1,1)$ output
therefore gives a fixed linear map $J:\Mat_h\to\Mat_h$ with
\[
 J(FHF)=F\Theta(H)F.
\]
At $F=I_h$, this says $J=\Theta$, so
\begin{equation}\label{eq:higher-jordan-identity}
 \Theta(FHF)=F\Theta(H)F.
\end{equation}
Replacing $F$ by $I_h+sX$ and taking the coefficient of $s$ gives
\[
 \Theta(XH+HX)=X\Theta(H)+\Theta(H)X.
\]
If $B_* =\Theta(I_h)$, setting $H=I_h$ yields
$\Theta(X)=(XB_*+B_*X)/2$.
Substitution back in the preceding identity gives
$[H,[X,B_*]]=0$ for every $H,X$.
Thus $[X,B_*]$ is central; its trace is zero, so it is zero.
Therefore $B_*$ is scalar and $\Theta(H)=\alpha H$.
The gauge $\tr E_{11}=0$ now gives
\begin{equation}\label{eq:higher-upper-blocks}
 E_{11}=0,\qquad G_{11}=\alpha I_h,
 \qquad E_{12}=G_{21}=0.
\end{equation}

It remains to determine the interactions with the padding. These
are governed by the blocks of degree one in $F,t$ in
\eqref{eq:higher-dual-fiber}. Suppose $k>0$. The upper-right block
of the smaller matrix is now
\[
 B'=HG_{12}+BG_{22}.
\]
For input $B$ alone, the top extra block of its image is
$FBG_{22}(F,t)$, whereas all input entries have total degree at most
one. Any positive-degree homogeneous part of $G_{22}$ would
produce degree at least two after multiplication by generic $F$
and $B$, and must vanish. Hence $G_{22}$ is constant; similarly
$E_{22}$ is constant. Denote them by $G_0,E_0$.

For input $H$, the second extra block is $-tHG_{12}(F,t)$.
It has degree at most two, so $G_{12}$ has degree at most one.
Write
\[
 G_{12}(F,t)=B_3+tB_2+\Gamma(F)
\]
with $\Gamma$ linear. Comparing the terms quadratic in $F$ in the
top extra block, the only input of that degree is again $FHF$.
Hence a fixed linear map $J_0:\Mat_h\to\Mat_{h,k}$ satisfies
\[
 J_0(FHF)=FH\Gamma(F).
\]
At $F=I_h$ this gives $J_0(H)=HB_1$, where $B_1=\Gamma(I_h)$.
Consequently $FH\Gamma(F)=FHFB_1$ for all $F,H$, so
$\Gamma(F)=FB_1$ on the invertible locus, and hence everywhere.
Therefore
\[
 G_{12}=FB_1+tB_2+B_3=K_LB_0,
 \qquad B_0=\begin{pmatrix}B_1\\-B_2\\B_3\end{pmatrix}
 \in\Mat_{r,k}.
\]
The row version of the same argument gives
$E_{21}=C_0K_R$ for a constant $C_0\in\Mat_{k,r}$.
We have proved that the fiber pair is necessarily
\begin{equation}\label{eq:higher-pure-parameters}
 E_{F,t}=\begin{pmatrix}0&0\\C_0K_R&E_0\end{pmatrix},
 \qquad
 G_{F,t}=\begin{pmatrix}\alpha I_h&K_LB_0\\0&G_0\end{pmatrix},
 \qquad h\alpha+\tr E_0+\tr G_0=0.
\end{equation}
For $k=0$ the same conclusion follows directly from
\eqref{eq:higher-upper-blocks}, with empty off-diagonal and lower
blocks and with $h\alpha=0$.

Conversely, every parameter choice in~\eqref{eq:higher-pure-parameters}
defines a symmetry. Set
\[
 M_0=\begin{pmatrix}0_{r\times r}&0\\C_0&E_0\end{pmatrix},
 \qquad
 N_0=\begin{pmatrix}\alpha I_r&B_0\\0&G_0\end{pmatrix}.
\]
They satisfy
$M_0B_R=B_RE_{F,t}$ and $B_LN_0=G_{F,t}B_L$.
The constant linear operator $W\mapsto M_0W+WN_0$ consequently
restricts on each dual fiber to \eqref{eq:higher-fiber-action}.
Equivalently its dual is the quotient operator
$Y\mapsto N_0Y+YM_0$; this is well defined because it sends
$L_{\mathrm{top}}$ into itself. In the chosen coordinates this is
exactly the vector field in the statement. The trace condition in
\eqref{eq:higher-pure-parameters} makes it a stabilizer.

Finally, the fiber parameters are unique in the fixed gauge, the
maps $B_0\mapsto K_LB_0$ and $C_0\mapsto C_0K_R$ are injective,
and the dual fibers span $\mathcal Y^*$. Thus the displayed
parameters are independent as operators. Their dimension is
\[
 rk+kr+k^2+k^2+1-1=2rk+2k^2,
\]
as claimed.
\end{proof}

\subsection{Stabilizer dimension and distinctness}

Write $y=(S,U,V,Z)$. Every stabilizer has $\delta x=T(x)$, as
proved above. In the kernel of this projection, write
$\delta y=B(x)+R(y)$. The two terms have different $x$-degrees after
differentiation, namely $2h+1$ and $2h$. Thus they vanish separately.
For a fixed generic $x$, surjectivity of compression and conciseness of
the determinant give $\pi_x B(x)=0$. These equations are precisely the
matrix syzygy for $B_S$, the left and right vector syzygies for the
$k$ columns and rows of $B_U,B_V$, and $B_Z=0$. By
Lemma~\ref{lem:higher-syzygies}, their total dimension is
$14h^2-4+2rk$. Lemma~\ref{lemma:higher-pure} gives $2k^2+2rk$ for
the pure $y$-action. Adding the image and these two kernels yields
\begin{align*}
 \dim\Stab(H_{h,k})
 &=(4h^2+3)+(14h^2-4+2rk)+(2k^2+2rk)\\
 &=18h^2+4rk+2k^2-1=2n^2-1.
\end{align*}
Since $H_{h,k}$ is a boundary point, the stabilizer dimension
$2n^2-1$ proves that its orbit closure is a boundary component.

The translation dimension in \eqref{eq:higher-invariants} recovers $h$,
so it distinguishes the higher-corank components from one another.
To distinguish them from all odd-skew components, use the codimension of
the linear span of the largest singular component. It is $2$ here.
For $r=3$ in the odd-skew family it is $3$; for every odd $r\ge5$ it is
$0$, since the rank-$\le r-3$ skew matrices contain all rank-two skew
matrices and therefore span the full skew space. Thus the two families remain distinct even when their translation
dimensions agree. Equal orbit closures would have equal dense orbits, so these
orbit invariants distinguish the components. This finishes the proof of
Theorem~\ref{thm:higher}.

\begin{corollary}\label{cor:strong-count}
For every $n\ge3$, the determinant boundary has at least
\[
 \left\lfloor\frac{n-1}{2}\right\rfloor+
 \left\lfloor\frac n3\right\rfloor-1
\]
distinct concise components. Including the traceless component gives
\[
 \#\operatorname{Irr}(\partial\overline{\Orb_n})\ \ge\
 \left\lfloor\frac{n-1}{2}\right\rfloor+
 \left\lfloor\frac n3\right\rfloor.
\]
\end{corollary}
The $h=1$ block construction recovers the $r=3$ odd-skew family and is
therefore counted only once. For $n=6,7,9,12$, the displayed total lower
bounds are respectively $4,5,7,9$. They remain lower bounds, not a
classification of all components.

\section{Boundary components from middle Koszul maps}
\label{sec:koszul}

The middle Koszul maps give a third family of boundary components.
The first seed has size ten, so arbitrary padding supplies every
size $n\ge10$. These are the forms used for the exact-complexity
bound in Theorem~\ref{thm:intro-gap}.

We apply the leading-coefficient construction of
\cite[Section~6.7.1]{LandsbergBook} to these classical spaces of
exterior-multiplication matrices. We prove the component and padding
statements uniformly in the exterior degree. Landsberg and Manivel
\cite[Section~3.1, Proposition~3]{LMConstant} identify the common
constant compression kernel in a more general rectangular setting.
We include an elementary proof in the square setting used here.

Recall that $\Lambda^p V$ has a basis indexed by $p$-element subsets
of a chosen basis of $V$. The exterior product is alternating on
vectors, so a product with a repeated vector is zero. Fix $p\ge2$
and put
\[
 d=2p+1,\quad V=\C^d,\quad E=\Lambda^pV,\quad
 r=\binom{2p+1}{p},\quad R=\binom{2p}{p},\quad
 c=r-R=\binom{2p}{p-1}.
\]
Here $d$ is the dimension of the seed space, $r$ is the matrix size,
$R$ is its rank at nonzero $v$, and $c$ is its corank. For $p=2$,
\[
 d=5,\qquad r=10,\qquad R=6,\qquad c=4.
\]
Choose a volume form on $V$. It identifies $\Lambda^{p+1}V$ with
$E^*$, so exterior multiplication by $v\in V$ becomes a square
matrix $A(v):E\to E^*$.
Equivalently, up to a fixed sign,
\[
 A(v)(\alpha,\beta)=\operatorname{vol}(v\wedge\alpha\wedge\beta),
 \qquad \alpha,\beta\in E.
\]
In the exterior-power bases, every entry is zero or a signed
coordinate of $v$. The transpose is $(-1)^pA(v)$.

For $v\ne0$, choose a basis with first vector $v$. A basis wedge is
killed by multiplication by $v$ exactly when it already contains
$v$. Counting wedges that do not contain it gives rank $R$; counting
those that do gives kernel dimension $c$. Thus the left and right
kernels are both
\[
 U_v=v\wedge\Lambda^{p-1}V,\qquad \dim U_v=c.
\]
Let $\mathcal L_p=\{A(v):v\in V\}$, a $d$-dimensional matrix space,
and choose a complement $\mathcal S_p$ in $\Mat_r$.
For $k\ge0$, set $n=r+k$ and let
$S\in\mathcal S_p$, $U\in\Mat_{r,k}$, $V_0\in\Mat_{k,r}$,
and $Z\in\Mat_k$ be independent coefficient variables. Define
\begin{equation}\label{eq:koszul-limit}
 \mathcal K_{p,k}(v,S,U,V_0,Z)
 =[t^c]\det\begin{pmatrix}A(v)+tS&U\\tV_0&Z\end{pmatrix}.
\end{equation}
The notation $[t^c]$ denotes the coefficient of $t^c$. The empty
blocks are omitted for $k=0$, and there are $n^2$ independent
variables. Balancing the entries selected from block rows and block
columns shows that this coefficient has degree $R$ in $v$ and
degree $c+k$ in the other variables.
In particular its total degree is $n$.

\begin{theorem}\label{thm:koszul-family}
For every $p\ge2$ and $k\ge0$, the polynomial $\mathcal K_{p,k}$
is concise in $n^2$ variables and satisfies
\[
 \dim\Stab(\mathcal K_{p,k})=2n^2-1.
\]
Its orbit closure is a determinant-boundary component.
For fixed $n$, the components obtained from all
$p\ge2$ with $\binom{2p+1}{p}\le n$ are pairwise distinct.
They are different from every component in the odd-skew and
higher-corank block families.
\end{theorem}

\subsection{Constant and linear kernel relations}

For each nonzero $v$, the coefficient matrix determines a bilinear
form on $E\oplus\C^k$. Restricting both of its arguments to
$U_v\oplus\C^k$ gives a smaller matrix of size $c+k$. This is the
compression used in the proof.

We determine the constant matrices, and the matrices linear in
$v$, whose restrictions vanish for every $v$. The constant kernel
is used to prove conciseness. The linear kernel determines the
stabilizer maps from $v$ to the coefficient variables.

\begin{lemma}\label{lemma:koszul-syzygies}
The common constant kernel of
$B\mapsto B|_{U_v\times U_v}$, for $B\in E^*\otimes E^*$,
is $\mathcal L_p$, of dimension $d$.
The space of matrix-valued linear maps $B(v)$ satisfying
$B(v)|_{U_v\times U_v}=0$ for every $v$ has dimension $2r^2-1$.
After quotienting by the $\mathcal L_p$-valued maps, its dimension is
$2r^2-d^2-1$.
The space of linear maps $b(v)\in E^*$ annihilating $U_v$ has
dimension $r$; the analogous left and right assertions coincide.
\end{lemma}
\begin{proof}
We begin with the constant matrices. A bilinear form $B$ on
$\Lambda^pV$ can be viewed as a function of $2p$ vector arguments,
alternating within each of two groups of $p$ arguments. Its vanishing on
$(v\wedge\alpha,v\wedge\beta)$, polarized in $v$, makes it
antisymmetric under exchanging the first arguments of the two blocks.
Together with alternation within each block, this generates all
transpositions. The tensor is therefore fully alternating.
Conversely every fully alternating tensor vanishes when the two
blocks share $v$. The common kernel is thus
$\Lambda^{2p}V^*$, identified with $\mathcal L_p$ by the volume form.
The same polarization argument for a tensor in
$V^*\otimes\Lambda^pV^*$ proves the vector assertion: its kernel is
$\Lambda^{p+1}V^*$, of dimension $r$.

For matrices $B(v)$ linear in $v$, we compute the kernel by
decomposing the relevant tensor spaces into irreducible
$\GL(V)$-modules.

Put $W=V^*$.
The map in question is the double exterior coproduct followed by
symmetric multiplication,
\begin{equation}\label{eq:koszul-cubic-map}
 W\otimes\Lambda^pW\otimes\Lambda^pW
 \longrightarrow
 \Sym^3W\otimes\Lambda^{p-1}W\otimes\Lambda^{p-1}W.
\end{equation}
Entrywise this is exactly $B(v)\mapsto K(v)^TB(v)K(v)$, where
$K(v):\Lambda^{p-1}V\to\Lambda^pV$ is exterior multiplication by $v$.
We write $\mathbb S_\lambda W$ for the irreducible $\GL(W)$
representation with symmetry type indexed by the partition $\lambda$.
The partition is pictured by rows of boxes of lengths $\lambda_i$;
its width is the largest row length. The Pieri rules describe what
happens to these types when one tensors with an exterior or symmetric
power~\cite[Section~8.1.3]{LandsbergBook}. Applied here, they first give
\[
 \Lambda^pW\otimes\Lambda^pW
 =\bigoplus_{j=0}^{p}
 \mathbb S_{(2^j,1^{2p-2j})}W.
\]
Tensoring with $W$ gives the multiplicity decomposition
\begin{equation}\label{eq:koszul-pieri-domain}
 \bigoplus_{j=1}^{p}\mathcal A_j
 \ \oplus\ 2\bigoplus_{j=1}^{p}\mathcal C_j
 \ \oplus\ \mathcal C_0,
\end{equation}
where
\[
 \mathcal A_j=\mathbb S_{(3,2^{j-1},1^{2p-2j})}W,
 \qquad
 \mathcal C_j=\mathbb S_{(2^j,1^{2p+1-2j})}W,
 \quad \mathcal C_0=\Lambda^dW.
\]
The map is $\GL(W)$-equivariant, so an irreducible summand can have
nonzero image only if the same representation occurs in the target.
The target of~\eqref{eq:koszul-cubic-map} has no summand of width
at most two: tensoring with $\Sym^3W$ adds a horizontal strip of
three boxes, using three columns. Hence every $\mathcal C_j$ maps
to zero.

Each $\mathcal A_j$ occurs once in the source. To show that its
restriction is nonzero, choose a basis $e_1,\ldots,e_d$ of $W$ and
put $W'=\langle e_2,\ldots,e_d\rangle$.
Choose a nonzero highest-weight vector $z_j$ of shape
$(2^{j-1},1^{2p-2j})$ in
$\Lambda^{p-1}W'\otimes\Lambda^{p-1}W'$, which occurs by Pieri.
Wedging $e_1$ into each factor gives
\[
 w_j=((e_1\wedge -)\otimes(e_1\wedge -))z_j.
\]
This is a highest-weight vector of shape $(2^j,1^{2p-2j})$:
the raising operators within $W'$ kill $z_j$, while a raising
operator introducing $e_1$ creates a repeated factor.
Consequently $e_1\otimes w_j$ is a highest-weight vector of type
$\mathcal A_j$. Its image under~\eqref{eq:koszul-cubic-map} has
$e_1^3$ coefficient $z_j\ne0$.
Every $\mathcal A_j$ occurs once in the source, so complete
reducibility now identifies the kernel as
$\mathcal C_0\oplus2\bigoplus_{j=1}^p\mathcal C_j$.
Finally, another application of Pieri gives
\[
 \Lambda^pW\otimes\Lambda^{p+1}W
       =\bigoplus_{j=0}^{p}\mathcal C_j.
\]
Both exterior powers have dimension $r$, and
$\dim\mathcal C_0=1$. The kernel dimension is therefore
$1+2(r^2-1)=2r^2-1$.
The space of linear maps $V\to\mathcal L_p$ has dimension $d^2$
and is contained in this kernel, giving the quotient assertion.
\end{proof}

In particular, the common constant kernel of restriction from
$\Mat_n$ to the spaces $U_v\oplus\C^k$ is exactly
$\mathcal L_p$ in the upper-left block. Indeed the seed assertion
was just proved, the spaces $U_v$ span $E$, and the remaining
$Z$ block is unrestricted. Thus the coefficient space can be
identified intrinsically for the calculation as
\[
 \mathcal Y=\Mat_n/(\mathcal L_p)_{\mathrm{top}}.
\]
The choice of a complement in~\eqref{eq:koszul-limit} does not affect
the resulting equivalence class.

\subsection{Local determinant form and multiplicity loci}

The compressed determinant gives a local description of the
polynomial and its multiplicity loci. The largest component of a suitable
multiplicity locus recovers the seed directions, as the singular
locus did for the preceding families.

All coefficients below $t^c$ in~\eqref{eq:koszul-limit} vanish.
On $\det Z\ne0$, this follows by Schur elimination and the
constant rank $R$ of $A(v)$ for $v\ne0$; it then holds identically.
The leading coefficient is nonzero. For $t\ne0$ the displayed
matrix parametrizes $\Mat_n$ invertibly, so its normalized limit
belongs to $\overline{\Orb_n}$.

Near a fixed nonzero $v$, choose bases on a Zariski open chart
that reduce $A(v)$ to an invertible $R\times R$ block and a zero
$c\times c$ block.
Eliminating the invertible block shows that
\begin{equation}\label{eq:koszul-local}
 \mathcal K_{p,k}(v,Y)=u(v)\det_{c+k}(\kappa_v(Y)),
 \qquad u(v)\ne0,
\end{equation}
where $\kappa_v$ is restriction to the left and right kernels,
including the padding coordinates.
It is surjective: restriction of an arbitrary bilinear form to
$U_v\oplus\C^k$ is surjective, and
$(\mathcal L_p)_{\mathrm{top}}$ is in its kernel.
The entries of $\kappa_v(Y)$ are therefore part of a local
coordinate system, in which the polynomial is a generic determinant
times a unit. The \emph{multiplicity} at a point is the least degree
of a nonzero term after translating that point to the origin. For
a determinant, this is the corank. Hence the multiplicity here is
\[
 (c+k)-\rank\kappa_v(Y).
\]

Since every monomial has degree $R$ in $v$, the multiplicity is
at least $R$ along $v=0$. Put $N=n^2$. Thus the linear space
\[
 M_p=\{v=0\}\times\mathcal Y,
 \qquad\dim M_p=N-d,
\]
is contained in its multiplicity-at-least-$R$ locus.
Outside $M_p$, the same multiplicity locus is empty if $c+k<R$,
and otherwise has
codimension $R^2$ by~\eqref{eq:koszul-local} and the usual rank
codimension for a square matrix. Since $R^2>d$ for $p\ge2$,
$M_p$ is the unique largest irreducible component of this
multiplicity locus. Every linear stabilizer preserves it.
Thus every infinitesimal stabilizer satisfies $\delta v=T(v)$,
with no coefficient-variable contribution to $\delta v$.

To prove conciseness, consider a constant directional derivative
annihilating the polynomial. Its $v$-direction has $v$-degree $R-1$ and the coefficient
directions have degree $R$, so they vanish separately.
Changing the basis of $V$ changes the exterior-power bases in a
compatible way. In particular, it gives a linear $\GL(V)$ action on the
coefficient quotient, under which $\mathcal K_{p,k}$ transforms by
a character. Hence its space of annihilating constant
$v$-directions is $\GL(V)$-stable. Since $V$ is irreducible, that
space is zero or all of $V$; the latter would contradict its
positive $v$-degree and nonvanishing.
For the remaining direction,~\eqref{eq:koszul-local} and
conciseness of the determinant imply $\kappa_v(Y_0)=0$ for every
$v\ne0$. The common constant kernel in the quotient is zero, so
$Y_0=0$.

The same local form also gives
\begin{equation}\label{eq:koszul-singular-dimension}
 \dim\Sing(\mathcal K_{p,k})=N-4.
\end{equation}
Indeed on $v\ne0$ the singular determinant locus has codimension
four, since $c+k\ge4$. The omitted locus $M_p$ has codimension
$d\ge5$ and cannot increase that dimension.

\subsection{The coefficient stabilizer under padding}
\label{subsec:koszul-padding-rigidity}

For each $v$, a coefficient stabilizer induces row and column
operations on the compressed determinant. The induced operations
must arise from a single linear map on the coefficient space.
Intersections of the kernel spaces impose the compatibility
conditions used below.

The next lemma also holds for $p=1$. Fix $p\ge1$ and $k\ge0$, and put
\[
 V=\C^{2p+1},\qquad E=\bigwedge^p V,\qquad
 r=\binom{2p+1}{p},\qquad
 c=\binom{2p}{p-1},\qquad T=\C^k.
\]
For $0\ne v\in V$, set
\[
 U_v=v\wedge\bigwedge^{p-1}V\subset E,\qquad P_v=U_v\oplus T.
\]
Then $\dim U_v=c$. A volume form on $V$ identifies the middle Koszul
map $A(v):\bigwedge^pV\to\bigwedge^{p+1}V$ with a bilinear form on $E$;
its left and right kernels are both $U_v$.
Let $\mathcal L=A(V)$, embedded in the upper-left block of
$\Mat_{r+k}$, and regard the coefficient space as
\[
 \mathcal Y=\Mat_{r+k}/\mathcal L.
\]
By the common-kernel calculation above, restriction of bilinear
forms to $U_v\times U_v$, for all $v\ne0$, has common kernel
$\mathcal L$. Since the $U_v$ span $E$, its padded version says that
the dual compression spaces
\begin{equation}\label{eq:koszul-dual-spaces}
 W_v=P_v\otimes P_v\subset\mathcal Y^*
 \quad\text{satisfy}\quad
 \sum_{v\ne0}W_v=\mathcal Y^*.
\end{equation}
Thus the images of the dual compression maps span the full dual
coefficient space. Splitting off the padding directions gives four
blocks. If
$W_0=\sum_{v\ne0}U_v\otimes U_v$, then
\begin{equation}\label{eq:koszul-dual-splitting}
 \mathcal Y^*=W_0\oplus(E\otimes T)\oplus(T\otimes E)
                  \oplus(T\otimes T).
\end{equation}

For each $v\ne0$, compression gives a surjection
$\pi_v:\mathcal Y\to P_v^*\otimes P_v^*$. The padded leading
determinant coefficient, denoted by $\mathcal K_{p,k}(v,Y)$, is a nonzero scalar
multiple of $\det_{c+k}(\pi_vY)$ when $v$ is fixed. The scalar can
depend on the choice of bases and on $v$; it has no effect on the
following infinitesimal statement.

\begin{lemma}\label{lemma:koszul-pure}
The space of linear operators $R:\mathcal Y\to\mathcal Y$ satisfying
\[
 D_Y\mathcal K_{p,k}(v,Y)[R(Y)]=0\qquad\text{for all }v,Y
\]
has dimension
\[
 \dim\mathfrak p_{p,k}=2k^2+2rk.
\]
More precisely, its elements are uniquely parametrized by
\[
 \alpha\in\C,\qquad C,D\in\operatorname{End}(T),\qquad
 \Phi,\Psi\in\operatorname{Hom}(E,T),\qquad
 c\alpha+\tr C+\tr D=0.
\]
Their dual action on \eqref{eq:koszul-dual-splitting} is the restriction of
\begin{equation}\label{eq:koszul-pure-operator}
 A\otimes\mathrm{id}+\mathrm{id}\otimes B,
 \qquad
 A=\begin{pmatrix}0&0\\\Phi&C\end{pmatrix},\qquad
 B=\begin{pmatrix}\alpha I_E&0\\\Psi&D\end{pmatrix},
\end{equation}
where the matrices act on $E\oplus T$.
For $k=0$, the trace condition forces $\alpha=0$, so the space is zero.
\end{lemma}

We use two elementary facts. First, the determinant Lie algebra
consists of the left--right actions $H\mapsto EH+HF$ with
$\tr E+\tr F=0$. The following argument gives the required
description directly. If $L$ is tangent to
the determinant hypersurface, fix nonzero left and right null vectors
$a,b$ and consider
\[
 Z_{a,b}=\{H:a^{\mathsf t}H=0,\ Hb=0\}.
\]
Rank-$(m-1)$ matrices are dense in this space. At these matrices the
adjugate is a nonzero multiple of $ba^{\mathsf t}$, so tangency gives
$a^{\mathsf t}L(H)b=0$ throughout $Z_{a,b}$. In particular,
\[
 L(xy^{\mathsf t})\in
 x\C^{m\,\mathsf t}+\C^m y^{\mathsf t}
\]
for all nonzero $x,y$. At a matrix unit $E_{ij}$, this means that
$L(E_{ij})$ is supported on its row and column. Applying the same
condition to $(e_i+e_s)e_j^{\mathsf t}$ and
$e_i(e_j+e_t)^{\mathsf t}$ makes the corresponding off-diagonal
coefficients independent of $i$ and $j$, respectively. Subtracting
the resulting left--right action leaves $L(E_{ij})=q_{ij}E_{ij}$.
The condition at $(e_i+e_s)(e_j+e_t)^{\mathsf t}$ gives
$q_{ij}-q_{it}-q_{sj}+q_{st}=0$, hence $q_{ij}=u_i+w_j$.
These are the diagonal left--right actions. Finally,
$D\det_H[EH+HF]=(\tr E+\tr F)\det H$ gives the trace condition.
The size-one case is immediate. Taking duals gives the same
description as $E\otimes\mathrm{id}+\mathrm{id}\otimes F$ on the
dual matrix space.

Second, a linear endomorphism of an exterior power that preserves
every decomposable line is scalar. Indeed, the wedge products of
basis vectors are eigenvectors. If two such products differ in
exactly one factor, their sum, with an appropriate choice of sign,
is also decomposable, so their eigenvalues agree. The graph of
subsets of fixed cardinality joined by these replacements is
connected. The assertion follows; exterior powers of dimension one
are included.

\begin{proof}[Proof of Lemma~\ref{lemma:koszul-pure}]
Let $R$ satisfy the stated identity. For fixed $v\ne0$, it preserves
$\ker\pi_v$. To see this, take $Y_0\in\ker\pi_v$, replace $Y$ by
$Y+sY_0$ in the identity, and extract the coefficient of $s$. It gives
\[
 D\det_H[\pi_vR(Y_0)]=0\qquad\text{for all }H\in\Mat_{c+k}.
\]
The first derivatives of the generic determinant are linearly
independent, so $\pi_vR(Y_0)=0$. Consequently $R^*$ preserves $W_v$,
and the determinant Lie-algebra calculation gives
\begin{equation}\label{eq:koszul-local-action}
 R^*|_{W_v}=a_v\otimes\mathrm{id}+\mathrm{id}\otimes b_v,
 \qquad \tr a_v+\tr b_v=0,
\end{equation}
for $a_v,b_v\in\operatorname{End}(P_v)$.

We compare the actions $a_v,b_v$ on intersections of the spaces
$W_v$ to obtain the constant block formula in the statement. For subspaces
$E_i$ of a common vector space,
\[
 \bigcap_i(E_i\otimes E_i)
 =\left(\bigcap_i E_i\right)\otimes
  \left(\bigcap_i E_i\right).
\]
This follows from $E\otimes E=(E\otimes H)\cap(H\otimes E)$,
where $H$ is the ambient space. Since $\bigcap_{v\ne0}U_v=0$, it follows
that $\bigcap_{v\ne0}W_v=T\otimes T$.
Thus $R^*$ preserves $T\otimes T$. In \eqref{eq:koszul-local-action},
projection to $(P_v/T)\otimes P_v$ and to $P_v\otimes(P_v/T)$
separately shows that both $a_v$ and $b_v$ preserve $T$.
When $T=0$, this statement is automatic.

If $v_1,\ldots,v_p$ are linearly independent and
$q=v_1\wedge\cdots\wedge v_p$, then
\[
 \bigcap_{i=1}^p U_{v_i}=\C q,
 \qquad
 \bigcap_{i=1}^p W_{v_i}=(\C q\oplus T)\otimes(\C q\oplus T).
\]
The first equality says that a degree-$p$ exterior form divisible by
each of $p$ independent vectors is divisible by their product.
Hence, for each fixed $v$ and each decomposable $q\in U_v$, both
$a_v$ and $b_v$ preserve $\C q\oplus T$: apply the same two
quotient projections to the invariant tensor square just displayed.
On $P_v/T=U_v\cong\bigwedge^{p-1}(V/\C v)$, the induced operators
therefore preserve every decomposable line. The second elementary
fact makes them scalar. This also covers $p=1$, when $U_v$ is a line.

The induced actions on $U_v$ are scalar. The decomposition
in~\eqref{eq:koszul-local-action} is unique up to
$(a_v,b_v)\mapsto(a_v+sI,b_v-sI)$. Use this freedom to make the
$U_v$-block of $a_v$ zero. We obtain unique local data
\begin{equation}\label{eq:koszul-local-blocks}
 a_v=\begin{pmatrix}0&0\\\phi_v&C_v\end{pmatrix},\qquad
 b_v=\begin{pmatrix}\alpha_v I_{U_v}&0\\\psi_v&D_v\end{pmatrix},
\end{equation}
where $\phi_v,\psi_v:U_v\to T$.

The scalar $\alpha_v$ is independent of $v$, also for $p=1$.
To prove this, project $R^*|_{W_0}$ to $W_0$ in
\eqref{eq:koszul-dual-splitting}. Every decomposable $p$-form $q$
belongs to some $U_v$, and \eqref{eq:koszul-local-blocks} gives
\[
 R^*_{00}(q\otimes q)=\alpha_v(q\otimes q).
\]
The cone of these nonzero tensors $q\otimes q$ is irreducible: it is
the image of the irreducible cone of decomposable $p$-forms under
$q\mapsto q\otimes q$. A linear operator has only finitely many
eigenvalues, and its eigenspaces are closed. This irreducible cone
must lie in a single eigenspace. Therefore every $\alpha_v$ has the
same value. No assertion about the span of decomposable squares in
$\operatorname{Sym}^2(U_v)$ is required here.

If $k=0$, all remaining blocks in \eqref{eq:koszul-local-blocks} are
absent. The trace condition is $c\alpha=0$, whence $\alpha=0$ and
$R^*=0$ by \eqref{eq:koszul-dual-spaces}. We may henceforth assume
$k>0$.

We next show that $C_v,D_v$ are constant. Project the restriction of
$R^*$ to $E\otimes T$ back to $E\otimes T$. For $q\in U_v$ and
$t\in T$, the result is
\[
 q\otimes D_vt.
\]
Choose a basis of $T$ and express this projected operator as a matrix
of linear operators $E\to E$. Each such operator preserves every
decomposable line in $E$, since every decomposable $q$ belongs to
some $U_v$. The second elementary fact makes every matrix entry a
scalar operator. Thus the projection is
$\mathrm{id}_E\otimes D$ for a single $D\in\operatorname{End}(T)$,
and $D_v=D$. Similarly the projection of $R^*|_{T\otimes E}$ onto
$T\otimes E$ acts locally by $(C_v+\alpha I_T)\otimes\mathrm{id}$.
It follows that $C_v=C$ for a single $C$.

Finally, the maps $\phi_v$ and $\psi_v$ extend to linear maps
$\Phi,\Psi:E\to T$. The following argument applies uniformly
for $p\ge1$. Let $F$ be the projection of
$R^*|_{E\otimes T}$ onto $T\otimes T$. Formula
\eqref{eq:koszul-local-blocks} gives
\[
 F(q\otimes t)=\phi_v(q)\otimes t
 \qquad(q\in U_v).
\]
Fix $0\ne t_0\in T$ and $\eta\in T^*$ with $\eta(t_0)=1$, and set
\[
 \Phi(u)=(\mathrm{id}_T\otimes\eta)F(u\otimes t_0).
\]
This is linear in $u$ and restricts to $\phi_v$ on every $U_v$.
Since the $U_v$ span $E$, the identity
$F(u\otimes t)=\Phi(u)\otimes t$ holds on all of $E\otimes T$.
The analogous projection from $T\otimes E$ gives a global
$\Psi$ restricting to every $\psi_v$.

Thus $R^*$ agrees with \eqref{eq:koszul-pure-operator} on every $W_v$,
and hence on all of $\mathcal Y^*$ by their spanning property.
The local trace condition becomes the single equation
\[
 c\alpha+\tr C+\tr D=0.
\]
Conversely, the operators in \eqref{eq:koszul-pure-operator} preserve
every $P_v\otimes P_v$, and this trace equation makes their induced
actions annihilate every compressed determinant. They therefore
give exactly the required pure coefficient stabilizers.

The parameters are unique. Indeed the local left--right ambiguity
has already been removed by the zero $U_v$-block of $a_v$; agreement
on all the $U_v$ determines $\Phi,\Psi$ because these subspaces span
$E$. There are $1+2k^2+2rk$ parameters and one nonzero linear trace
condition, proving the dimension formula.
\end{proof}

\subsection{Stabilizer dimension and distinctness}

The stabilizer dimension is the sum of three contributions: the
action on $v$, the maps from $v$ to the coefficient variables, and
the coefficient action.
The image of the stabilizer on the $v$-variables has dimension at
most $d^2$, and all of $\mathfrak{gl}(V)$ occurs.
Indeed, if $G=\Lambda^p g$ for $g\in\GL(V)$, exterior equivariance
gives
\[
 A(gv)=L_gA(v)R_g,\qquad
 L_g=(\det g)G^{-T},\qquad R_g=G^{-1}.
\]
The changes $S\mapsto L_gSR_g$, $U\mapsto L_gU$,
$V_0\mapsto V_0R_g$, $Z\mapsto Z$ descend to the coefficient
quotient and multiply~\eqref{eq:koszul-limit} by
$\det L_g\det R_g=(\det g)^{r-2c}$.
This scalar character is cancelled infinitesimally by a dilation
of $Y$, because its coefficient degree is $c+k>0$.

In the kernel of this projection, the pieces
$\delta Y=B(v)$ and $\delta Y=R_0(Y)$ have different $v$-degrees
in the stabilizer identity and are independent.
By~\eqref{eq:koszul-local}, the first kind is a stabilizer exactly
when $\kappa_v(B(v))=0$ for every $v$.
Lemma~\ref{lemma:koszul-syzygies} gives
$2r^2-d^2-1$ dimensions in the $S$ block and $r$ in each of the
$k$ added columns and rows. The $Z$ block must vanish.
This contributes $2r^2-d^2-1+2rk$ dimensions.
Lemma~\ref{lemma:koszul-pure} gives $2k^2+2rk$ for the pure
coefficient-variable stabilizer. Therefore
\[
 \dim\Stab(\mathcal K_{p,k})
 =d^2+(2r^2-d^2-1+2rk)+(2k^2+2rk)
 =2(r+k)^2-1.
\]
The orbit has codimension one in $\overline{\Orb_n}$ and is
outside the determinant orbit, whose stabilizer has dimension
$2n^2-2$. Its closure is therefore a boundary component.

The dimensions of the multiplicity loci distinguish the components
for different parameter values.

Fix $n$ and let $p<q$ be two allowed indices.
Write $R_p=\binom{2p}{p}$ and $d_p=2p+1$.
The multiplicity-at-least-$R_p$ locus of the $p$-form has
dimension $N-d_p$.
For the $q$-form, its part with nonzero seed vector has
codimension $R_p^2$ or is empty, by its local determinant form;
the remaining part has dimension $N-d_q$.
Thus its dimension is at most
\[
 \max\{N-d_q,\ N-R_p^2\}<N-d_p,
\]
because $d_q>d_p$ and $R_p^2>d_p$.
This invariant separates the two linear-equivalence classes.
If their orbit closures were the same component, their dense
orbits would coincide, so the components are distinct.
Finally~\eqref{eq:koszul-singular-dimension} separates this family
from the odd-skew and higher-corank block families, whose largest
singular components have dimension $N-3$.

Consequently this family adds
\[
 \kappa(n)=\#\left\{p\ge2:\binom{2p+1}{p}\le n\right\}
\]
distinct concise components. Its first seed sizes are $10$, $35$,
$126$, and $462$.

\section{The component count}

The three families are pairwise distinct. The largest singular
components of the odd-skew and block families have codimension
three. The codimensions of their linear spans separate the two
families, while their translation dimensions recover the parameter
within each family. The Koszul family has singular locus of
codimension four, and its multiplicity loci recover its parameter.
These invariants are preserved by linear changes of variables.
The omitted endpoint cases $h=1$ and $p=1$ both recover the
$3\times3$ skew construction.

\begin{corollary}\label{cor:combined-count}
For every $n\ge3$, put
$\kappa(n)=\#\{p\ge2:\binom{2p+1}{p}\le n\}$.
The determinant boundary has at least
\[
 \left\lfloor\frac{n-1}{2}\right\rfloor+
 \left\lfloor\frac n3\right\rfloor+\kappa(n)-1
\]
distinct concise components, and at least
\[
 \#\operatorname{Irr}(\partial\overline{\Orb_n})\ \ge\
 \left\lfloor\frac{n-1}{2}\right\rfloor+
 \left\lfloor\frac n3\right\rfloor+\kappa(n)
\]
components in total after including the traceless component.
\end{corollary}

\begin{center}
\begin{tabular}{rrrrrr}
\toprule
$n$&Odd-skew&Block family&Koszul family&Traceless&Total bound\\
\midrule
6&2&1&0&1&4\\
7&3&1&0&1&5\\
10&4&2&1&1&8\\
12&5&3&1&1&10\\
35&17&10&2&1&30\\
100&49&32&2&1&84\\
\bottomrule
\end{tabular}
\end{center}
The count concerns explicitly constructed components; it does not assert
that these exhaust the boundary. In particular, known extra components
at small sizes need not be represented by this uniform lower bound.

\section{A general component criterion}
\label{sec:all-corank-framework}

The preceding stabilizer calculations admit a general formulation.
Two quantities determine the excess orbit codimension: first-order
deformations of the singular matrix space beyond row and column
operations, and additional block-diagonal symmetries of its leading
determinant coefficient. Under the recognition and conciseness
hypotheses below, the orbit codimension is one plus the sum of
these two excesses.

Let $L\subset\Mat_r$ be a $d$-dimensional linear space of singular
matrices with generic rank $R<r$. Put $c=r-R$, and assume $d<rR$.
Choose $\Mat_r=L\oplus S_0$, write $A$ and $S$ for its two generic
matrices, and set
\[
 P(A,S)=[t^c]\det(A+tS).
\]
The form $P$ has degree $R$ in $A$ and degree $c$ in $S$.
Assume that it is concise in its $r^2$ variables and that every
linear vector field annihilating $P$ preserves the $S_0$-directions.
Explicitly, its $A$-part has the form $\delta A=T(A)$, with no
$S$-to-$A$ term.

Let $\mathcal F_{d,R}$ be the closed subscheme of the Grassmannian
$\operatorname{Gr}(d,\Mat_r)$ defined by the vanishing of all
$(R+1)$-minors on the parametrized matrix space. Thus it parametrizes
$d$-dimensional linear spaces of matrices of rank at most $R$.
This is the \emph{Fano scheme} used below.

A tangent vector at $[L]$ is a first-order perturbation
$A\mapsto A+\varepsilon F(A)$, with $\varepsilon^2=0$, satisfying
the linearized rank equations. Define
\begin{align*}
 f&=\dim T_{[L]}\mathcal F_{d,R},\\
 \mathfrak n(L)&=\{(C,D)\in\Mat_r\oplus\Mat_r:
                  CA+AD\in L\text{ for every }A\in L\},
 &\nu&=\dim\mathfrak n(L),\\
 b&=\dim\{(T,J)\in\operatorname{End}(L)\oplus
             \operatorname{End}(S_0):
             D_AP[T(A)]+D_SP[J(S)]=0\}.
\end{align*}
The space $\mathfrak n(L)$ is the normalizer Lie algebra for row
and column operations. The number $b$ is the dimension of the
block-diagonal stabilizer, whose elements act separately on $A$
and $S$.

There are $2r^2$ parameters in an arbitrary pair of row and column
changes. Thus the orbit of $L$ under these changes has dimension
$f_0=2r^2-\nu$. Since its tangent space is contained in the Fano
tangent space, $\delta=f-f_0\ge0$.

\begin{theorem}\label{thm:all-corank}
Under the stated hypotheses, $b\ge\nu-1$.
Put $\rho=b-(\nu-1)\ge0$. Then $P$ belongs to the determinant
boundary and
\[
 \dim\Stab(P)=f+b=2r^2-1+\delta+\rho,
 \qquad
 \operatorname{codim}_{\overline{\Orb_r}}\overline{\GL_{r^2}\cdot P}
       =1+\delta+\rho.
\]
Its orbit closure is a boundary component if and only if
$\delta=\rho=0$. Equivalently, the Fano tangent space consists
exactly of infinitesimal row and column changes, and the
block-diagonal stabilizer has the minimum dimension $\nu-1$.
\end{theorem}

\begin{proof}
The seed-to-coefficient stabilizers are the tangent vectors counted
by $f$. At a generic $A\in L$ of rank
$R$, let $\kappa_A(B)$ be the
induced map $\ker A\to\operatorname{coker}A$, of matrix size $c$.
A perturbation $B$ preserves rank at most $R$ to first order exactly
when it maps $\ker A$ into $\im A$. This is the condition
$\kappa_A(B)=0$, so the tangent space to the rank-at-most-$R$ variety
at $A$ is $\ker\kappa_A$. Since $L$ lies in that variety,
$\kappa_A(L)=0$, and hence $\kappa_A:S_0\to\Mat_c$ is surjective.
After choosing bases locally on the rank-$R$ locus,
\[
 P(A,S)=u(A)\det_c\kappa_A(S),\qquad u(A)\ne0.
\]
A tangent vector to $\operatorname{Gr}(d,\Mat_r)$ is a map
$F:L\to\Mat_r/L$, represented in $S_0$.
Linearizing the defining minors shows that it lies in
$T_{[L]}\mathcal F_{d,R}$ exactly when
$\kappa_A(F(A))=0$ for every generic $A\in L$.
It is enough to test the generic locus because these are
polynomial identities; at smaller rank all first derivatives
of the $(R+1)$-minors already vanish.
By surjectivity of $\kappa_A$ and conciseness of the determinant,
the same condition is equivalent to
$D_SP(A,S)[F(A)]=0$ for all $A,S$.
Thus the $A$-to-$S$ stabilizers have dimension exactly $f$.

By the recognition hypothesis, every stabilizer has
$\delta A=T(A)$ and $\delta S=F(A)+J(S)$.
The $F$-part has bidegree $(R+1,c-1)$ in the stabilizer identity,
while the other two parts have bidegree $(R,c)$.
They vanish separately, proving $\dim\Stab(P)=f+b$.

To prove $b\ge\nu-1$, we combine the action of the trace-zero
normalizer with the weighted Euler action. We first compute the
kernel of the normalizer action. For $(C,D)\in\mathfrak n(L)$ let
$N(X)=CX+XD$ and impose $\tr C+\tr D=0$.
This is one independent condition on $\mathfrak n(L)$, since
$(I_r,0)$ belongs to it. Write
\[
 N(A)\in L,\qquad N(S)=H(S)+J(S),
 \qquad H(S)\in L,\quad J(S)\in S_0.
\]
The determinant identity
$D\det_X[N(X)]=(\tr C+\tr D)\det X=0$, applied to $X=A+tS$,
shows at its coefficient of $t^c$ that
$\delta A=N(A)$, $\delta S=J(S)$ is a block-diagonal stabilizer
of $P$.
Suppose this induced field vanishes. Then $N(A)=0$ and
$N(S)=H(S)\in L$. In that case
\[
 0=D\det_{A+tS}[tH(S)]
   =tD_A\bigl(\det(A+tS)\bigr)[H(S)].
\]
The coefficient of $t^{c+1}$ is $D_AP(A,S)[H(S)]=0$.
The recognition hypothesis forces the pure $S$-to-$A$ field
$H$ to vanish. Hence $N=0$ on all of $\Mat_r$, whose only pair
representatives are $(C,D)=(\lambda I_r,-\lambda I_r)$.
The trace-zero normalizer therefore has an image of dimension
$\nu-2$ in the block-diagonal stabilizer.

Every such image has ordinary trace zero on
$L\oplus S_0=\Mat_r$: its trace equals that of the upper triangular
ambient operator $N$, namely $r(\tr C+\tr D)=0$.
The bidegree $(R,c)$ of $P$ also gives the weighted Euler stabilizer
\[
 \delta A=cA,\qquad\delta S=-RS.
\]
Its ordinary trace is
$cd-R(r^2-d)=r(d-rR)\ne0$.
It is therefore independent of the normalizer image, giving
$b\ge(\nu-2)+1=\nu-1$.

The normalized leading coefficient belongs to
$\overline{\Orb_r}$, since $A+tS$ is an invertible linear
parametrization of $\Mat_r$ for $t\ne0$.
Using $f=2r^2-\nu+\delta$ and $b=\nu-1+\rho$ gives
$\dim\Stab(P)=2r^2-1+\delta+\rho$.
This is strictly greater than the determinant stabilizer dimension
$2r^2-2$, so $P$ is outside the orbit, and the displayed
codimension formula follows. Finally, the determinant boundary
is pure of codimension one~\cite[Corollary~5.4.1]{Chen}, so the
orbit closure is a component exactly when $\delta+\rho=0$.
Both terms are nonnegative, proving the criterion.
\end{proof}

The equality $\delta=0$ expresses \emph{infinitesimal rigidity}:
every first-order deformation within the bounded-rank locus is
induced by row and column operations. The equality $\rho=0$ gives
the minimum block-diagonal stabilizer dimension. Both equalities
are required, together with the stated conciseness and recognition
hypotheses.

\subsection{Infinitesimal rigidity of the block and Koszul spaces}
\label{subsec:seed-infinitesimal-rigidity}

We verify infinitesimal rigidity for the block and Koszul spaces.
Every first-order deformation preserving the rank bound is induced
by row and column operations. In the Koszul case, compare the
rigidity of the associated homogeneous bundles discussed
in~\cite{LMConstant}.

For this proof write $\mathfrak F_{d,R}(r)$ for the parameter
scheme just described: it is the closed subscheme of
$\operatorname{Gr}(d,\Mat_r)$ parametrizing $d$-dimensional vector spaces
of matrices of rank at most $R$. Thus its equations say that every
$(R+1)\times(R+1)$ minor vanishes identically on the parametrized space.
The left--right group $G=\operatorname{GL}_r\times\operatorname{GL}_r$
acts on this scheme by row and column operations.

\begin{proposition}\label{prop:seed-infinitesimal-rigidity}
The following points of the indicated Fano schemes are smooth, and their
left--right orbits are open in the unique irreducible components through
them:
\begin{enumerate}
\item the block space $\mathcal L_h$ of Section~\ref{sec:higher}, with
      $h\ge2$, $r=3h$, $d=h^2+2$, and $R=2h$;
\item the middle Koszul space $\mathcal L_p$ of Section~\ref{sec:koszul},
      with $p\ge2$, $r=\binom{2p+1}{p}$, $d=2p+1$, and
      $R=\binom{2p}{p}$.
\end{enumerate}
Their tangent dimensions are, respectively,
\[
 \dim T_{[\mathcal L_h]}\mathfrak F_{d,R}(r)=14h^2-4,
 \qquad
 \dim T_{[\mathcal L_p]}\mathfrak F_{d,R}(r)=2r^2-d^2-1.
\]
In particular, every first-order deformation of either seed as a
bounded-rank linear space is induced by row and column operations.
\end{proposition}

\begin{proof}
For a matrix space $\mathcal L$ of generic rank $R$, a Grassmannian tangent
vector is a linear map $B:\mathcal L\to\Mat_r/\mathcal L$.
Choose a lift and write the corresponding deformation as
$A(x)+\epsilon B(x)$, where $\epsilon^2=0$. At a rank-$R$ matrix, the
linearized determinantal equations are exactly
\begin{equation}\label{eq:fano-compression-tangent}
 K_L(x)B(x)K_R(x)=0,
\end{equation}
where $K_R$ and $K_L$ span its right and left kernels. Equivalently,
$B(x)$ maps the right kernel into the image of $A(x)$.
The rank-$R$ locus is dense in $\mathcal L$, so these identities there
are equivalent to the vanishing of the linearizations of all restricted
$(R+1)$-minors as polynomials in $x$. Adding an $\mathcal L$-valued map
to the lift of $B$ only changes the parametrization of the seed.
Consequently~\eqref{eq:fano-compression-tangent} computes the full
scheme tangent space. Lemma~\ref{lem:higher-syzygies} and
Lemma~\ref{lemma:koszul-syzygies} give the two dimensions in the statement.

It remains to show that row and column operations account for these
entire tangent spaces. Write the infinitesimal left--right action as
$A\mapsto CA+AD$, absorbing the usual sign of the second group factor
into $D$. Its normalizer Lie algebra is
\[
 \mathfrak n(\mathcal L)=
 \{(C,D)\in\Mat_r^2: CA+AD\in\mathcal L
                   \text{ for every }A\in\mathcal L\}.
\]
The kernel of the induced action on $\mathcal L$ consists of pairs
with $CA+AD=0$ for every $A\in\mathcal L$. We show that this kernel
is exactly
\begin{equation}\label{eq:seed-pointwise-normalizer}
 (C,D)=(-\lambda I_r,\lambda I_r),\qquad \lambda\in\C.
\end{equation}

Such a pair must preserve every right kernel under $D$. For the block
space, work on $a=1$ and put $b=t$. The right kernels are the graphs
\[
 \operatorname{im}\begin{pmatrix}F\\-tI_h\\I_h\end{pmatrix}.
\]
Write $D=(D_{ij})$ in $h\times h$ blocks. Preservation of these graphs
gives
\[
\begin{aligned}
 D_{11}F-tD_{12}+D_{13}
     &=F(D_{31}F-tD_{32}+D_{33}),\\
 D_{21}F-tD_{22}+D_{23}
     &=-t(D_{31}F-tD_{32}+D_{33}).
\end{aligned}
\]
Comparison in $F,t$ first gives
$D_{21}=D_{23}=D_{31}=D_{32}=0$ and $D_{22}=D_{33}$.
The first equation then gives $D_{12}=D_{13}=0$ and
$D_{11}F=FD_{33}$ for every $F$. Thus all three diagonal blocks are
the same scalar matrix, and $D=\lambda I_r$.

For the Koszul space, $D$ preserves every
$U_v=v\wedge\Lambda^{p-1}V$. If $v_1,\ldots,v_p$ are independent,
\[
 \bigcap_{i=1}^p U_{v_i}
   =\C(v_1\wedge\cdots\wedge v_p).
\]
Hence $D$ preserves every decomposable line in $\Lambda^pV$ and is
scalar. One elementary verification is that the basis wedges are
eigenvectors, and the sum of two basis wedges differing in one factor
is decomposable; the connected graph of these replacements makes all
eigenvalues equal.

In both cases the images of the matrices in $\mathcal L$ span the
entire target. For $\mathcal L_h$, the matrices obtained by varying
$a$ and $b$ alone already have this property. For $\mathcal L_p$, the
images $v\wedge\Lambda^pV$ span $\Lambda^{p+1}V$ before the volume-form
identification. Thus $(C+\lambda I_r)A=0$ for every $A$ implies
$C=-\lambda I_r$, proving~\eqref{eq:seed-pointwise-normalizer}.

The coordinate actions on the block seed admit explicit row--column
lifts. Write $E_{ij}(B)$ for a block matrix with $B$ in position $(i,j)$.
The change $\delta F=PF+FQ$ is induced by
$C=\operatorname{diag}(-Q,P,P)$ and
$D=\operatorname{diag}(-P,Q,Q)$ in $\delta A=CA+AD$.
The changes $\delta F=aE$ and $\delta F=bG$ are induced respectively by
$(C,D)=(-E_{31}(E),-E_{13}(E))$ and
$(C,D)=(E_{21}(G),E_{12}(G))$.
The off-diagonal generators $\delta a=b$ and $\delta b=a$ are induced by
$(E_{23}(I_h),E_{32}(I_h))$ and
$(E_{32}(I_h),E_{23}(I_h))$.
Finally, the diagonal generators $\delta a=a$ and $\delta b=b$ use
$(\operatorname{diag}(0,0,-I_h),\operatorname{diag}(I_h,I_h,0))$ and
$(\operatorname{diag}(0,-I_h,0),\operatorname{diag}(I_h,0,I_h))$.
Thus every coordinate action counted in
\eqref{eq:higher-T-dim} is induced by row and column operations.

For $\mathcal L_h$, the rank-drop-locus calculation and explicit
row--column lifts leading to~\eqref{eq:higher-T-dim} show that the
induced normalizer action on the seed coordinates has dimension
$4h^2+3$. For $\mathcal L_p$, exterior equivariance realizes all of
$\mathfrak{gl}(V)$, so the induced action has dimension $d^2$.
Adding the one-dimensional kernel in
\eqref{eq:seed-pointwise-normalizer} gives
\[
 \dim\mathfrak n(\mathcal L_h)=4h^2+4,
 \qquad
 \dim\mathfrak n(\mathcal L_p)=d^2+1.
\]
The corresponding orbit dimensions are therefore
\[
 2(3h)^2-(4h^2+4)=14h^2-4,
 \qquad
 2r^2-(d^2+1)=2r^2-d^2-1.
\]
They equal the full Fano tangent dimensions. The differential of the
orbit map is consequently surjective onto the tangent space.

Equality of these dimensions also proves smoothness. The smooth orbit through
either point supplies a local dimension lower bound equal to the
tangent dimension. The reverse
inequality holds for every finite-type scheme, so the local dimension
equals the embedding dimension, namely the tangent-space dimension.
Its local ring is regular, and over
$\C$ the Fano scheme is smooth at the point. There is a unique
irreducible component through it; the orbit has the same dimension as
that component and is locally closed, hence open in it.
\end{proof}

In the notation of Theorem~\ref{thm:all-corank}, the proposition gives
$\delta=0$ for both seed families. Their block-diagonal stabilizer
calculations give $b=\nu-1$, hence $\rho=0$ as well. Thus both excess
terms in the general component criterion vanish for these seeds.

\section{A criterion for corank-one padding}
\label{sec:general-padding}

For a seed of generic corank one, the orbit codimension after
padding has an exact formula under the hypotheses below. The
formula depends on the linear relations among the left and right
kernel coordinates. Relations beyond those supplied by the seed
matrix produce additional stabilizers for each added row and
column. The resulting criterion applies both to the odd-skew family
and to components whose positive paddings cease to be components.

Let $L\subset\Mat_r$, $r\ge3$, consist of singular matrices and have
generic rank $r-1$. Write $A=A(x)$ for its generic matrix, choose a
complement $K$ with $\Mat_r=L\oplus K$, and let $S$ be generic in $K$.
Set
\[
 C(x)=\adj A(x)=g(x)u(x)v(x)^T,
 \qquad P(x,S)=\tr(C(x)S).
\]
Here $g$ is a nonzero polynomial, and $u,v$ are primitive right
and left polynomial kernel vectors. Primitive means that the
coordinates of each vector have no common nonconstant factor.
We impose the following hypotheses:
\begin{enumerate}\setlength{\itemsep}{0pt}
\item[(H1)] $P$ is concise in its $r^2$ coordinates: every one of
these variables is essential.
\item[(H2)] The vectors $u(x)$, as $x$ varies, span $\C^r$, and so
do the vectors $v(x)$. Equivalently, both projective kernel images
span $\mathbb P^{r-1}$.
\item[(H3)] Every linear vector field annihilating $P$ has
$\delta x=T(x)$, with no $S$-to-$x$ part.
\end{enumerate}
Condition (H3) requires the coefficient directions to be preserved
by the seed stabilizer. In the preceding families, these directions
were determined by the singular or multiplicity locus. No hypothesis
on the padded stabilizer is imposed.
Hypothesis (H1) and differentiation of $\det A=0$ give
\begin{equation}\label{eq:general-annihilator}
 \{D\in\Mat_r:\tr(C(x)D)=0\text{ for all }x\}=L.
\end{equation}
Indeed this space contains $L$ and intersects $K$ trivially.
Let $\pi:\Mat_r\to K$ be projection along $L$; thus
$\tr(C\pi(D))=\tr(CD)$.

Define the left and right linear syzygy spaces by
\[
 \Sigma_L=\{b:L\to\C^r\text{ linear}:v(x)^Tb(x)=0\},\qquad
 \Sigma_R=\{c:L\to(\C^r)^*\text{ linear}:c(x)u(x)=0\},
\]
Put $s_L=\dim\Sigma_L$ and $s_R=\dim\Sigma_R$, and define the
excess syzygy dimension
\[
 \epsilon=s_L+s_R-2r.
\] The columns and rows of $A$ provide $r$
independent members of the respective spaces. For otherwise $A$
would have a common kernel or cokernel vector, contrary to (H2).
Consequently $s_L,s_R\ge r$ and $\epsilon\ge0$.
The equality $\epsilon=0$ holds precisely when all linear syzygies
are constant linear combinations of these columns or rows.
Equivalently, adjoining a column or row linear in the same $x$
variables without increasing generic rank gives only a redundant
extension. This is the linear-syzygy condition related to
unexpandability in~\cite[Section~2.3]{HL}; no assertion about broader
conventions for simultaneous extensions or additional variables is needed.

The other term in the codimension formula is the dimension of the
seed stabilizers that extend to the padded polynomial. The extension
condition is an identity involving only the seed.

By (H3), every seed stabilizer has the form
$\delta x=T(x)$, $\delta S=F(x)+R(S)$. Let $\mathfrak h$ be the
subspace for which there are constant $E,F_0\in\Mat_r$ satisfying
\begin{equation}\label{eq:general-lift}
 dC_x[T(x)]+F_0C(x)+C(x)E=0,
 \qquad b_{\mathrm{lift}}=\dim\mathfrak h.
\end{equation}
For an element of $\mathfrak h$, hypothesis (H1) forces
$R(S)=\pi(ES+SF_0)$; the remaining $F(x)$ satisfies
$\tr(C(x)F(x))=0$. All pure $x$-to-$S$ seed stabilizers belong to
$\mathfrak h$.

\begin{theorem}\label{thm:general-padding}
Assume (H1)--(H3). For $k\ge1$, set $n=r+k$ and
\begin{equation}\label{eq:general-padded}
 Q_k=\det Z\,P(x,S)-\tr(C(x)U\adj ZV),
 \qquad U\in\Mat_{r,k},\ V\in\Mat_{k,r},\ Z\in\Mat_k.
\end{equation}
Then $Q_k$ is concise in $n^2$ variables, belongs to the determinant
boundary, and
\begin{align}
 \dim\Stab(Q_k)
 &=b_{\mathrm{lift}}+2k^2+2rk+k(s_L+s_R),
 \label{eq:general-stabilizer}\\
 \operatorname{codim}_{\overline{\Orb_n}}\overline{\GL_{n^2}\cdot Q_k}
 &=b_{\mathrm{lift}}-2r^2+2+k\epsilon.
 \label{eq:general-codimension}
\end{align}
Here $\Orb_n$ denotes the affine orbit of $\det_n$.
\end{theorem}

\begin{proof}
The coefficient of $t$ in
$\det\left(\begin{smallmatrix}A+tS&U\\tV&Z\end{smallmatrix}\right)$
is $Q_k$. For $t\ne0$ its entries parametrise $\Mat_n$ invertibly,
so the normalized limit lies in $\overline{\Orb_n}$.
Directional derivatives distinguish $Q_k$ from the determinant
orbit. Every nonzero constant
directional derivative of $\det_n$ is irreducible. To see this, for $D\ne0$ restrict
$f_D(X)=\tr(\adj(X)D)$ to $\GL_n$:
\[
 f_D(X)=\det(X)\tr(X^{-1}D).
\]
Inversion identifies its zero locus there with the nonempty open
part of a linear hyperplane, hence an irreducible reduced
hypersurface. An additional component would have to be the
determinant hypersurface, impossible since $\deg f_D=n-1$;
reducedness on $\GL_n$ also excludes a repeated factor.
But a nonzero $S$-direction of $Q_k$ has derivative
$\det Z\,\tr(CD)$, a product of two positive-degree nonzero factors.
Thus $Q_k\notin\Orb_n$.

For conciseness, a constant $x$-directional derivative has $x$-degree
$r-2$, whereas all other directions have degree $r-1$.
Setting $U=V=0$ and using (H1) forces the $x$-direction to vanish.
The coefficient involving $S$ then forces the $Z$-direction to
vanish. The coefficient linear in $V$, at $Z=I_k$, gives
$v(x)^TU_0Vu(x)=0$ for all $x,V$, forcing $U_0=0$ by (H2).
Similarly $V_0=0$, and then (H1) forces the $S$-direction to vanish.

The stabilizer calculation uses the same sequence of projections as
in the odd-skew case. We determine the exact dimension of each
image and kernel. Cofactor independence also applies for $k=1$,
when the unique cofactor is $1$.

\emph{Step 1: preservation of the coefficient directions.}
Let $\delta Q_k=0$. The part
$\delta x=C_S(S)+C_U(U)+C_V(V)+C_Z(Z)$ has $x$-degree $r-2$
and therefore satisfies a separate identity.
At $U=V=0$, its $S^2$ part says that the seed vector field
$\delta x=C_S(S)$, $\delta S=0$ annihilates $P$.
Hypothesis (H3) gives $C_S=0$. Its $SZ$ part and (H1) give
$C_Z=0$. Next set $V=0$, and then $U=0$, to obtain $C_U=C_V=0$.
Hence every padded stabilizer has $\delta x=T(x)$.

\emph{Step 2: the action on $Z$.}
Terms of $x$-degree $r$ involving $S$ force $\delta Z(x)=0$;
terms quadratic in $S$ force $\delta Z(S)=0$.
The $SU$ terms are
\begin{equation}\label{eq:general-mixed}
 P(x,S)\tr(\adj Z\,\delta Z(U))
   =g(x)v(x)^TU\adj Z\,\delta V(S)u(x).
\end{equation}
For generic $x$, restrict to $v(x)^TU=0$ and choose $S$ with
$P(x,S)\ne0$. The functional of $U$ on the left annihilates this
kernel for every such $x$. The annihilators corresponding to two
nonproportional values of $v$ are
$v_1\otimes(\C^k)^*$ and $v_2\otimes(\C^k)^*$, with zero
intersection. Thus cofactor independence gives $\delta Z(U)=0$.
Equation~\eqref{eq:general-mixed} and (H2) then give
$\delta V(S)=0$. Similarly $\delta Z(V)=\delta U(S)=0$.
We have $\delta Z=J(Z)$, and the coefficient of $S$ gives
$D(\det)_Z[J(Z)]=c\det Z$.
The determinant similitude description gives
$J(Z)=L_ZZ+ZR_Z$, an image of dimension $2k^2-1$.
Every such image occurs: with $c=\tr L_Z+\tr R_Z$, take
\[
 \delta Z=L_ZZ+ZR_Z,\quad
 \delta U=UR_Z-cU,\quad \delta V=L_ZV,\quad
 \delta S=-cS,\quad\delta x=0.
\]
Direct differentiation of $\det Z$ and $\adj Z$ verifies this lift.

\emph{Step 3: the kernel of the $Z$-projection.}
Set $\delta Z=0$. The induced pair
$(\delta x,\delta S(x)+\delta S(S))$ is a seed stabilizer.
The $x$-linear pieces of $\delta U,\delta V$ belong independently
to $\Sigma_L$ in each column and $\Sigma_R$ in each row; they
are genuine stabilizers and contribute $k(s_L+s_R)$ dimensions.
Subtract them. Terms involving only $Z$ give $\delta S(Z)=0$.
The terms linear in $V$ give, on invertible $Z$,
\[
 \delta S(V)=\pi\bigl(\delta U(Z)Z^{-1}V\bigr).
\]
The map $B\mapsto[V\mapsto\pi(BV)]$ is injective: otherwise
$v(x)^TBVu(x)=0$ for every $x,V$, and (H2) forces $B=0$.
Thus $\delta U(Z)Z^{-1}=B_0$ is constant, giving the stabilizers
\[
 \delta U=B_0Z,\quad\delta S=\pi(B_0V);
 \qquad
 \delta V=ZC_0,\quad\delta S=\pi(UC_0),
\]
with $2rk$ parameters in total. Subtract these as well.
Terms quadratic in $V$ give, by cofactor independence,
$(v^T\delta U(V))_a(Vu)_b=0$ for all $a,b$.
The polynomial ring is a domain and (H2) holds, so
$\delta U(V)=0$. Similarly $\delta V(U)=0$.

Only $\delta U=P_U(U)$ and $\delta V=P_V(V)$ remain in these
blocks. The $UV$ equation, after equating cofactors of $Z$, is
the matrix identity
\begin{equation}\label{eq:general-uv}
 VdC_x[T(x)]U+VC(x)P_U(U)+P_V(V)C(x)U=0.
\end{equation}
Write the $j$th output column of $P_U$ as
$\sum_\ell E_{j\ell}U_\ell$.
In entry $(i,j)$ of~\eqref{eq:general-uv}, a term involving
$U_\ell$ with $\ell\ne j$ occurs only as
$V_iCE_{j\ell}U_\ell$.
Thus $CE_{j\ell}=0$ for all $x$, and the spanning $v$-image
forces $E_{j\ell}=0$. Dually $P_V$ acts separately on each row.
Write $P_U(U)_j=E_jU_j$ and $P_V(V)_i=V_iF_i$; then
\[
 dC_x[T(x)]+CE_j+F_iC=0\qquad\text{for every }i,j.
\]
Taking differences and using the two spanning kernel images
gives $E_j=E$ and $F_i=F_0$, independent of the indices.
This is precisely~\eqref{eq:general-lift}.
Conversely every seed element satisfying that identity lifts by
$\delta U=EU$, $\delta V=VF_0$ and $\delta Z=0$.
Hence the seed projection has image exactly $\mathfrak h$.

\emph{Step 4: the residual scalar action.}
After removing the $k(s_L+s_R)$ syzygies and $2rk$ shears,
the kernel of the seed projection satisfies $F_0C+CE=0$.
Since $C=guv^T$, every $u(x)$ is an eigenvector of $F_0$ and
every $v(x)$ an eigenvector of $E^T$.
Each projective kernel-image closure is irreducible.
An irreducible spanning set of eigenvectors lies in one of the
finitely many eigenspaces, which must be the entire space.
Thus $F_0=-\lambda I_r$ and $E=\lambda I_r$.
The residual kernel is one-dimensional, with
$\delta U=\lambda U$, $\delta V=-\lambda V$.
The successive images and kernels therefore have total dimension
\[
 (2k^2-1)+b_{\mathrm{lift}}
       +k(s_L+s_R)+2rk+1,
\]
proving~\eqref{eq:general-stabilizer}. Subtracting the determinant
stabilizer dimension $2n^2-2$ gives~\eqref{eq:general-codimension}.
\end{proof}

Suppose the seed orbit closure is a component, and let $\eta$ be
the codimension of the liftable seed stabilizers. The orbit
codimension after padding is $1+k\epsilon-\eta$, with
$0\le\eta\le\epsilon$. Since $\eta$ is independent of $k$,
componenthood for $k=2$ is equivalent to componenthood for every
positive padding.

\begin{corollary}\label{cor:general-syzygy-criterion}
Assume additionally that the seed orbit closure is a determinant-boundary
component. Put
$\eta=2r^2-1-b_{\mathrm{lift}}$. Then $0\le\eta\le\epsilon$ and
\[
 \operatorname{codim}_{\overline{\Orb_n}}\overline{\GL_{n^2}\cdot Q_k}
       =1+k\epsilon-\eta.
\]
The following are equivalent:
\begin{enumerate}\setlength{\itemsep}{0pt}
\item $Q_2$ represents a boundary component;
\item $s_L=s_R=r$;
\item every $Q_k$, $k\ge1$, represents a boundary component.
\end{enumerate}
If all seed stabilizers lift, the codimension is $1+k\epsilon$.
\end{corollary}
\begin{proof}
The seed stabilizer has dimension $2r^2-1$, so $\eta\ge0$.
Theorem~\ref{thm:general-padding} gives the displayed formula.
Since $Q_1$ is in the boundary, its codimension is at least one,
giving $\eta\le\epsilon$.
The determinant boundary is pure of codimension one
\cite[Corollary~5.4.1]{Chen}: its stabilizer is reductive, hence its
orbit is affine, and the complement of a dense affine open has
pure codimension one. Thus a contained orbit closure is a
component exactly when its codimension is one.
If $\epsilon=0$, necessarily $\eta=0$, and every padding has
codimension one. Conversely componenthood of $Q_2$ gives
$2\epsilon=\eta\le\epsilon$, forcing $\epsilon=0$.
\end{proof}

For the odd-skew seed, the Pfaffian kernel syzygies give
$s_L=s_R=r$, and the intrinsic coefficient-space recognition
already established verifies (H3). The corollary recovers all its
paddings from the seed component theorem. The next example shows
why the syzygy condition cannot be omitted.

\subsection{A component whose positive paddings are not components}

The following seed defines a boundary component and has excess
syzygy dimension two. Its orbit codimension therefore increases
by two for each added row and column.

Consider the known second concise $4\times4$ seed
\cite[Section~8.4 and Proposition~8.4.1]{Hut},
\[
 A=\begin{pmatrix}
 a&b&0&0\\0&0&a&b\\c&0&d&0\\0&c&0&d
 \end{pmatrix},\qquad
 u=\begin{pmatrix}-bd\\ad\\bc\\-ac\end{pmatrix},\qquad
 v=\begin{pmatrix}c\\d\\-a\\-b\end{pmatrix}.
\]
Then $\adj A=uv^T$, and both kernel images span $\C^4$.
Write $E=\langle a,b\rangle$ and $F=\langle c,d\rangle$.
The coefficients of $P=v^TSu$, viewed as a linear polynomial in
$S$, are all cubic monomials containing two variables from one
of $E,F$ and one from the other. In tensor notation their span is
\[
 W=\Sym^2E\otimes F\ \oplus\ E\otimes\Sym^2F,
 \qquad\dim W=12,
\]
so $P$ is concise in sixteen variables.
Its multiplicity-at-least-three locus is exactly $x=0$ with
arbitrary $S$: mixed $S,x$ second derivatives include
$a^2,b^2,c^2,d^2$, and the converse follows from $x$-degree three.
This proves (H3).

The four entries of $v$ are independent linear forms. Their usual
pairwise alternating relations give six independent linear syzygies,
so $s_L=6$. The entries of $u$ are the four products of coordinates
of $E$ and $F$. Their syzygies split into
$\bigwedge^2E\otimes F$ and $E\otimes\bigwedge^2F$, each of
dimension two; hence $s_R=4$ and $\epsilon=2$.
All seed stabilizers lift, as can be checked directly.
The $x$-to-$S$ stabilizers are the kernel of the surjective
multiplication map
\[
 (E\oplus F)\otimes W\longrightarrow
 \Sym^3E\otimes F\ \oplus\ \Sym^2E\otimes\Sym^2F
       \ \oplus\ E\otimes\Sym^3F.
\]
Its dimension is $48-(8+9+8)=23$.
The remaining $x$-actions preserve $W$: their off-diagonal blocks
would produce forbidden $E^3$ or $F^3$ terms, so they are exactly
$\mathfrak{gl}(E)\oplus\mathfrak{gl}(F)$, of dimension eight.
Their action on both $u$ and $v$ is linear in the corresponding
four coordinates, so they satisfy~\eqref{eq:general-lift}.
Thus $b_{\mathrm{lift}}=23+8=31$, and $\eta=0$.
For every $k\ge1$, with $n=4+k$, the exact conclusions are
\[
 \dim\Stab(Q_k)=2n^2-1+2k,
 \qquad
 \operatorname{codim}_{\overline{\Orb_n}}\overline{\GL_{n^2}\cdot Q_k}
       =2k+1.
\]
Consequently no positive padding of this concise seed component
is a boundary component. This is a second application of the codimension formula.

%% file: affine_planes.tex
\section{Bounds from affine level sets}
\label{sec:growing-dc-gap}

A nonzero level set of a polynomial can contain large affine spaces.
Under an injective affine determinantal representation, each such
space maps to an affine space of invertible matrices. Classical
bounds on the dimension of matrix spaces therefore give lower bounds
for exact determinantal complexity. This method does not require
free cokernels and applies to all three component families.

We write an affine space as $a+T$, where $T$ is its linear space of
directions. Its dimension is $\dim T$.

\subsection{Affine level sets and nilpotent matrix spaces}

We use Gerstenhaber's theorem~\cite{Gerstenhaber}: a complex linear
space of nilpotent $m\times m$ matrices has dimension at most
$\binom m2$. Recall that a matrix is nilpotent if some power of it
is zero, equivalently if all of its eigenvalues are zero.
The corresponding bounds for affine spaces of invertible matrices
and their use with injective affine matrix maps appear in
Meshulam~\cite{Meshulam}. The result below is this classical
obstruction, stated in the form needed for our families.

\begin{theorem}\label{thm:affine-level-bound}
Let $f$ be a nonzero polynomial concise in $N$ variables. Suppose
that an affine space $a+T\subseteq\C^N$ satisfies
$f|_{a+T}=\gamma$ for a nonzero constant $\gamma$.
Then every affine-linear determinantal representation of $f$ of size
$m$ satisfies
\[
 \dim T\le\binom m2.
\]
\end{theorem}

\begin{proof}
Write the representation as $f(x)=\det(M_0+L(x))$, with $L$ linear.
First, $L$ must be injective. If $L(h)=0$, then the representing
matrix, and hence $f$, is unchanged under every translation by $th$.
Conciseness says that $f$ has no such nonzero constant direction, so
$h=0$.

The matrix $M_a=M_0+L(a)$ is invertible because
$f(a)=\gamma\ne0$. Normalize the representation by multiplying
by $M_a^{-1}$. For $h\in T$ and every scalar $t$,
\[
 \det\bigl(I_m+tM_a^{-1}L(h)\bigr)
 =\frac{f(a+th)}{f(a)}=1.
\]
This identity says that the characteristic polynomial of
$M_a^{-1}L(h)$ is $z^m$. Thus every matrix in
$M_a^{-1}L(T)$ is nilpotent. This linear space has dimension
$\dim T$, because $L$ is injective. Gerstenhaber's theorem supplies
the stated bound.
\end{proof}

The use of conciseness is to make the linear map $L$ injective.
Without this hypothesis, the proof bounds the dimension only after
quotienting by constant directions of $f$.

\subsection{Applications to determinant fibers}

For each component construction, fixing the seed variables at a
point of maximal rank leaves a generic determinant after a surjective
linear substitution in the coefficient variables. This gives the
following consequence of the affine-level-set bound.

\begin{corollary}\label{cor:compressed-fiber-gap}
Let $f(x,y)$ be concise in $d+D$ variables, with
$x\in\C^d$ and $y\in\C^D$. Suppose that for some $x_0$ there are
a nonzero constant $u$ and a surjective linear map
$\kappa:\C^D\to\Mat_q$ such that
\[
 f(x_0,y)=u\det_q\kappa(y).
\]
Then every affine-linear representation of $f$ of size $m$ satisfies
\begin{equation}\label{eq:compressed-fiber-exact-bound}
 m(m-1)\ge 2D-q(q+1).
\end{equation}
In particular,
\begin{equation}\label{eq:compressed-fiber-integer-bound}
 \dc(f)\ge
 \left\lceil
 \frac{1+\sqrt{1+8D-4q(q+1)}}{2}
 \right\rceil.
\end{equation}
\end{corollary}

\begin{proof}
A matrix of the form $I_q+U$, with $U$ strictly upper triangular,
has determinant one. Let $\mathfrak u_q$ be the linear space of
these strictly upper triangular matrices; it has dimension
$\binom q2$. Choose $y_0$ with $\kappa(y_0)=I_q$. Then $f$ is
the nonzero constant $u$ on
\[
 \{x_0\}\times\bigl(y_0+\kappa^{-1}(\mathfrak u_q)\bigr).
\]
There are $D-q^2$ directions in $\ker\kappa$ and
$\binom q2$ additional directions mapping to
$\mathfrak u_q$. The displayed affine space therefore has dimension
\[
 D-q^2+\binom q2=D-\frac{q(q+1)}2.
\]
Apply Theorem~\ref{thm:affine-level-bound} and rearrange. Solving the
quadratic inequality in $m$ gives the integer bound. Surjectivity
implies $D\ge q^2$, so its square root is defined.
\end{proof}

The estimate compares the number $D$ of coefficient variables with
the order $q$ of the determinant obtained after fixing the seed.
The kernel of the compression map contributes $D-q^2$ independent
directions to every exact representation of the original concise
polynomial.

\subsection{A bound from partial degree}

The same inequality follows from a partial-degree hypothesis.
Suppose $f$ is concise, $T$ is a $D$-dimensional space of
input directions, $f(a)\ne0$, and
\[
 \deg_t f(a+th)\le q\qquad\text{for every }h\in T.
\]
Then every size-$m$ affine representation satisfies
\[
 D\le\binom m2+\binom{q+1}{2}.
\]
This is equivalent to \eqref{eq:compressed-fiber-exact-bound}.

Normalize the representing matrix at $a$ to $I_m$, as in the proof
of Theorem~\ref{thm:affine-level-bound}. The images of the directions form a
$D$-dimensional linear space of matrices $B$ with
$\deg_t\det(I_m+tB)\le q$. When $0<q<m$, every such matrix has
zero as an eigenvalue of algebraic multiplicity at least $m-q$.
Atkinson's theorem~\cite{AtkinsonZeros}, also recorded in
Loewy--Radwan~\cite[Theorem~2.7]{LoewyRadwan}, bounds the dimension
of this matrix space by
\[
 mq+\frac{(m-q)(m-q-1)}2
 =\binom m2+\binom{q+1}{2}.
\]
For $q=0$, this is Gerstenhaber's nilpotent-space bound. For
$q\ge m$, injectivity gives $D\le m^2$, which already implies
the assertion.

Thus low degree in a large space of directions through a nonzero
value suffices. In our constructions, the smaller determinant also
makes the constant-value affine spaces explicit.

\subsection{The trace-pairing bound}

A short linear-algebra argument gives the slightly different bound
\begin{equation}\label{eq:compressed-fiber-trace-bound}
 m^2\ge 2D-q^2.
\end{equation}
This bound uses the Hessian at a point where the polynomial is
nonzero, rather than at a point of its zero set.

For $f(a)\ne0$, define
\[
 H_{\log f}(a)=
 \frac{\operatorname{Hess}f(a)}{f(a)}
 -\frac{\nabla f(a)\nabla f(a)^T}{f(a)^2}.
\]
This rational expression is the logarithmic Hessian; it requires no
choice of a logarithm. For a representation $f=\det(M_0+L)$, put
$M_a=M_0+L(a)$. Differentiating the determinant twice gives
\[
 H_{\log f}(a)(h,h')
 =-\tr\bigl(M_a^{-1}L(h)M_a^{-1}L(h')\bigr).
\]
The expression on the right is a restriction of the nondegenerate
bilinear form $(B,C)\mapsto\tr(BC)$ on $\Mat_m$.

If a nondegenerate
bilinear form on an $m^2$-dimensional space has rank $s$ on a
$D$-dimensional subspace $W$, then its kernel within $W$ has
dimension $D-s$. That kernel lies in the orthogonal complement
$W^\perp$, which has dimension $m^2-D$. Hence
\[
 D-s\le m^2-D,\qquad\text{or}\qquad m^2\ge2D-s.
\]

Use the coefficient directions at
$a=(x_0,y_0)$ from Corollary~\ref{cor:compressed-fiber-gap}.
Their image has dimension $D$, by conciseness. Since
$\kappa(y_0)=I_q$, the logarithmic Hessian on these directions is
the pullback under $\kappa$ of the nondegenerate trace form on
$\Mat_q$. Its rank is exactly $q^2$. Substituting $s=q^2$ proves
\eqref{eq:compressed-fiber-trace-bound}.

The same reasoning without a restricted direction space gives
\[
 \dc(f)^2\ge2N-\rank H_{\log f}(a)
\]
for every concise $N$-variable polynomial and every $a$ with
$f(a)\ne0$. A small rank on a large space forces a large ambient
matrix space. Both this argument and the constant-value-space argument
use an actual injective matrix representation of $f$. They do not
assert these bounds for border determinantal complexity.

\subsection{Bounds for the three component families}

Each family is concise in $n^2$ variables. At a seed value of maximal
rank, the coefficient variables give the smaller determinant required
by Corollary~\ref{cor:compressed-fiber-gap}. The relevant dimensions
are
\begin{center}
\begin{tabular}{llll}
\toprule
Family & Seed variables $d$ & Coefficient variables $D$ & Small order $q$\\
\midrule
$Q_{r,k}$ & $\binom r2$ & $n^2-\binom r2$ & $k+1$\\
$H_{h,k}$ & $h^2+2$ & $n^2-h^2-2$ & $h+k$\\
$\mathcal K_{p,k}$ & $2p+1$ & $n^2-(2p+1)$ & $c+k$\\
\bottomrule
\end{tabular}
\end{center}
For the first two rows, use the compressed determinants displayed in
the constructions. For the third, use the local determinant formula
\eqref{eq:koszul-local}. The compression map is surjective in each
case. Thus \eqref{eq:compressed-fiber-integer-bound} gives an explicit
integer lower bound for every parameter choice in all three families.
Their border complexity is exactly $n$: the construction gives the
upper bound, and their degree $n$ gives the lower bound.

\paragraph{Odd-skew padding.}
For odd $r\ge3$, $k\ge0$, and $n=r+k$, substitution gives
\begin{equation}\label{eq:odd-growing-gap}
 \dc(Q_{r,k})\bigl(\dc(Q_{r,k})-1\bigr)
 \ge2n^2-r(r-1)-(k+1)(k+2).
\end{equation}
Choose an allowed odd $r$ within a bounded distance of $n/2$, and
put $k=n-r$. The leading term on the right is $\tfrac32n^2$.
Therefore
\[
 \dc(Q_{r,k})\ge\sqrt{\frac32}\,n-O(1),
\]
and
\[
 \dc(Q_{r,k})-\bdc(Q_{r,k})
 \ge\left(\sqrt{\frac32}-1\right)n-O(1).
\]
The additive gap grows linearly even though these seeds have corank
one. This supplies a bound for padded odd-skew forms to which the
stronger free-cokernel theorem was not applied.

\paragraph{The block family.}
For $h\ge2$, $k\ge0$, and $n=3h+k$, the same substitution gives
\begin{equation}\label{eq:block-growing-gap}
 \dc(H_{h,k})\bigl(\dc(H_{h,k})-1\bigr)
 \ge15h^2+10hk+k^2-h-k-4.
\end{equation}
Take $h=\lfloor n/3\rfloor$ and $k=n-3h\in\{0,1,2\}$.
This choice is available for every $n\ge6$. The right side of
\eqref{eq:block-growing-gap} is
\[
 \frac53n^2-\frac13n-\frac23k(k+1)-4.
\]
Consequently
\[
 \dc(H_{h,k})\ge\sqrt{\frac53}\,n-O(1),
\]
and
\[
 \dc(H_{h,k})-\bdc(H_{h,k})
 \ge\left(\sqrt{\frac53}-1\right)n-O(1).
\]
An explicit affine representation in Section~\ref{sec:dc} also gives
$\dc(H_{h,k})\le n+4h$. The gap for this choice of parameters is
therefore bounded both below and above by positive constant multiples
of $n$ for large $n$:
\[
 \dc(H_{h,k})-\bdc(H_{h,k})=\Theta(n).
\]
Thus the additive gap for this family has linear order of growth.

\paragraph{The Koszul family.}
Put $r=\binom{2p+1}{p}$ and $c=\binom{2p}{p-1}$. For every
$k\ge0$, with $n=r+k$, we get
\begin{equation}\label{eq:koszul-growing-gap}
 \dc(\mathcal K_{p,k})\bigl(\dc(\mathcal K_{p,k})-1\bigr)
 \ge2n^2-2(2p+1)-(c+k)(c+k+1).
\end{equation}
For the unpadded sequence $n=r$,
\[
 \frac cn=\frac{p}{2p+1}\longrightarrow\frac12,
 \qquad \frac{2p+1}{n^2}\longrightarrow0.
\]
Hence
\[
 \dc(\mathcal K_{p,0})
 \ge\left(\frac{\sqrt7}{2}-o(1)\right)n,
\]
and
\[
 \dc(\mathcal K_{p,0})-\bdc(\mathcal K_{p,0})
 \ge\left(\frac{\sqrt7}{2}-1-o(1)\right)n.
\]
For this family the kernel-degree result of
Corollary~\ref{cor:koszul-degree-gap} is stronger. The present
argument remains useful because the same simple calculation also
covers every odd-skew padding and the block family.

\subsection{Asymptotics for a fixed seed}

Fix the seed and let the amount of padding grow. Let $d$ be the number of seed
variables and $R$ the degree in those variables. In all three
families, the smaller determinant has order $q=n-R$, while there
are $D=n^2-d$ coefficient variables. The common inequality becomes
\[
 m(m-1)\ge n^2+(2R-1)n-R^2+R-2d.
\]
Compare its right side with the value for $m=n+R-1$. Their difference
is
\[
 2\bigl(n-(R-1)^2-d\bigr).
\]
Thus, if
\begin{equation}\label{eq:fixed-seed-padding-threshold}
 n>(R-1)^2+d,
\end{equation}
exact size $n+R-1$ is impossible. Hence
\begin{equation}\label{eq:fixed-seed-padding-gap}
 \dc(f)\ge n+R\quad\text{when }n>(R-1)^2+d.
\end{equation}
The seed degrees are
\[
 R=r-1\text{ for }Q_{r,k},\qquad
 R=2h\text{ for }H_{h,k},\qquad
 R=\binom{2p}{p}\text{ for }\mathcal K_{p,k}.
\]
This is a lower bound for arbitrary affine representations. It does
not determine the exact complexity or claim the best possible
threshold.

All bounds in this section are applications of classical
nilpotent-space and affine-invertible-space theory~\cite{Gerstenhaber,Meshulam}.
The family-specific ingredient is the smaller determinant left after
fixing the seed. Seed corank alone does not supply the lower bound;
what matters here is the number of coefficient directions compared
with the order of that remaining determinant.

%% file: regular_determinant.tex
\section{Regular determinantal representations}\label{sec:regular-det}

Restriction to a nonzero seed value relates the exact complexity of
a Koszul form to the regular determinantal complexity of a generic
determinant. A representation $g(X)=\det M(X)$ is regular at $X=0$
if $M(0)$ has corank one, and $\rdc(g)$ denotes the least size of such
a representation. This is a condition at one point; the freeness
hypothesis in Theorem~\ref{thm:degree} is a global condition.

\subsection{Restriction to a generic determinant}

For the Koszul family, write
\[
 d_0=2p+1,\qquad r=\binom{2p+1}{p},\qquad
 c=\binom{2p}{p-1},\qquad n=r+k.
\]
Fix a nonzero value of the seed vector $v$. As a polynomial in the
remaining coefficient variables, $\mathcal K_{p,k}$ becomes a nonzero
scalar times the determinant of a $(c+k)\times(c+k)$ matrix. The map
from the coefficient variables to that matrix is onto. Choose a linear
section of this map: this means choosing coefficient values that
produce each desired small matrix, linearly in its entries. Restricting
to this section leaves exactly a generic determinant, up to the
nonzero scalar.

Let $M$ be any exact representation of $\mathcal K_{p,k}$.
Theorem~\ref{thm:koszul-universal-regularity} gives corank one at
every zero of the polynomial. In particular, after the restriction
just described, it has corank one at the zero small matrix. Absorbing
the nonzero scalar by a constant row rescaling gives a regular
representation of the generic determinant. The matrix size has not
increased. Therefore
\begin{equation}\label{eq:rdc-reduction}
 \dc(\mathcal K_{p,k})\ge\rdc(\det_{c+k}).
\end{equation}
The inequality applies to arbitrary exact representations, with no
assumption on their matrix form.

For a generic determinant, regularity at the zero matrix has a special
consequence. The classification recalled in~\cite[Section~4.4]{BL}
leaves three possible rank-one maximal Cohen--Macaulay modules: the
free module and two nonfree classes. For a determinant of order $q$,
each of the latter has a $q$-dimensional fiber at the zero matrix.
A regular representation has a one-dimensional cokernel fiber there,
so, for $q\ge2$, only the free module is possible. The kernel-degree
lemma consequently gives
\[
 \rdc(\det_q)\ge2q-1.
\]
For $q=1$ the same bound is immediate. On the other hand, a known
regular construction has size $(q^3-q)/3+1$~\cite{IL}. Since $c+k$ is a
constant fraction of $n=r+k$, a superlinear lower bound for
$\rdc(\det_q)$ would give one for the Koszul family through
\eqref{eq:rdc-reduction}.

\subsection{A spectral formulation}

There is an equivalent formulation that uses only a linear map between
matrix spaces. Suppose
\[
 \det M(X)=\det_q X,\qquad M(X)=M_0+L(X),
\]
where $M$ has size $m$ and $L$ is linear. At $X=I_q$ the determinant
is one, so $M(I_q)$ is invertible. Normalize by this matrix and define
\[
 B(T)=M(I_q)^{-1}L(T).
\]
The determinant identity at $I_q+tT$ then becomes
\[
 \det(I_m+tB(T))=\det(I_q+tT).
\]
Equivalently, for every $T$,
\[
 \det(\lambda I_m-B(T))
   =\lambda^{m-q}\det(\lambda I_q-T).
\]
Thus $B$ preserves the eigenvalues of $T$, with the additional $m-q$
eigenvalues all zero, counting algebraic multiplicity.

The regularity condition also has a direct meaning. Since
\[
 M(I_q)^{-1}M(0)=I_m-B(I_q),
\]
regularity is equivalent to $I_m-B(I_q)$ having corank one. The
characteristic-polynomial identity says that the eigenvalue $1$ of
$B(I_q)$ has algebraic multiplicity $q$. Corank one says that its
eigenspace is one-dimensional. In Jordan normal form, it must
therefore have one Jordan block, of length $q$.

Conversely, suppose a linear map $B:\Mat_q\to\Mat_m$ satisfies the
characteristic-polynomial identity above, and the eigenvalue $1$ of
$B(I_q)$ has a single Jordan block. Then
\[
 M(X)=I_m-B(I_q)+B(X)=I_m+B(X-I_q)
\]
is regular at zero. Substituting $T=X-I_q$ in the determinant identity
shows that $\det M(X)=\det_q X$.

Thus regular determinantal complexity is equivalently the minimum
$m$ for which a linear map $B$ satisfies this characteristic-polynomial
identity and the stated Jordan-block condition. A superlinear lower
bound for this minimum would give a superlinear bound for the Koszul
family through~\eqref{eq:rdc-reduction}. Such a bound remains open.

%% file: residue_criterion.tex
\section{A criterion for polynomial realizations of rational matrices}

The Koszul construction admits a formula in terms of a rational
projection matrix. We study polynomial realizations of this matrix,
first proving a general criterion in terms of its residue module.
The criterion concerns a prescribed rational matrix; its implications
for scalar determinant representations are discussed below.

Let $P=X/q$, where $X$ is a polynomial matrix and $q$ is irreducible.
A polynomial realization is an expression
\[
 P=A+CD^{-1}B
\]
with polynomial matrices of compatible sizes and $D$ square.
The determinant of $D$ controls the denominator of $CD^{-1}B$.
We prescribe this determinant, but impose no bound on the size of
$D$ or on the degrees of the entries. These requirements are weaker
than those for an affine determinantal representation of a scalar
polynomial.

\subsection{The residue module}

Reduce $X$ modulo $q$. Its columns now live on the hypersurface $q=0$,
and their span records which output directions occur in the numerator
at that hypersurface. Write this image as $\im\bar X$. Away from
points where its rank drops, it is a vector bundle: locally it is a
fixed number of independent columns varying algebraically with the
point.

We will use its \emph{reflexive extension} $E$. Concretely, take the
double dual of the image module and regard it inside the ambient free
module. Over a normal ring this is the extension determined by what
happens away from subsets of codimension at least two. Thus isolated
exceptional behavior in higher codimension does not change $E$.
The relevant question is whether $E$ has full depth at those exceptional
points. The standard name for this condition is maximal
Cohen--Macaulay (MCM), as defined earlier. It is exactly the condition
needed for a short free resolution over the ambient polynomial ring.

\begin{theorem}\label{thm:residue}
Let $T=\C[v_1,\ldots,v_d]$, let $q\in T$ be irreducible, and suppose
$S=T/(q)$ is normal. Let $X$ be an $r\times s$ polynomial matrix and
put $P=X/q$. Suppose $\bar X$ has rank $a>0$ and constant rank $a$
away from a closed subset $Z\subset\Spec S$ of codimension at least two.
Let $E\subset S^r$ be the reflexive extension of $\im\bar X$.
The following are equivalent:
\begin{enumerate}
\item There are polynomial matrices $A,B,C,D$ of compatible sizes,
with $D$ square, such that
\[
 P=A+CD^{-1}B,\qquad \det D=u q^a,\quad u\in\C^\times.
\]
\item The module $E$ is maximal Cohen--Macaulay at every local ring of $S$.
\end{enumerate}
The matrices need not be homogeneous or affine linear. The equivalence
is an existence criterion, with no upper bound on the size of $D$.
\end{theorem}

The exponent $a$ is the minimum permitted by the generic residue
rank. At a general point of $q=0$, at least $a$ directions contribute
a pole. Equality in the exponent forces all of these poles to be
simple. The proof uses this observation and the extension of the
residue module across the exceptional set.

\begin{proof}
\emph{Step 1: Smith exponents.}
Suppose the realization in part (1) exists. Localize the polynomial
ring at the prime ideal $(q)$. This is a discrete valuation ring: it
measures the power of $q$ dividing a polynomial. Smith normal form
writes $D$, up to invertible row and column operations over this local
ring, with diagonal entries $q^{e_i}$ times units. The integers
$e_i\ge0$ are its Smith exponents, and
\[
 \sum_i e_i=a
\]
because $\det D=u q^a$.

Let $t$ be the number of positive exponents. Only these $t$ directions
can contribute singular terms to $CD^{-1}B$, so its singular part has
rank at most $t$. Multiplying that part by $q$ makes it regular: the
whole expression $q(P-A)=X-qA$ is regular, and the Smith blocks with
exponent zero also contribute regular terms. Reducing modulo $q$
therefore gives a matrix of rank at most $t$. Its rank is $a$, the
rank of $\bar X$. We obtain $a\le t\le\sum_i e_i=a$.
Consequently there are exactly $a$ positive exponents, and all are one.

\emph{Step 2: the cokernel of $D$.}
Set $N=qD^{-1}$. This is a polynomial matrix. Indeed the only possible
denominator factor of $D^{-1}$ is $q$, while Step 1 shows that
multiplication by $q$ removes its poles at $(q)$. We have
\[
 DN=ND=qI.
\]
It follows that $F=\coker_T D$ is annihilated by $q$, so it is an
$S$-module. Its rank over $S$ is $a$. The exact sequence
\[
 0\longrightarrow T^m\xrightarrow{D}T^m\longrightarrow F
 \longrightarrow0
\]
shows, by the depth formula over the regular ring $T$, that $F$ is
MCM over $S$.

\emph{Step 3: identification with the residue module.}
The map
\[
 F\longrightarrow S^r,\qquad [z]\longmapsto CNz\bmod q
\]
is well-defined, because $CND=qC$. Its image contains $\im\bar X$:
multiplying the realization by $q$ gives $X=qA+CNB$. Both $F$ and
$\im\bar X$ have generic rank $a$, so this map is generically
injective. An MCM module over our normal domain is torsion-free;
hence the map is injective everywhere.

Away from $Z$, the constant-rank image of $\bar X$ is a subbundle of
$S^r$. In particular the quotient of $S^r$ by that image is locally
free, and therefore torsion-free. The quotient of the image of $F$
by $\im\bar X$ is a submodule of this torsion-free quotient. But it
has rank zero, so it is torsion. It must vanish. Thus $F$ and
$\im\bar X$ agree away from $Z$. Both $F$ and $E$ are reflexive,
and the omitted set has codimension at least two. They therefore
agree globally. This proves that $E\simeq F$ is MCM.

\emph{Step 4: construction from an MCM residue module.}
Conversely, assume part (2). The local Auslander--Buchsbaum formula
over $T$ gives a projective resolution of length one. Quillen--Suslin
makes its projective terms free. Thus we can write
\[
 0\longrightarrow T^m\xrightarrow{D}T^m\longrightarrow E
 \longrightarrow0.
\]
The two ranks are equal because $E$ is torsion as a $T$-module. Its
support is $q=0$, and its generic length there is $a$. Hence
$\det D=u q^a$ for a nonzero constant $u$. Since $qE=0$, the matrix
$N=qD^{-1}$ is polynomial and satisfies $DN=ND=qI$.

It remains to recover the desired numerator $X$. Lift the inclusion
$E\hookrightarrow S^r$ to a polynomial map $H:T^m\to T^r$. The
composition $HD$ is zero modulo $q$, so define
\[
 C=HD/q.
\]
Then $CN=H$. Lift the columns of $\bar X$, viewed as elements of $E$,
through the surjection $T^m\to E$; collect these lifts in a polynomial
matrix $B$. The matrix $X-CNB$ is now divisible by $q$. Define
\[
 A=(X-CNB)/q.
\]
Since $N=qD^{-1}$, these definitions give
$A+CD^{-1}B=X/q$. All four matrices are polynomial, as required.
\end{proof}

\section{The residue obstruction for the Koszul projection}

We apply the criterion to the unpadded Koszul form. Equip
$V=\C^{2p+1}$ with its standard nondegenerate symmetric bilinear form,
and write
\[
 p\ge2,\qquad q(v)=\sum_i v_i^2,\qquad
 r=\binom{2p+1}{p},\qquad
 a=\binom{2p-1}{p-1}=\tfrac12\binom{2p}{p}.
\]
There are two basic maps on exterior powers. Exterior multiplication
by $v$, denoted $K(v)$, sends $\Lambda^{p-1}V$ to $\Lambda^pV$.
Contraction by $v$, denoted $C_0(v)$, goes in the opposite direction.
Their product has the identities
\[
 X(v)=K(v)C_0(v),\qquad X(v)^2=q(v)X(v),\qquad
 P(v)=X(v)/q(v).
\]
For $q(v)\ne0$, the last identity says that $P^2=P$: it is a
projection. Its image is the kernel space
$U_v=v\wedge\Lambda^{p-1}V$ used in the Koszul construction.

\subsection{The projection formula}

Normalize the unpadded Koszul form by a fixed nonzero scalar. Then
\begin{equation}\label{eq:scalar}
 f(v,S)=q(v)^a\det_r\bigl(I_r+P(v)(S-I_r)\bigr).
\end{equation}
In particular, the value at $S=I_r$ is $q^a$, and the dependence on
the coefficient variables is expressed through $P$.

Here we may choose the coefficient complement to contain $I_r$.
Alternatively let $S$ range over all of $\Mat_r$: adding a member
of the original seed space to $S$ does not change the formula.
To verify \eqref{eq:scalar}, first work where $q(v)\ne0$. There is
an orthogonal splitting
\[
 \Lambda^pV=U_v\oplus U_v^\perp.
\]
On these summands $P$ is respectively the identity and zero. The
determinant on the right of \eqref{eq:scalar} is therefore the
determinant of the restriction of $S$ to $U_v$.
The middle Koszul map vanishes on $U_v$. On its complementary summand,
which has dimension $2a$, its determinant is a fixed nonzero scalar
times $q(v)^a$. This last assertion follows by taking
$v=t e_{2p+1}$ and then using equivariance under the orthogonal group.
The leading coefficient of $\det(A(v)+tS)$ is the product of these
two determinants. Thus the equality holds on $q\ne0$, and hence as
a polynomial identity everywhere.

\begin{proposition}\label{prop:koszul-no}
For $p\ge2$, there is no polynomial realization
\[
 P=A+CD^{-1}B,\qquad \det D=u q^a,
\]
regardless of the size of the square matrix $D$.
\end{proposition}

\begin{proof}
We check the hypothesis and the obstruction in
Theorem~\ref{thm:residue}. The only exceptional point will be the
vertex $v=0$ of the quadric cone.

\emph{The residue has constant rank $a$ away from the vertex.}
Let $v\ne0$ satisfy $q(v)=0$. Choose a hyperbolic partner $w$ and
a nondegenerate complement $W$ of dimension $2p-1$. In the resulting
exterior-power decomposition, $KC_0$ maps
$w\wedge\Lambda^{p-1}W$ isomorphically onto
$v\wedge\Lambda^{p-1}W$ and kills the other summands. Its rank is
therefore $\binom{2p-1}{p-1}=a$.

\emph{The associated bundle has an intermediate cohomology group.}
Pass from the punctured cone to the smooth projective quadric
\[
 Q=\{q=0\}\subset\mathbf P^{2p}.
\]
Let $L=\mathcal O_Q(-1)$ be the line spanned by $v$, and let
$\mathcal H=L^\perp/L$. The image bundle just computed is
\[
 \mathcal E=L\otimes\Lambda^{p-1}\mathcal H
 =\Lambda^{p-1}T_Q(-p)=\Omega_Q^p(p-1).
\]
The identities use $T_Q=L^*\otimes\mathcal H$ and
$\det T_Q=\mathcal O_Q(2p-1)$. Thus the bundle can be expressed in
terms of differential forms on the quadric, whose cohomology supplies
the required test.

In fact
\[
 H^p\bigl(Q,\mathcal E(1-p)\bigr)
   =H^p(Q,\Omega_Q^p)\ne0.
\]
One way to see nonvanishing is to take the $p$-th power of the
hyperplane class. If it were zero, its product with the remaining
hyperplane powers would also be zero. That product has top degree
two on a quadric. Since $\dim Q=2p-1$ and $p\ge2$, the cohomology
degree $p$ is strictly between zero and the dimension.

\emph{This is a depth defect at the vertex.}
Let $E=(\im\bar X)^{**}$ be the reflexive residue module on the
affine cone, and let $\mathfrak m$ be the vertex ideal. Its associated
sheaf on $Q$ is $\mathcal E$. The standard correspondence between
sheaf cohomology and graded local cohomology gives
\[
 H_{\mathfrak m}^{p+1}(E)_{1-p}
 \simeq H^p(Q,\Omega_Q^p)\ne0,
 \qquad p+1<2p=\dim S.
\]
Full depth would force these local cohomology groups to vanish below
$\dim S$. Hence $E$ is not MCM. The realization criterion now proves
the proposition.
\end{proof}

The restriction $p\ge2$ is necessary. When $p=1$, the projection is
$P=vv^T/q$, and the scalar matrix $D=[q]$ already gives a realization.

\subsection{Realizations with a larger denominator}

The exponent in Proposition~\ref{prop:koszul-no} is essential.
With a larger exponent of $q$ in $\det D$, the same projection has
an explicit realization of size linear in $r$.

Put
\[
 b=\binom{2p+1}{p-1},\qquad h=2^p,\qquad
 N=h\lceil b/h\rceil.
\]
Choose complex Clifford matrices $\Gamma_1,\ldots,\Gamma_{2p+1}$
of size $h$, with
\[
 \Gamma_i\Gamma_j+\Gamma_j\Gamma_i=2\delta_{ij}I_h.
\]
For example, these follow from tensor products of the Pauli matrices.
The anticommutation relation is the only property we use: it makes
$\bigl(\sum v_i\Gamma_i\bigr)^2=q(v)I_h$. Repeating this pencil
in diagonal blocks gives a size-$N$ linear matrix $\Gamma(v)$ with
\[
 \Gamma(v)^2=q(v)I_N.
\]
Let $J:\C^b\hookrightarrow\C^N$ be the coordinate inclusion.
Consider the affine-linear matrix
\[
 \mathcal M(v,S)=
 \begin{pmatrix}
 I_r&K(v)J^T&0&0\\
 0&\Gamma(v)&-I_N&0\\
 0&0&\Gamma(v)&JC_0(v)\\
 S-I_r&0&0&I_r
 \end{pmatrix}.
\]
The inverse of
the middle triangular block has upper-right block
$\Gamma(v)^{-2}=q(v)^{-1}I_N$. Eliminating that block leaves the
Schur complement
\[
 \begin{pmatrix}I_r&-P\\S-I_r&I_r\end{pmatrix}.
\]
The eliminated block has determinant $q^N$. We therefore obtain
\begin{equation}\label{eq:quadratic-factor-upper}
 \det\mathcal M(v,S)
 =q(v)^N\det_r\bigl(I_r+P(v)(S-I_r)\bigr)
 =q(v)^{N-a}f(v,S).
\end{equation}
The calculation first holds where $q\ne0$; both sides are
polynomials, so it holds everywhere. The size is $2r+2N<6r$, because
$b<r$ and $h\le r$. Also $N\ge b>a$, so the added power of $q$ is
strictly positive.

This gives a linear-size representation of $q^{N-a}f$. It does not
give a way to remove the extra factor. The same rational projection
$P$ has been realized with $\det D=q^N$, whereas
Proposition~\ref{prop:koszul-no} excludes $\det D=q^a$.

\subsection{Relation to scalar determinantal representations}

An exact determinant for the scalar polynomial $f$ need not arise
from this particular projection formula. The proposition rules out
one prescribed rational-matrix realization; it does not rule out
other exact representations of $f$.

Exact scalar representations of $f$ exist at finite size. Therefore,
no normal-form theorem can convert every such representation into
a realization excluded by Proposition~\ref{prop:koszul-no}. The next
section constructs small scalar matrices with the same base value
and first coefficient derivatives, without realizing the prescribed
projection in that form.

%% file: local_limits.tex
\section{First-order coefficient data}

We compare the residue obstruction with the information obtained by
differentiating an arbitrary scalar determinant representation at
$S=I$. We then construct an affine matrix of linear size whose
determinant has the same base value and all first coefficient
derivatives as the Koszul form. For $p\ge3$, the construction can
also preserve the degree in each group of variables.

Throughout this section, use the unpadded Koszul parameters
\[
 d_0=2p+1,\qquad r=\binom{2p+1}{p},\qquad
 a=\binom{2p-1}{p-1},\qquad c=\binom{2p}{p-1},\qquad r=2a+c.
\]
As before, $q(v)=\sum_i v_i^2$, $X=KC_0$, and $P=X/q$.
The scalar identity \eqref{eq:scalar} gives
$f(v,I)=q(v)^a$. We first construct a regular representation of this
base value and then incorporate the coefficient derivatives.

\subsection{Regular representations of quadratic powers}

Let $h=2^p$ and $N=h\lceil a/h\rceil$. Use a block repetition of
the Clifford pencil from the preceding section, so that
\[
 \Gamma(v)^2=q(v)I_N.
\]
Let $J:\C^a\hookrightarrow\C^N$ be the coordinate inclusion, and
let $T_a$ be the nilpotent Jordan block of order $a$. Define
\begin{equation}\label{eq:qpower}
 M_a(v)=
 \begin{pmatrix}
 -T_a&J^T\Gamma(v)\\
 -\Gamma(v)J&I_N
 \end{pmatrix}.
\end{equation}
Eliminate the identity block. Its Schur complement is
$qI_a-T_a$, because $J^TJ=I_a$ and $\Gamma^2=qI_N$. Therefore
\[
 \det M_a=q^a,\qquad
 \operatorname{size}(M_a)=a+N<2a+h.
\]
Moreover, at every zero of $q$, this Schur complement is $-T_a$,
which has corank one. The full matrix has corank one there as well,
including at the vertex $v=0$.

If $h$ divides $a$, the size is exactly $2a$, equal to the degree of
$q^a$. There is no conflict with the kernel-degree lemma: for $a>1$
the polynomial $q^a$ is reducible, whereas that lemma requires
irreducibility.

For the Koszul values above, the size is linear in $r$. The rounding
overhead is only $O(2^p)=o(r)$. Thus the large multiplicity of the
base value $q^a$ does not, by itself, force an additional factor such
as $\log r$ or $\sqrt{\log r}$ in the matrix size.

\subsection{Derivatives of a scalar determinant representation}

Suppose an arbitrary affine matrix $M(v,S)$ represents the full scalar
polynomial $f(v,S)$. Set $D(v)=M(v,I)$. Since the entries of $M$ are
affine linear, its change in a coefficient direction $T$ is a constant
linear map $L(T)$, independent of $v$.
Differentiating the determinant at $S=I$ gives
\[
 \det D=q^a,\qquad
 \tr\bigl(D^{-1}L(T)\bigr)=\tr\bigl(P(v)T\bigr).
\]
These equalities hold as rational functions where $q\ne0$.
By varying $T$, one recovers the entries of $P$ as linear combinations
of entries of $D^{-1}$. But arbitrary linear combinations of those
entries need not have the factorized form $CD^{-1}B$. The latter
form is the requirement in Theorem~\ref{thm:residue}.

One can see the difference at the generic point of $q=0$. The global
corank-one theorem for $f$ makes $D$ have corank one there. Its Smith
normal form consequently has one positive exponent, equal to $a$.
By contrast, a realization of the prescribed projection with
$\det D=q^a$ would require $a$ positive exponents, all equal to one,
as the proof of Theorem~\ref{thm:residue} showed.

Thus the trace measurements in a scalar representation can cancel
higher-order poles. The residue criterion does not measure the cost
of those cancellations. No lower bound for that cost is established
here.

\subsection{A simultaneous realization of the first derivatives}

For a coefficient direction $T$, write the quadratic polynomial
\[
 G_T(v)=\tr(X(v)T)=v^TH(T)v,
\]
where $H(T)$ is symmetric and depends linearly on $T$.
Equation~\eqref{eq:scalar} says that the required first-order expansion
is
\[
 f(v,I+T)=q^a+q^{a-1}G_T(v)+O(T^2).
\]
Here $O(T^2)$ means terms of total coefficient degree at least two in
$T$; no analytic estimate is intended.

Consider the small affine matrix and scalar factor
\[
 A(v,T)=
 \begin{pmatrix}0&v^T\\-v&I_{d_0}-H(T)\end{pmatrix},
 \qquad l(T)=1+\tr H(T).
\]
Expanding its determinant to first order gives
\[
 \det A=q+G_T-q\tr H+O(T^2),\qquad
 l(T)\det A=q+G_T+O(T^2).
\]
The scalar factor cancels the unwanted trace term.
Multiplying this determinant by the regular $q^{a-1}$ block from
\eqref{eq:qpower} now gives the right base value and the right first
derivative.

We also preserve corank one along the entire base $q=0$. The block
diagonal product has corank two there, because both main blocks are
singular. A constant upper block reduces this corank to one without
changing the determinant. The following choice gives the required
coupling. Take
$T_{a-1}$ with ones on the superdiagonal. On $q=0$, a left kernel
vector of its quadratic-power block has top coordinate $a-1$ equal
to one. A right kernel vector of $A(v,0)$ is $(1,v)^T$, with first
coordinate one. Put a single constant $1$ in the upper block joining
these two coordinates. The pairing of the chosen left and right
kernel vectors through this block is then one. This is exactly the
condition that removes one of the two kernel dimensions.

Including the scalar block $l$, which equals one on the base, gives
an affine matrix with
\[
 \det M(v,T)=q^a+q^{a-1}G_T(v)+O(T^2),
 \qquad \operatorname{size}(M)<2a+d_0+h=O(r).
\]
It has corank one at every point with $q=0$ and $T=0$.
Because $H$ is linear in the whole matrix $T$, this construction
matches every coefficient direction simultaneously.

\subsection{Preservation of bidegree}

The construction can be modified to have the same bidegree as the
Koszul form, while retaining the base value and first derivatives.

Assume $p\ge3$, so that $c\ge d_0$. Choose a linear functional
$\ell$ on coefficient matrices with $\ell(I)=1$. Since
$G_I=cq$, the symmetric matrix representing this quadratic form is
$H(I)=cI_{d_0}$. Define
\[
 \widetilde H(S)=H\bigl(S-\ell(S)I\bigr),\qquad
 \widehat A(v,S)=
 \begin{pmatrix}
 0&v^T\\
 -v&\ell(S)I_{d_0}-\widetilde H(S)
 \end{pmatrix},
\]
and
\[
 \widehat l(S)=\ell(S)+\tr\widetilde H(S).
\]
Use the same constant coupling with a regular $q^{a-1}$ block, and
add $c-d_0$ scalar blocks $\ell(S)$. The resulting determinant is
\[
 g(v,S)=q^{a-1}\det\widehat A(v,S)\,
              \widehat l(S)\,\ell(S)^{c-d_0}.
\]
Its degree in $v$ is $2a$, and its degree in $S$ is $c$, exactly as
for the Koszul form. These two degrees are its \emph{bidegree}.

We check the first derivative as well. Write $S=I+T$ and
$\sigma=\ell(T)$. Homogenizing the preceding first-order construction
changes its quadratic first derivative to
$G_T+(d_0-c)q\sigma$. The first derivative of the additional factor
$\ell^{c-d_0}$ cancels the extra term. Consequently
\begin{equation}\label{eq:imitation}
 g(v,I+T)=q^a+q^{a-1}G_T(v)+O(T^2).
\end{equation}
For the quadratic-power block put $s=a-1$ and
$N=h\lceil s/h\rceil$. The full matrix size is
\[
 (d_0+1)+(s+N)+1+(c-d_0)
 =c+a+1+N\le r+h-1=r+o(r).
\]
The chosen coupling still gives corank one everywhere on $q=0$ at
$S=I$.

The polynomial $g$ need not equal the Koszul form: its higher
coefficient terms may differ. No claim of irreducibility, conciseness,
or corank one away from $S=I$ is made. The construction shows that
the scalar base value, all first coefficient derivatives, the target
bidegree, and corank one along the base can hold simultaneously at
matrix size $r+o(r)$. Hence these data alone do not imply a superlinear
lower bound.

\section{Limitations of two further reductions}

Two further reductions would be relevant to extending the
kernel-degree method: bounding the span of homogeneous pieces of
kernel coordinates, and deducing simultaneous triangularizability
from a homogeneous determinant identity. The following examples show
that neither conclusion follows from the conditions considered here.

\subsection{Homogeneous decompositions of kernel coordinates}

Let
\[
 F=\sum_{i=1}^d y_i x_i^2,\qquad d\ge3.
\]
Start with the size-$2d+1$ affine representation
\[
 \begin{pmatrix}
 0&x^T&0\\
 0&I&-\operatorname{diag}(y)\\
 -x&0&I
 \end{pmatrix}.
\]
Append a determinant-one triangular block that computes the prefix
products $\prod_{i=1}^j(1+x_i)$. This gives a representation of size
$3d+1$ whose kernel along $y=0$ contains the coordinates
$1,x_1,\ldots,x_d$ and those prefix products.

There are only linearly many displayed coordinates. Nevertheless,
if one separates all of them into homogeneous pieces, their span has
dimension
\[
 1+\frac{d(d+1)}2.
\]
Indeed, in each degree $j=1,\ldots,d$, the nonzero homogeneous parts
of the prefix products contribute $d-j+1$ independent polynomials;
summing these dimensions and adding the constant gives the formula.
Thus taking homogeneous pieces can turn a linear coordinate count
into a quadratic one.

The added states can be removed, so this is not a counterexample to
an invariant minimized over all possible representations. It does
show that coefficient purity and factoriality alone do not justify
a bound of $O(m)$ for the span obtained by grading an arbitrary
size-$m$ representation.

\subsection{Homogeneity and simultaneous triangularization}

Consider
\[
 N=\begin{pmatrix}0&x&0\\y&0&x\\0&-y&0\end{pmatrix},
 \quad D=\operatorname{diag}(N,N),\quad
 b^T=(z,0,0,0,0,s),\quad c=(0,0,w,t,0,0)^T.
\]
The direct identities $N^3=0$ and $b^Tc=b^TDc=0$ give
\[
 \det\begin{pmatrix}0&b^T\\c&I_6-D\end{pmatrix}
 =st\,y^2-zw\,x^2.
\]
Thus the determinant is a concise irreducible homogeneous quartic,
represented by an affine matrix of size seven. For example,
irreducibility follows by treating it as a primitive polynomial
linear in $s$; the coefficients $ty^2$ and $-zwx^2$ are relatively
prime.

The two coefficient matrices of $N$ have no common invariant line.
They are nilpotent, and their kernels are distinct coordinate lines,
so no line can be invariant under both. The same absence of a common
invariant line holds for their doubled matrices in $D$. In particular
these coefficient matrices cannot be simultaneously triangularized.
A pure-degree determinant identity therefore does not, by itself,
supply a common invariant flag.

\subsection{Computational verification}

Finite-field calculations over $\mathbb F_{1009}$ check the
quadratic-power determinants and corank-one assertions for
$p=1,2,3,4$. The corresponding pairs $(a,\text{matrix size})$ are
\[
 (1,3),\quad(3,7),\quad(10,26),\quad(35,83).
\]
For $p=3$, the construction matching the first derivative and
bidegree has size $42$. Checks in three independent coefficient
directions verify the base value and the complete derivative in each
of those directions, as well as the bidegree scaling. A nonzero
isotropic point checks corank one on the base. The source bundle
contains the code and its results.

These computations check the explicit matrix identities in small
cases. They do not replace the proofs above or the commutative-algebra
arguments for the main lower bound.

%% file: certificates.tex
\section{Computational verification}\label{sec:certificates}

The source archive contains exact computations verifying the
formulas and symmetry dimensions in several small cases. This appendix
describes the certificates and their verifiers. These computations are
independent checks of the constructions; the theorems for all
parameters are proved by the algebraic arguments in the text.

\subsection{Odd-skew examples}

To measure the infinitesimal orbit of a polynomial $Q$, form the
polynomials $X_j\partial_{X_i}Q$ and list their coefficients in a
matrix. Its rank is the dimension of the orbit. For two padded
odd-skew forms, the resulting matrices and certified ranks are
\begin{center}
\begin{tabular}{rrrrrr}
\toprule
$r$&$k$&$n$&Matrix shape&Certified rank&Minor modulo $101$\\
\midrule
3&2&5&$4682\times625$&576&23\\
5&2&7&$305804\times2401$&2304&55\\
\bottomrule
\end{tabular}
\end{center}
Here a nonzero minor modulo $101$ proves that the same integer minor
is nonzero over $\C$. It therefore gives a rigorous lower bound on
the complex rank. The fact that the form lies on the determinant
boundary gives the matching upper bound, so the rank is determined
exactly.

The archive includes the selected row monomials and column pairs,
the integer polynomial generator, and a verifier using only the Python
standard library. The verifier uses row elimination; the original
certificate search used column elimination. For $n=5$, a separate
calculation expands the full determinant and performs a dense modular
verification as well.

\subsection{Block-space examples}

Two further certificates concern the $h=2$ block seed, written in
coordinates obtained from Huang--Landsberg's case (III) by row
operations. They check both the orbit dimension and dependence on
all variables:
\begin{center}
\begin{tabular}{rrrrrr}
\toprule
$n$&Variables&Terms&Orbit rank&Minor modulo $101$&Partial rank\\
\midrule
6&36&285&1225&79&36\\
7&49&953&2304&65&49\\
\bottomrule
\end{tabular}
\end{center}
``Partial rank'' is the rank of the first partial derivatives. Its
being equal to the number of variables certifies conciseness.
The corresponding minors are $16$ and $85$ modulo $101$.

The independent verifier expands the entire determinant using monomial
exponent vectors. It checks that all coefficients before the claimed
leading coefficient vanish, compares that leading coefficient exactly,
and verifies the chosen orbit minors by row elimination. These checks
are separate from the uniform proof over $\C$.

\subsection{Koszul examples}

For the exterior-algebra construction, exact rational calculations and
finite-field calculations check the constant and degree-one
compression maps. The numbers in the first two cases are
\begin{center}
\begin{tabular}{rrrrr}
\toprule
$p$&Seed size&Constant rank/domain&Linear rank/domain&Quotient relations\\
\midrule
2&10&$95/100$&$301/500$&174\\
3&35&$1218/1225$&$6126/8575$&2400\\
\bottomrule
\end{tabular}
\end{center}
The numerator records the rank of a linear map; the denominator records
the dimension of its domain. The last column counts the remaining
linear relations after passing to the quotient used in the component
proof. These are the same dimensions obtained in that proof for all
$p$ by representation theory. Splitting the calculation into weight
spaces makes the second example manageable: its largest block has
only $140$ columns.

These certificates do not replace the proofs for general parameters,
and they do not check the commutative-algebra arguments used for the
exact-complexity lower bound. Their role is to make the displayed
constructions and several decisive dimension counts independently
reproducible.